\documentclass[preprint,12pt]{elsarticle}
\usepackage[utf8]{inputenc}
\usepackage{amsmath}
\usepackage{amsfonts}
\usepackage{amsthm}
\usepackage{amssymb}
\usepackage{graphicx}
\usepackage[dvipsnames]{xcolor}
\usepackage{bm}
\usepackage{dsfont}
\usepackage{color}  
\usepackage{fullpage}
\usepackage[section]{placeins}
\usepackage{sistyle}
\usepackage{url}

\graphicspath{{figures_datamixing/}}

\theoremstyle{definition}

\theoremstyle{definition}

\journal{Journal of Computational Physics}
\begin{document}
\begin{frontmatter}
\title{Lagrangian description and quantification of scalar mixing in fluid flows from particle tracks}
\author[leuphana]{Anna Klünker}
\ead{anna.kluenker@leuphana.de}
\affiliation[leuphana]{organization={Leuphana Universität Lüneburg, Institute of Mathematics and its Didactics, Applied Mathematics},
addressline={Universitätsallee 1},
postcode={21335},
postcodesep={},
city={Lüneburg},
country={Germany}}

\author[haw]{Alexandra von Kameke}
\ead{Alexandra.vonKameke@haw-hamburg.de}
\affiliation[haw]{organization={Hamburg University of Applied Sciences, Department of Mechanical Engineering and Production Management},
addressline={Berliner Tor 21},
postcode={20099},
postcodesep={},
city={Hamburg},
country={Germany}}
\author[leuphana]{Kathrin Padberg-Gehle\corref{cor1}}
\ead{kathrin.padberg-gehle@leuphana.de}
\cortext[cor1]{Corresponding author}

\begin{abstract} 
Understanding, quantifying and controlling transport and mixing processes are central in the study of fluid flows. Many different Lagrangian approaches have been proposed for detecting organizing flow structures that determine material transport, including recent data-based methods that aim to identify such coherent objects directly from tracer trajectories. These methods have helped to gain a better understanding of the underlying dynamics. However, the quantification of scalar mixing has not been the focus. Here, we develop a data-driven description and quantification of transport and mixing of scalar quantities by combining a diffusion map approach for the extraction of coherent flow structures with aspects of deterministic particle methods.   
\end{abstract}

%%Research highlights
\begin{highlights}
\item Particle tracks are key for Lagrangian studies of fluid flows.  
\item A novel data-driven transport description of scalar quantities is proposed.
\item The approach combines diffusion maps and deterministic particle methods. 
\item Advective-diffusive mixing is studied based on particle tracks.
\end{highlights}


\end{frontmatter}

%%%%%%%%%%%%%%%%%%%%%%%%%%%%%%%%%%%%%%

\section{Introduction}

Transport and mixing processes play an important role in the study of fluid flows, with applications ranging from atmospheric flows to process engineering. 
Much attention has recently been placed on the definition and identification of Lagrangian coherent flow structures such as coherent sets.
These are time-dependent material regions in the physical domain of the fluid flow that minimally mix with the surrounding fluid and thus act as organizers of transport and mixing processes. A number of established mathematical frameworks from nonlinear dynamics and ergodic theory for detecting coherent behavior in fluid flows exist, see e.g.\   \cite{Allshouse2015,Haller2015,Hadjighasem2017,Badza2023,Balasuriya2018} for reviews and comparative studies. In this context, there has been an increasing interest in making direct use of tracer trajectories calculated from velocity fields derived from numerical simulations or experimental data in two or three dimensions such as in   \cite{Huhn2012a,Huhn2012b,Kelley2013,Kameke2019,Aksamit2024,FroylandPadberg2015,weiland2023,Schoeller2025,Curbelo2023}. Among the recently developed purely data-based approaches for the identification of coherent flow structures are spectral clustering methods that extract coherent sets as tight groups of tracer trajectories in space-time. The basis for these approaches is a weighted network with trajectories serving as nodes and links weighted according to proximity or similarity of trajectories, see for example \cite{hadjighasem2016spectral, padberg2017network,schlueter2017, froyland2018robust,Filippi2021,Schneide2022,Vieweg2024}. Notably, in \cite{banisch2017understanding} the method of diffusion maps \cite{coifman2006diffusion,nadler2006diffusion, lafon2006diffusion} is extended to sparse trajectory data to estimate finite-time coherent sets. This approach is closely related to transfer operator methods from numerical ergodic theory \cite{Froyland2015}. 

The data-based methods have been merely applied to oceanographic \cite{FroylandPadberg2015,Filippi2021}, atmospheric \cite{padberg2017network,Schoeller2025} or turbulent convection systems \cite{Vieweg2021,Schneide2022,Vieweg2024}. However, recently, the Lagrangian view on transport and mixing processes has also reached the field of chemical process engineering \cite{Kameke2019,weiland2023}, realizing that the fluid dynamic processes cannot be reliably described by time-averaged velocity fields or instantaneous measurement of Eulerian quantities. 
Especially mixing dynamics on the timescales of the reactions need to be identified to become manipulable and guide the process engineering with regard to reactor and flow adjustments. The trend of analyzing the material behavior is likely to gain importance due to the massive improvement of time-resolved trajectory measurements in terms of Particle Tracking Velocimetry (4D-PTV) \cite{Schanz2016,Schroeder2023,Tan2020}.

In addition to the identification and extraction of coherent flow structures, the quantification of transport and mixing is of particular interest in the different scientific communities.  In \cite{klunker2022open} we proposed a transfer operator framework for time-periodic fluid systems with in- and out-flow (i.e.\ open systems) to model the evolution of scalar quantities and to measure mixing of two differently colored fluids.  

In experiments, the simultaneous time-resolved measurement of particle tracks and of scalar quantities (e.g. dye) in three-dimensional volumes is often not possible. Moreover, if we want to change the input, such as location or time of a scalar quantity in simulations or experiments, we would have to repeat the experiment or redo the numerical solution of the partial differential equation of the tracer concentration in the simulated or measured velocity field. However, in the experiments, the underlying flow and thus the tracer trajectories would also be different from the previous run due to turbulence. Therefore, it is desirable to develop a data-based model that allows us to conduct mixing experiments \emph{in silico} by using the measured particle trajectories. 

This is the motivation for the present work. We extend the space-time diffusion map approach \cite{banisch2017understanding} for the identification of coherent sets in order to model the evolution of a scalar quantity under the action of advection and an effective diffusion by means of given Lagrangian tracer trajectories. The resulting data-based model is similar in spirit to classical particle methods \cite{degond1989weighted,eldredge2002particles,Chertock2017} for the solution of partial differential equations and we draw explicit connections. However, unlike in these numerical approaches, one might be faced with the problem of having only very sparse trajectory information on the domain of interest, with the underlying velocity field often not known. Moreover, trajectory data might be incomplete (e.g.\ gaps in observations) and we have to care about that as well.

The remainder of the paper is organized as follows: 
In section \ref{sec:theory} we motivate our research (section \ref{sec:problem}) and briefly review the concept of deterministic particle methods with particle strength exchange for the numerical approximation of advection-diffusion equations (section \ref{sec:particlemethods}) before discussing the popular diffusion map framework (section \ref{sec:diffmap}) and its application to the identification of coherent flow structures. In section \ref{sec:datamethod} we present our trajectory-based advection-diffusion approach. Its main aspect concerns the data-based propagation of densities using a combination of particle methods and diffusion maps (section \ref{sec:trajdiffmap}). We also address numerical issues resulting from open systems and missing data (section \ref{sec:opensystem}) and briefly discuss measures for mixing quantification (section \ref{sec:mixing}).
Our approach is tested in a number of carefully chosen example systems in section \ref{sec:traex}, including a simulation of a stirred tank reactor, where we quantify the impact of coherent flow structures. We summarize our findings in section \ref{sec:conclusion} and discuss open topics.

%%%%%%%%%%%%%%%%%%%%%%%%%%%%%%%%%%%%

\section{Theoretical background}
\label{sec:theory}

In this section we define our problem and give the necessary theoretical background on particle methods and diffusion maps.
\subsection{Problem and definitions}
\label{sec:problem}
 We consider the most simple setting of an incompressible flow (zero divergence), where a passive scalar quantity $c$ evolves according to an advection-diffusion equation
\begin{equation}\label{eq:advectiondiffusion}
\frac{\partial c}{\partial t} (t, \bm{x})+ \bm{u}(t, \bm{x})\cdot \nabla c(t, \bm{x}) = D\nabla^2 c(t,\bm{x}).
\end{equation}
Here $D \geq 0$  is an effective diffusion constant (isotropic diffusion) which incorporates molecular diffusion but might also represent a much stronger dispersion due to small scale turbulence that is not resolved in the velocity field data \cite{Neufeld2009,Bakunin2008} and $\bm{u}\in \mathbb{R}^d$ the underlying possibly only large scale velocity field in the physical domain of the fluid (i.e.\ $d=2$ or $d=3$), with position $\bm{x} \in \mathbb{R}^d$ and time $t\in \mathbb{R}$. 

In case that $D =0$, i.e.\ there is no diffusion, neither of molecular nor turbulent kind, the evolution of the scalar quantity is exactly described by the underlying particle trajectories, which are solutions of the ordinary differential equation
\begin{equation}
\frac{{\rm{d}}{\bm{x}(t)}}{{\rm{d}}t} =  \bm{u}(t,\bm{x}(t)).\label{eq:advection}  
\end{equation}
In particular,
$c$ is conserved along trajectories: Fix an initial time $t_0$, a scalar field $c_0$ and an initial position $\bm{x}_0$ of a tracer.  Let $\bm{x}(t):=\bm{x}(t; t_0, \bm{x}_0)$ be the tracer position at time $t$. Then $c(t,\bm{x}(t))=c_0(\bm{x}_0)$.

In this contribution, we have the following setting: We are given particle trajectories from a numerical simulation or from particle tracking (e.g.\ 4D-PTV), i.e.\ we only have information on the advective dynamics as in \eqref{eq:advection}, possibly only on a larger scale if the particle trajectories are obtained via time resolved large eddy simulation (LES) or particle tracking, where small scale velocity fluctuations might not be well resolved in dependence of cut-off scale in LES or particle seeding density in tracking experiments. Based on this, we aim to describe the evolution of a scalar field $c$ as in equation \eqref{eq:advectiondiffusion} for $D >0$ and study the mixing dynamics in a purely data-based manner. To this end, we borrow ideas from deterministic particle methods for the solution of partial differential equations and diffusion maps.

For the sake of uniform representation and notation, we introduce the Gaussian kernel
$$
k: \mathbb{R}^d \times \mathbb{R}^d \to \mathbb{R}, \; \; \; k(\bm{x}, \bm{y})=\frac{1}{(\sqrt{\pi})^d}e^{-||\bm{x}-\bm{y}||^2}.
$$
where
$$
\int_{\mathbb{R}^d} k(\bm{x}, \bm{y}) \mathrm{d}\bm{y}=1 \; \; \mbox{ and } \; \; \int_{\mathbb{R}^d} (\bm{x}\cdot \bm{e}_i)^2 k(\bm{x}, \bm{y}) \mathrm{d}\bm{y}=\frac{1}{2}.
$$
Here, $\bm{e}_i \in \mathbb{R}^d$ denotes the $i$-th canonical basis vector and $\bm{x}\cdot \bm{e}_i$ the dot product, resulting in the $i$-th entry of $\bm{x}$. For some scaling parameter $\epsilon >0$ we form the scaled kernel
$$
k_{\epsilon}: \mathbb{R}^d \times \mathbb{R}^d \to \mathbb{R}, \; \; \; k_{\epsilon}(\bm{x}, \bm{y})=\frac{1}{(\epsilon \sqrt{\pi})^d}e^{-\frac{||\bm{x}-\bm{y}||^2}{\epsilon^2}}.
$$
It follows similarly,
$$
\int_{\mathbb{R}^d} k_{\epsilon}(\bm{x}, \bm{y}) \mathrm{d}\bm{y}=1 \; \; \mbox{ and } \; \; 
\int_{\mathbb{R}^d} \frac{1}{\epsilon^2}(\bm{x}\cdot \bm{e}_i)^2 k_{\epsilon}(\bm{x}, \bm{y}) \mathrm{d}\bm{y}=\frac{1}{2}.
$$
Note that $k_{\epsilon}(\bm{x}, \cdot)$ approaches the Dirac measure $\delta_{\bm{x}}$ as $\epsilon \to 0$.

\subsection{Deterministic particle methods}
\label{sec:particlemethods}
The starting point of deterministic particle methods (see e.g.\ \cite{degond1989weighted,eldredge2002particles,Chertock2017}) is to substitute the Laplacian operator $\nabla^2$ in \eqref{eq:advectiondiffusion} by an integral operator $Q_{\epsilon}$, where
$$
Q_{\epsilon}f(\bm{x})=\frac{1}{\epsilon^2}\int_{\mathbb{R}^d} (f(\bm{y})-f(\bm{x}))k_{\epsilon}^{\mathrm{lap}}(\bm{x},\bm{y})\mathrm{d}\bm{y}.
$$
The Laplacian kernel $k_{\epsilon}^{\mathrm{lap}}$ is generally obtained from some kernel $\kappa$, which is scaled as to fulfill certain moment conditions \cite{degond1989weighted,eldredge2002particles}. These determine the order of the approximation, $Q_{\epsilon}f(\bm{x})=\nabla^2f(\bm{x})+\mathcal{O}(\epsilon^r)$. Based on the Gaussian kernel $k_{\epsilon}$ introduced above, we obtain 
a second order integral approximation of the Laplacian operator (i.e. $r=2$) via 
 $$k^{\mathrm{lap}}_{\epsilon}(\bm{x}, \bm{y})=4 k_{\epsilon}(\bm{x}, \bm{y}) $$ see \cite{eldredge2002particles}. The original advection-diffusion equation \eqref{eq:advectiondiffusion} is then approximated by 
\begin{equation}\label{eq:advectiondiffusionepsi}
\frac{\partial c_{\epsilon}}{\partial t} (t, \bm{x})+ \bm{u}(t, \bm{x})\cdot \nabla c_{\epsilon}(t, \bm{x})= D Q_{\epsilon}(t) c_{\epsilon}(t,\bm{x}).
\end{equation}

The idea of deterministic particle methods is to replace the integro-differential equation \eqref{eq:advectiondiffusionepsi} by a set of ordinary differential equations in the following way.
The value of the quantity of interest $c_{\epsilon}$ at a particle position $\bm{x}_i(t)$ is identified with a particle strength $w_i(t)$. The approximate field $c_{\epsilon}^N(t, \bm{x})$ is then composed of the collection of particles \cite{Chertock2017}
\begin{equation}
c_{\epsilon}^N(t, \bm{x})=\sum_{i=1}^N w_i(t)k_{\epsilon}(\bm{x},\bm{x}_i(t)).\label{eq:scalarfield}
\end{equation}
Thus the value $c_{\epsilon}^N(t, \bm{x})$ is obtained as a (distance-) weighted average of the particle strengths at time $t$. 

Positions of particles and their strengths evolve according to
\begin{eqnarray}
\frac{{\rm{d}}\bm{x}_i(t)}{{\rm{d}}t} &= & \bm{u}(t,\bm{x}_i(t))\label{eq:padvection}\\
\frac{{\rm{d}}{w_i(t)}}{{\rm{d}}t} &= & \frac{D}{\epsilon^2}\sum_{j=1}^N V_j(w_j(t)-w_i(t))k_{\epsilon}^{\rm{lap}}(\bm{x}_i(t),\bm{x}_j(t)) \label{eq:pdiffusion}
\end{eqnarray}
where $V_j$ denotes the fluid volume that the $j$-th particle represents. This is independent of time in our case as we assume a divergence-free system. In case that the initial particles are given on a regular grid with spacing $h$, one would set $V_j=h^d$ for all $j=1, \ldots, N$. 
In the particle strength exchange framework \cite{degond1989weighted,eldredge2002particles} the initial strengths are approximated as
$w_i(t_0)=c_0(\bm{x}_i(t_0))$.
Note that 
$$ 
\frac{1}{\epsilon^2}\sum_{j=1}^N V_j (w_j(t)-w_i(t))k_{\epsilon}^{\rm{lap}}(\bm{x}_i(t),\bm{x}_j(t))
$$
is obtained from a midpoint quadrature discretization of $Q_{\epsilon}$ \cite{Chertock2017}, in particular, $$\frac{1}{4}\sum_{j=1}^N V_j k_{\epsilon}^{\rm{lap}}(\bm{x}_i(t),\bm{x}_j(t)) \approx \frac{1}{4}\int_{\mathbb{R}^d} k_{\epsilon}^{\mathrm{lap}}(\bm{x}_i(t),\bm{y}) \mathrm{d}\bm{y} = 1,$$
due to the conditions on $k_{\epsilon}^{\mathrm{lap}}$. Moreover, in this setting the evolution of particles in \eqref{eq:padvection} is completely independent of the diffusive transport of the scalar quantity $c$, which is described in \eqref{eq:pdiffusion} in terms of particle strengths.

\subsection{Diffusion maps}\label{sec:diffmap}
We briefly review the diffusion maps framework (e.g.\ \cite{coifman2006diffusion,lafon2006diffusion,nadler2006diffusion,coifman2014changing}), which is a spectral approach to learning the underlying geometry of data (see also \cite{belkin2003laplacian}).
Let $(X, \mathcal{B}, \mu)$ be a measure space, consisting of data points $X$ that are distributed according to the measure $\mu$. 
We consider a symmetric, positivity-preserving kernel $\kappa: X \times  X \to \mathbb{R}$ that encodes how similar two points are (e.g. $\kappa$ is a Gaussian kernel). Compute 
$$
d(x)=\int \kappa(\bm{x}, \bm{y}) \mathrm{d}\mu({\bm{y}})
$$
and obtain the new kernel $p: X \times X \to \mathbb{R}$ by
$$
p(\bm{x},\bm{y})=\frac{\kappa(\bm{x},\bm{y})}{d(x)}.
$$
Although this new kernel is no longer symmetric, it has a useful preservation property 
$$
\int_X p(\bm{x},\bm{y}) \mathrm{d}\mu({\bm{y}})=1
$$
and can thus be interpreted as a transition kernel of a Markov chain on $X$. We can define the integral operator $\mathcal{P}:L^2(X, \mu) \to L^2(X, \mu)$
$$
\mathcal{P}f(\bm{x})=\int_X p(\bm{x},\bm{y}) f(\bm{y})\mathrm{d}\mu({\bm{y}})
$$
for all $f \in L^2(X, \mu)$. $\mathcal{P}$ is a diffusion or averaging operator and preserves constant functions \cite{coifman2006diffusion}.  In the context of diffusion maps, which is a famous dimension-reduction tool, this integral operator is estimated from data.
In the following, we assume that the underlying measure $\mu$ is (normalized) Lebesgue measure and that data points are uniformly distributed. We revisit the rotation-invariant scaled kernel $k_{\epsilon}$ and include a cutoff. We define
\begin{equation}
    \hat{k}_{\epsilon}(\bm{x}, \bm{y}) = \frac{1}{(\epsilon\sqrt{\pi})^d} e^{-\frac{\|\bm{x}-\bm{y}\|_2^2}{\epsilon^2}} \mathbf{1}_{\|\bm{x}-\bm{y}\|_2\leq r_{\epsilon}} \label{eq:khat_diffmap}
\end{equation}
 on $\mathbb{R}^d$, where
$$
\mathbf{1}_{\|\bm{x}-\bm{y}\|_2\leq r_{\epsilon}} =\begin{cases} 1, &  \text{if } \|\bm{x}-\bm{y}\|_2\leq r_{\epsilon} \\ 0, & \text{otherwise} \end{cases},
$$
and $r_{\epsilon}$ is the cutoff radius. 

Suppose we are given $N$ data points $\bm{x}_i$, $i=1, \ldots, N$.
Then we obtain the symmetric weight matrix $\bm{K}_{\epsilon}\in \mathbb{R}^{N \times N}$ with entries $k_{ij}=\hat{k}_{\epsilon}(\bm{x}_i, \bm{x}_j)$, $i,j =1, \ldots, N$.
Compute $d_i=\sum_{j=1}^N k_{ij}$, which is a density estimate at $\bm{x}_i$, and for some $\alpha \in [0,1]$ build a new symmetric matrix $\tilde{\bm{K}}_{\epsilon}$ with entries
\begin{equation}
  \tilde{k}_{ij}=\frac{k_{ij}}{d_i^{\alpha} d_j^{\alpha}}.  \label{eq:alpha_diffmap}
\end{equation}

$\alpha$ can be used for fine-tuning the influence of the density of points by renormalizing the rotation-invariant kernel into an anisotropic weight. Typical choices are $\alpha=0$ (isotropic diffusion), $\alpha=\frac{1}{2}$ (Fokker-Planck diffusion) and $\alpha=1$ (heat kernel), which make a crucial difference when the underlying density is not uniform, see \cite{coifman2006diffusion} for the analysis of the respective operators and convergence results. In what follows we will use $\alpha=1$ in \eqref{eq:alpha_diffmap} throughout, as for this parameter choice the influence of the particle density is minimized due to dividing $k_{ij}$ by the density estimates $d_i$ and $d_j$.
In the last step, a stochastic transition matrix $\bm{P}_{\epsilon}$ is obtained by row normalizing $\tilde{\bm{K}}_{\epsilon}$: Set $\tilde{d}_i=\sum_{j=1}^N \tilde{k}_{ij}$ and obtain the matrix $\bm{P}_{\epsilon}$ with entries 
\begin{equation}
p_{ij}=\frac{\tilde{k}_{ij}}{\tilde{d}_i}. \label{eq:definition_pij}
\end{equation} 
$\bm{P}_{\epsilon}$ is a row-stochastic matrix and serves as a transition matrix of a finite-state Markov chain, where the states are the underlying data points. The transition probabilities take into account the distance between points, but given in terms of a diffusion distance rather than Euclidean distance. This is crucial when the data points are sampled from a low-dimensional manifold in a high-dimensional space and by careful tuning of $\epsilon$ and $r_{\epsilon}$ the diffusion process respects the geometry of the data set. Moreover, by means of the dominant eigenvectors of $\bm{P}_{\epsilon}$ (or a reversibilised version thereof) a low-dimensional data representation  is obtained that preserves the diffusion distance \cite{coifman2006diffusion,nadler2006diffusion,lafon2006diffusion}. 

Here, we are interested in another property of the transition matrix $\bm{P}_{\epsilon}$: A diffusive process of a scalar quantity $c_0$ on the data points $\bm{x}_i$ can now be modeled as follows. We initialize a vector $\bm{z}^0\in \mathbb{R}^N$ with $z^{0}_i=c_0(\bm{x}_i)$, $i=1, \ldots, N$, and obtain
$\bm{z}^{k+1}=\bm{P}_{\epsilon}\bm{z}^k$. In particular, if $\bm{z}^0$ is a multiple of the all-ones vector, we have that $\bm{P}_{\epsilon}\bm{z}^0=\bm{z}^0$, so uniform densities are preserved.

\subsection{Space-time diffusion maps and coherent sets}\label{sec:spacetimediff}
We briefly describe the basic idea of the diffusion map framework for the identification of coherent sets from trajectories as proposed by \cite{banisch2017understanding}. Coherent sets are time-dependent fluid regions that minimally mix with the surrounding fluid. 
In what follows, we assume that we are given $N$ trajectories in terms of solutions $\bm{x}_i(t)$, $i=1, \ldots, N$, of \eqref{eq:advection} over the time interval given by $\mathbb{T}$ with respect to initial values $\bm{x}_{0,i}=\bm{x}_i(t_0)$. Each trajectory is evaluated at $T+1$ time slices $t_k \in \mathbb{T}$, $k=0,\ldots, T$. The notion of coherent sets in such a trajectory-based setting is the following \cite{FroylandPadberg2015}: Coherent sets, such as vortices, are made up of trajectories that are ``close'' or ``similar'' in some sense over the time interval under consideration given by $\mathbb{T}$. Many different concepts of ``closeness'' have been proposed, such as based on minimal distance \cite{padberg2017network}, average distance \cite{hadjighasem2016spectral}, dynamic similarity \cite{schlueter2017},  number of encounters \cite{Schneide2022} and diffusion distance \cite{banisch2017understanding}.

For the latter construction we revisit the data-based kernel from the previous section \ref{sec:diffmap}:
 $$\hat{k}_{\epsilon}(\bm{x}, \bm{y}) = \frac{1}{(\epsilon \sqrt{\pi})^d}e^{-\frac{\|\bm{x}-\bm{y}\|_2^2}{\epsilon^2}} \mathbf{1}_{\|\bm{x}-\bm{y}\|_2\leq r_{\epsilon}}. 
 $$
The cutoff radius $r_{\epsilon}$ is chosen large enough so that 
$\int \frac{1}{\epsilon^2}(\bm{x}\cdot\bm{e}_i)^2 k_{\epsilon}(\bm{x}, \bm{y}) \mathrm{d}\bm{x} \approx \frac{1}{2}$, where $\bm{e}_i$ is again the $i$-th unit vector. In this work, we use $r_{\epsilon}=3\epsilon$ throughout.

For each time slice $t\in \mathbb{T}$, we compute the instantaneous kernel matrix $\bm{K}_{\epsilon}(t)$ with entries:
\begin{equation*}
k_{ij}(t)= \hat{k}_{\epsilon}(\bm{x}_i(t), \bm{x}_j(t))
\end{equation*}
and from that we form the transition matrix $\bm{P}_{\epsilon}(t)$ with entries $p_{ij}(t)$ based on equations \eqref{eq:alpha_diffmap} and \eqref{eq:definition_pij} as described in section \ref{sec:diffmap}. In \cite{banisch2017understanding} a time averaged matrix is built as
\begin{equation}
\bm{Q}_{\epsilon}:= \frac{1}{T+1}\sum_{k=0}^T \bm{P}_{\epsilon}(t_k). \label{eq:Qepsilon}
\end{equation}
An entry $q_{ij}$ of $\bm{Q}_{\epsilon}$ is large if the trajectories $\bm{x}_i$ and $\bm{x}_j$ are close on $\mathbb{T}$. The problem of finding coherent sets has been reduced to that of considering a graph with the trajectories serving as nodes and the links with weights $q_{ij}$. In particular, coherent sets manifest themselves as subgraphs or clusters that are closely connected within but only loosely connected to other parts of the graph. Finding such subgraphs is the well-established task of spectral clustering methods \cite{vonLuxburg2007}. 
These proceed as follows: For some sufficiently large $n$ compute the largest $n$ eigenvalues $\lambda_i$, $i=1, \ldots, n$ of $\bm{Q}_{\epsilon}$ and corresponding eigenvectors $v_i$. Note that $\lambda_1=1$ and $v_1=\mathbf{1}$ by construction of $\bm{Q}_{\epsilon}$.  
Identify a spectral gap, i.e. find $k$ such that $\lambda_{k}-\lambda_{k+1}$ is large. 
Extract $k$ clusters from the eigenvectors $v_1, \ldots, v_k$ using a hard-clustering approach such as $k$-means or a soft-clustering method such as SEBA \cite{Froyland2019}.

For completeness, we note that in \cite{banisch2017understanding} the theoretical framework is developed based on forward-back time matrices $\bm{B}_{\epsilon}(t)$ obtained from kernel evaluations and combinations from $\hat{k}_{\epsilon}$ to form the time-averaged matrix $\bm{Q}_{\epsilon}$ as in equation \ref{eq:Qepsilon}. Such a construction is necessary in order to be able to draw analytical connections to the dynamic Laplacian framework \cite{Froyland2015}. However, the computations of $\bm{B}_{\epsilon}(t)$ require significant effort, and the authors propose to use $\bm{Q}_{\epsilon}$
based on the matrices $\bm{P}_{\epsilon}(t)$ as described above instead, arguing that this simpler construction can render the original matrix to good accuracy \cite{banisch2017understanding}.

\section{Data-based advection-diffusion dynamics}\label{sec:datamethod}
Now we combine the ideas of the previous sections to derive a data-driven reconstruction of the advection-diffusion dynamics and also describe different approaches to quantify mixing. Our construction is similar to that of \cite{banisch2017understanding} and heavily relies on the instantaneous diffusion matrices $\bm{P}_{\epsilon}(t)$.

In the first subsection \ref{sec:trajdiffmap}, we describe the construction of the \textit{diffusion map transition matrices} as well as a simple numerical scheme for the propagation of densities.  
We consider then the case of open systems and missing data (section \ref{sec:opensystem}). Finally, we discuss the application of different measures to quantify mixing (section \ref{sec:mixing} ).  

\subsection{Construction of diffusion matrices and density propagation} \label{sec:trajdiffmap} 

In what follows, we assume that we are given $N$ trajectories in terms of solutions $\bm{x}_i(t)$, $i=1, \ldots, N$, of \eqref{eq:advection} over the time interval $\mathbb{T}$ with respect to initial values $\bm{x}_{0,i}=\bm{x}_i(t_0)$. Each trajectory is evaluated at $T+1$ time slices $t_k \in \mathbb{T}$, $k=0,\ldots, T$. For simplicity, we assume that the time steps $t_{k+1}-t_k=\tau$ are constant. We want to describe the advective-diffusive transport of a scalar quantity $c$ in a particle-oriented approach, given that we have information on the particle tracks (or tracer trajectories).

The challenge is now to estimate for each particle position $\bm{x}_i(t_k)$ the respective value of the scalar quantity $c(t_k,\bm{x}_i(t_k))$. To this end, the diffusion part \eqref{eq:pdiffusion} in the particle method has to be adapted accordingly. 
The idea of spacetime diffusion maps \cite{banisch2017understanding} is to obtain information of the global dynamics (here in form of coherent sets) by using only local information in form of distances of neighboring particles. This is done by introducing a diffusion process on the trajectory data as described in the previous section.  

We revisit the instantaneous matrices $\bm{P}_{\epsilon}(t)$ with entries $p_{ij}(t)$ that form the basis of the diffusion map approach for the extraction of coherent sets.  These are now used in the diffusion part of the particle method and we obtain an alternative description of the dynamics of particle weights
\begin{equation}
\frac{{\rm{d}}{w_i(t)}}{{\rm{d}}t} =  \frac{4D}{\epsilon^2}\sum_{j=1}^N p_{ij}(t)(w_j(t)-w_i(t)) \label{eq:diffmapdiffusion}
\end{equation}

Here $p_{ij}(t)$ approximates $ \frac{1}{4}V_j k_{\epsilon}^{\mathrm{lap}}(\bm{x}_i(t),\bm{x}_j(t))$, and, in particular, $\sum_{j=1}^N p_{ij}(t)=1$. 
The particle volume $V_j$ that is required from the midpoint rule approximation of the integral in \eqref{eq:advectiondiffusionepsi} is no longer included but eliminated by (i) the symmetrical normalization in \eqref{eq:alpha_diffmap}, where the density estimate $d_i$ at a tracer $\bm{x}_i$ can be interpreted as an inverse volume $d_i\approx \frac{1}{V_i}$ for unit mass, and (ii) the subsequent row normalization to form the stochastic transition matrix $\bm{P}_{\epsilon}(t)$.  This approximation and substitution is advantageous as the volumes are not reliably available from scattered data anyway. For the initialization of the weights $w_i$ we can thus use $w_i(t_0)=c_0(\bm{x}_i(t_0))$ and assume each particle to represent a unit volume.

As we are only given the trajectory positions at discrete times, we
approximate the time evolution of the particle strengths via an explicit Euler scheme. Starting with the initial weight $w_i^0=w_i(t_0)$ we obtain
\begin{equation}
w_i^{k+1} = w_i^k+\tau \frac{4D}{\epsilon^2}\sum_{j=1}^N p_{ij}(t_k)(w_j^k-w_i^k)   \label{eq:diffmap_update}
\end{equation}
where $w_i^{k+1}\approx w_i(t_k)$.

We can now write this in vectorized form. Let $c_0$ be an initial scalar quantity and we initialize the respective particle-based description by a vector $\bm{w}^0$ with entries $w_{i}^0$ as defined above.
The coevolved vector after one time step $\tau$ is then given by
\begin{equation}\label{eq:onestep}
\bm{w}^1=(1-\tilde{D})\bm{w}^0+ \tilde{D}\bm{P}_{\epsilon}(t_0)\bm{w}^0.
\end{equation}
Here
\begin{equation}
\tilde{D}=\frac{4D\tau}{\epsilon^2} \label{eq:diffusionconstant}
\end{equation}
where $D$ is the original effective or turbulent diffusion coefficient, $\tau$ the time step and $\epsilon$ the scaling parameter of the diffusion kernel. $D$ is fixed, representing the effective diffusion coefficient of a specific scalar quantity, like a chemical concentration, resulting from a stochastic process on unresolved scales caused by molecular or turbulent motion. The time step $\tau$ is constrained by the length of the data.
According to equation \eqref{eq:onestep} $\epsilon$ should chosen such that $\tilde{D}\leq 1$, see also \cite{degond1989weighted,eldredge2002particles} for respective discussions on the stability of time stepping schemes. This condition also ensures that $\epsilon$ is large compared to the effective diffusion constant $D$. Moreover, depending on the distribution of particles, the magnitude of $\epsilon$ should be such as to allow for an interaction of neighboring particles.

Given the family of computed diffusion matrices $\{\bm{P}_{\epsilon}(t_k) \}_{k=0, \ldots, T}$, we obtain
\begin{equation}
\bm{w}^{k+1}=(1-\tilde{D})\bm{w}^k+ \tilde{D}\bm{P}_{\epsilon}(t_k)\bm{w}^k, \; \; k=0, \ldots, T-1. \label{eq:coevolution}
\end{equation}
As the final vector $\bm{w}^{T}$ is obtained by means of $\bm{P}_{\epsilon}(t_{T-1})$ in the above scheme, the last diffusion matrix  $\bm{P}_{\epsilon}(t_{T})$ is not required in practice.
 Sometimes, it is of practical interest to increase the time step $\tau$ to some multiple $\tau'=C\tau$, $C \in \mathbb{N}$. To this end, one may simply use a time-averaged transition matrix over each time span of length $\tau'$ obtained from the respective $C$ transition matrices computed for the original time step $\tau$. Note that all constructions can also be adapted for variable step sizes $\tau_k$, accordingly.

\subsection{Open systems and missing data}\label{sec:opensystem}
In the practically relevant case of an open system, at every time slice tracers may enter or leave the domain of interest. Thus not all observed trajectories are available for the whole time span and we have to deal with incomplete trajectory data. In particular, such gaps in observation are very frequent in experimental data obtained from particle tracking. As mentioned in \cite{banisch2017understanding}, the construction of $\bm{Q}_{\epsilon}$ works also for incomplete trajectory data. A natural way to deal with incomplete trajectories is to assign the distance of the $i$-th particle to others to $\infty$ at a time slice $t$ when $\bm{x}_i(t)$ is missing (or the particle is not in the domain $A$) as recorded in the instantaneous matrix $\bm{P}_{\epsilon}(t)$ with entries $p_{ij}(t)$. The corresponding transition probabilities for that trajectory to others at that time slice $t$ is then $p_{ij}(t)=0$, $i\neq j$. If we set $p_{ii}(t)=1$, $\bm{P}_{\epsilon}(t)$ and $\bm{Q}_{\epsilon}$ are stochastic. 

This construction is simple in general but has the disadvantage that all trajectories, no matter how long and when they are available, are considered at every single time slice. In systems with in- and outflow, where one has a continuous exchange of particles, this would quickly blow up the size of the (albeit sparse) matrices. Moreover, this construction does not fully solve the problem of gaps in observation and the corresponding update of the coevolved vector as it is not clear how to determine the current vector entry corresponding to a newly observed tracer. 

Therefore, we propose the following scheme that only takes into account those particles that are present in the respective time slice.
Let $\{i_1^{k-1}, \ldots, i_n^{k-1}\} \subseteq \{1, \ldots, N\}$ be the indices (or IDs) of the particles present in the $(k-1)$-th time step and
$\{i_1^{k}, \ldots, i_m^{k}\}$ those for the $k$-th time step, for simplicity we assume that the indices are ordered. Let $\bm{w}^{k} \in \mathbb{R}^n$ be the vector evolved from time step $k-1$ to $k$ by \eqref{eq:coevolution}, based on the $n$ particles available in time step $k-1$. 
Let $\bm{P}_{\epsilon}(t_{k}) \in \mathbb{R}^{m,m}$ be transition matrix at time step $k$, obtained from the diffusion map construction of the $m$ particles present at time step $k$. In order to compute  $\bm{w}^{k+1}$ a vector $\hat{\bm{w}}^{k} \in \mathbb{R}^m$ has to be evolved.
We construct $\hat{\bm{w}}^{k}$ from $\bm{w}^{k}$ in the following way.  

Initialize $\hat{\bm{w}}^{k}\in \mathbb{R}^m$ and for each current particle with ID $i^{k}_j$, $j=1, \ldots, m$ that was already there in the  previous time step, i.e. there is $l\in \{1, \ldots n\}$ such that  $i^{k}_j=i^{k-1}_l$ set $\hat{w}^{k}_j=w_l^{k}$. New particles get the value NaN.
We will deduce the vector entries of new particles (i.e.\ that have currently the value NaN) from the transition matrix $\bm{P}_{\epsilon}(t_{k})$.  
Let $l \in \{1, \ldots, m \}$ with $\hat{w}^k_l =\text{NaN}$, i.e. the particle with ID $i^k_l$ is new. We propose three different ways to determine the value of $\hat{w}^k_l$. 
\begin{enumerate}
\item[(i)] \textbf{Nearest neighbors:} Set $\hat{w}^k_l=\hat{w}^k_j\neq \text{NaN}$, where $i^k_j$ is the ID of an old particle closest to the new particle with ID $i^k_l$ (this corresponds to $j\in \{1, \ldots, m\}$ with $j \neq l$ for which $p_{l,j}$ is maximal, given that $i^k_j \in \{i_1^{k-1}, \ldots, i_n^{k-1}\}$. 
\item[(ii)] \textbf{Weighted average:} 
$\hat{w}^k_l=\sum_{j=1}^m \hat{p}_{l,j}\hat{w}^k_j$, where the matrix row $\hat{p}_{l,\cdot}$ is obtained from $p_{l,\cdot }(t_k)$ by first setting the entries corresponding to new particles to zero and then normalizing the row. This in spirit of the data-based definition of a scalar field as in equation \eqref{eq:scalarfield}.
\item[(iii)] \textbf{Constant input:} In some cases, for instance, in systems with a constant inflow, we can set the vector entry for a new particle to a predefined value.
\end{enumerate}
To sum up, an initial vector $\bm{w}^0$ corresponding to particles observed at the initial time step $t_0$, is evolved according to
$$
\bm{w}^1 =(1-\tilde{D})\bm{w}^0+ \tilde{D}\bm{P}_{\epsilon}(t_0)\bm{w}^0 \; \mapsto \; \hat{\bm{w}}^1 \; \mapsto \;  \bm{w}^2 =(1-\tilde{D})\hat{\bm{w}}^1+ \tilde{D}\bm{P}_{\epsilon}(t_1)\hat{\bm{w}}^1  \; \mapsto \;  \hat{\bm{w}}^2 \; \mapsto \;  \bm{w}^3 = \ldots
$$
\subsection{Mixing quantification} \label{sec:mixing}
In the following, we will discuss different notions of mixing measures to quantify mixing. For this, let us consider again $N$ trajectories $\{\bm{x}^i(t)\}, i=1,\ldots, N$, evaluated at $T+1$ time slices $t\in \mathbb{T}=\{t_0 \ldots ,t_{T}=\}$ within a fixed time step of length $\tau=t_{k+1}-t_k$, $k=0,\ldots, T-1 $ on a domain $A\subset \mathbb{R}^d$. For simplicity of exposition we assume that there are no gaps in observation, i.e.\ all particles can be observed in each time slice (the extension to the case of missing data is straightforward).

Let $\bm{w}^0 \in \mathbb{R}^N$ be the vector that is obtained from the initial distribution $c_0$ of a scalar quantity $c$. $c_0$ could be non-negative (density) or signed, e.g.\ modeling two differently colored fluids. The system is fully mixed when the scalar quantity is uniformly distributed and thus homogeneous. 

\paragraph{Sample variance} As a simple mixing measure one can use the sample variance $V_k$ of the vector $\bm{w}^k$ 
\begin{equation}
    V_k=\frac{1}{N-1} \sum_{i=1}^{N} (w_{i}^k-\overline{w}^k)^2,\label{eq:samplevariance}
\end{equation}
where $\overline{w}^k= \frac{1}{N}\sum_{i=1}^N w^k_{i}$ is the mean. In the perfectly homogeneous mixed case the variance is 0. 

We note that the variance is only meaningful as a mixing measure when the scalar quantity is subjected to diffusion as in our study. In particular, the advection (eventually only on the large scale) of the flow does enhance the overall mixing and leads to a faster decay of $V_k$ than in a purely diffusive setting. 

In the case of advection (stirring) without an effective diffusion the variance is constant, i.e. $V_k=V_0$ for $k=1, \ldots, T$ due to the conservation property of scalar quantities along trajectories in divergence-free flows. For this case sophisticated multi-scale mix-norms \cite{mathew2005multiscale,Thiffeault2012} have been proposed, which are in principle also applicable to measure scalar mixing under advection and diffusion. Since we explicitly deal with an effective diffusion (on small, molecular or unresolved scales) in the present work, we will use the sample variance due to its simple construction and interpretation.

\paragraph{Node degree} A different approach to study mixing is motivated by the network interpretation of the diffusion map transition matrices.
The family of transition matrices $\bm{P}_{\epsilon}(t_k)$ defines a time-evolving network, where the particle trajectories are the nodes and the links are weighted according to similarity (i.e.\ transition probabilities).
We build the time averaged matrix $\bm{Q}_{\epsilon}$ with entries $q_{ij}$  from the family of instantaneous network transition matrices $\bm{P}_{\epsilon}(t_k)$ via 
$$
\bm{Q}_{\epsilon}= \frac{1}{T+1}\sum_{k=0}^T \bm{P}_{\epsilon}(t_k)
$$
as in \eqref{eq:Qepsilon} and form a new matrix $\bm{A}$ by setting $a_{ij}=1$ if $q_{ij} \neq 0$ and else $a_{ij}=0$. $\bm{A}$ is the adjacency matrix of the time-averaged network and encodes which particles have come close to each other and thus have interacted over the time span defined by $\mathbb{T}$.
The local node degree $\bm{d}$ with 
\begin{equation}
    d_i=\sum_{j=1}^N a_{ij} \label{eq:nodegree}
\end{equation} 
takes large values for particles that have encountered many other particles and thus have been particularly involved in mixing processes \cite{padberg2017network,Banisch2019}. In open systems a high node degree also corresponds to high residence times, which can be observed for particles near the stable manifold of a chaotic saddle \cite{klunker2022open}.
The average node degree $\bar{d}=\frac{1}{N}\sum_{i=1}^N d_i$ can be used as a mixing measure in order to compare the general mixing properties of the underlying flow (e.g.\ in parameter studies), where a large value of $\bar{d}$ indicates strong mixing. 

\paragraph{Sign-based node degree} In order to explicitly account for a given scalar quantity to be mixed, we propose a modified mixing measure based on the node degree. For this let $\bm{w}^0$ be the initial vector (i.e.\ the initial scalar field evaluated at the particle positions) and $\bar{w}^0$ the mean. We form a new signed vector $\bm{z}^0$ by subtracting the mean from $\bm{w}^0$: $z^0_i=w^0_i-\bar{w}^0$, $i=1, \ldots, N$. Heuristically, to achieve good mixing, especially particles corresponding to entries of $\bm{z}^0$ of different sign have to interact.
We thus form a modified, sign-based adjacency matrix $\bm{A}^s$ from $\bm{A}$, where $a_{ij}^s = a_{ij}$ if $\text{sign}(z_i^0)\neq \text{sign}(z_j^0)$ and $a_{ij}^s=0$ otherwise. So only those network connections between particles are kept that have an impact on mixing. We again form the node degree $\bm{d}^s$ from $\bm{A}^s$ with \begin{equation}
    d_i^s=\sum_{j=1}^N a_{ij}^s \label{eq:signnodegree}
\end{equation} 
and its average $\bar{d}^s=\frac{1}{N}\sum_{i=1}^N d_i^s$. $\bar{d}^s$ is used as a heuristic mixing measure. It takes high values if there is a lot of mixing/interaction predominantly between particles whose initial scalar field values are on opposite sides of the mean. \\[2mm]

\noindent Finally, we note that whereas the sample variance is only meaningful as a mixing measure when diffusive processes are included as discussed above, the two proposed constructions of node degrees are also applicable in the context of stirring (i.e.\ only advection). They may thus provide a simple alternative to the multi-scale mix-norms \cite{mathew2005multiscale,Thiffeault2012}.

%%%%%%%%%%%%%%%%%%%%%%%%%%%%%%%%%%%

\section{Example systems} \label{sec:traex}

We will apply the data-based framework introduced in the previous section to different example systems. As already mentioned, we will restrict to the choices $r=3\epsilon$ for the cut-off radius \eqref{eq:khat_diffmap} and $\alpha=1$ in \eqref{eq:alpha_diffmap}  in the construction of the diffusion map matrices.  We will first demonstrate the proposed framework to a simple vortex flow and compare the data-based results with those from the corresponding numerical solution of the partial differential equation (section \ref{sec:cellflow}). We then move on to studying mixing in the well known double gyre flow, both in the closed setting (section \ref{sec:cdg}) and the extension to an open flow (section \ref{sec:odg}) as considered in \cite{klunker2022open}. These two-dimensional flows under consideration are defined via a stream function
$$
\Psi(t,\cdot):\mathbb{R}^2 \to \mathbb{R}
$$ 
and the velocity field $\bm{u}(t, \bm{x})$ is obtained as
$$
\bm{u}(t, \bm{x})=\left(\frac{\partial \Psi}{\partial y}(t, \bm{x}), -\frac{\partial \Psi}{\partial x}(t, \bm{x})\right)
$$
where $\bm{x}=(x,y) \in \mathbb{R}^2$. As these toy models are given in terms of dimensionless units of time and space, also the chosen effective diffusion constants will be given without units. 

In view of studying transport and mixing in realistic applications from process engineering, in our final example, we will analyze mixing in a three-dimensional simulated lab-scale stirred tank reactor (section \ref{sec:str}). From these velocity fields the Lagrangian trajectories of passive particles are obtained according to equation \eqref{eq:padvection} for further processing.

\subsection{Cellular flow}\label{sec:cellflow}

\begin{figure}[htb]
\centering
\begin{tabular}{ccc}
\includegraphics[height=0.2\textheight]{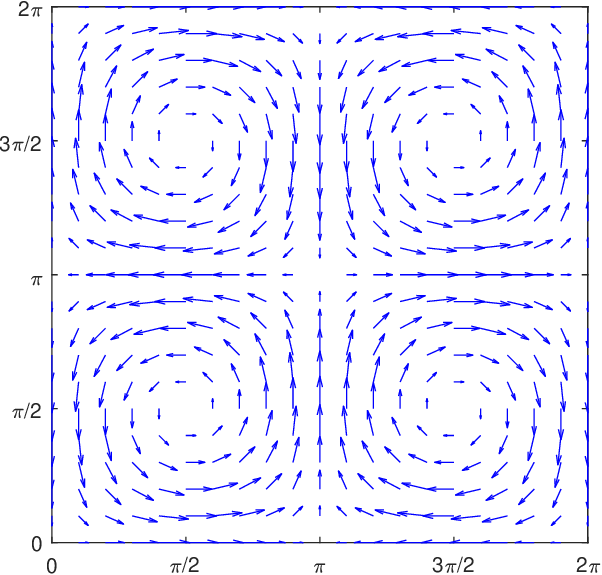} &
\includegraphics[height=0.2\textheight]{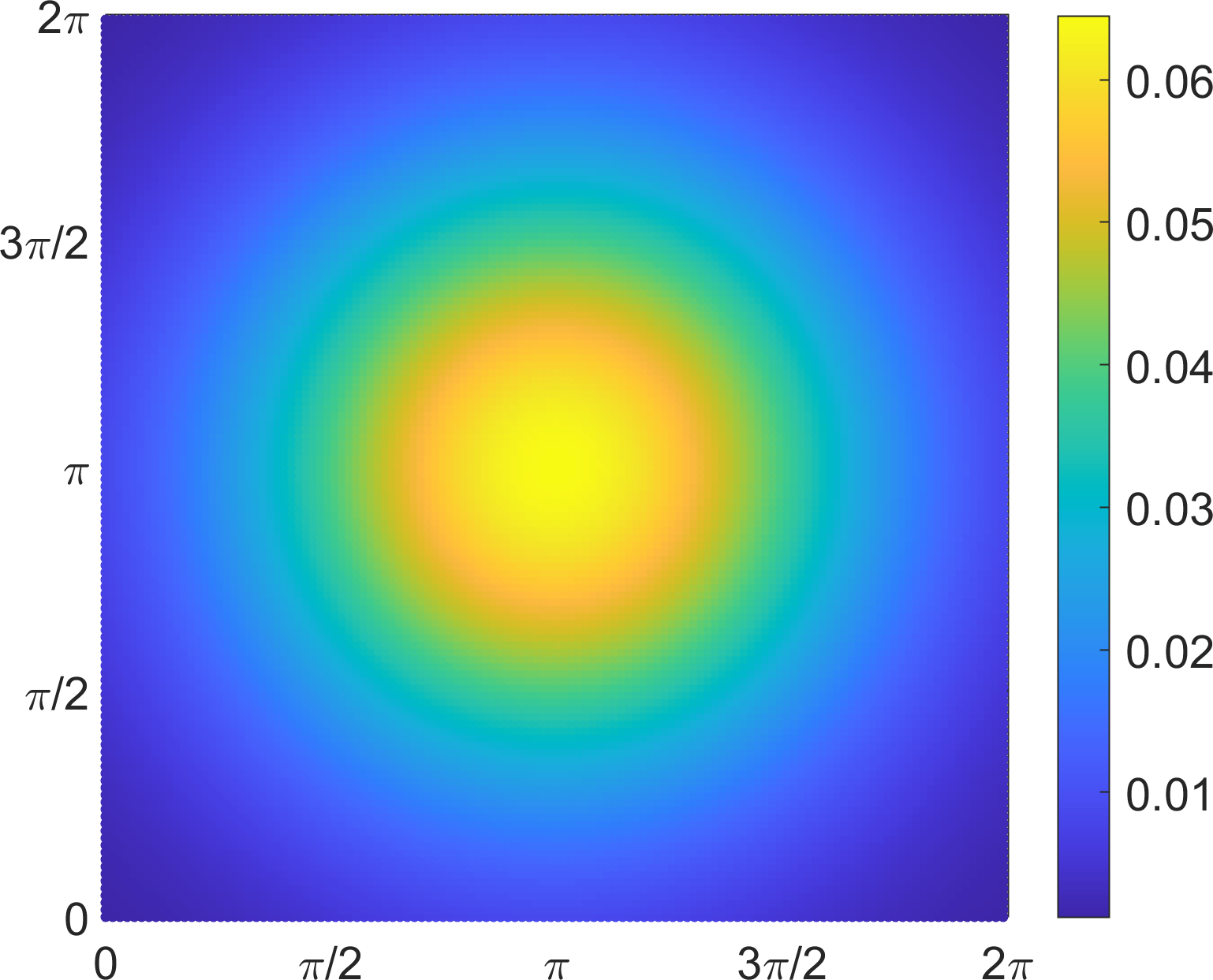} &
\includegraphics[height=0.2\textheight]{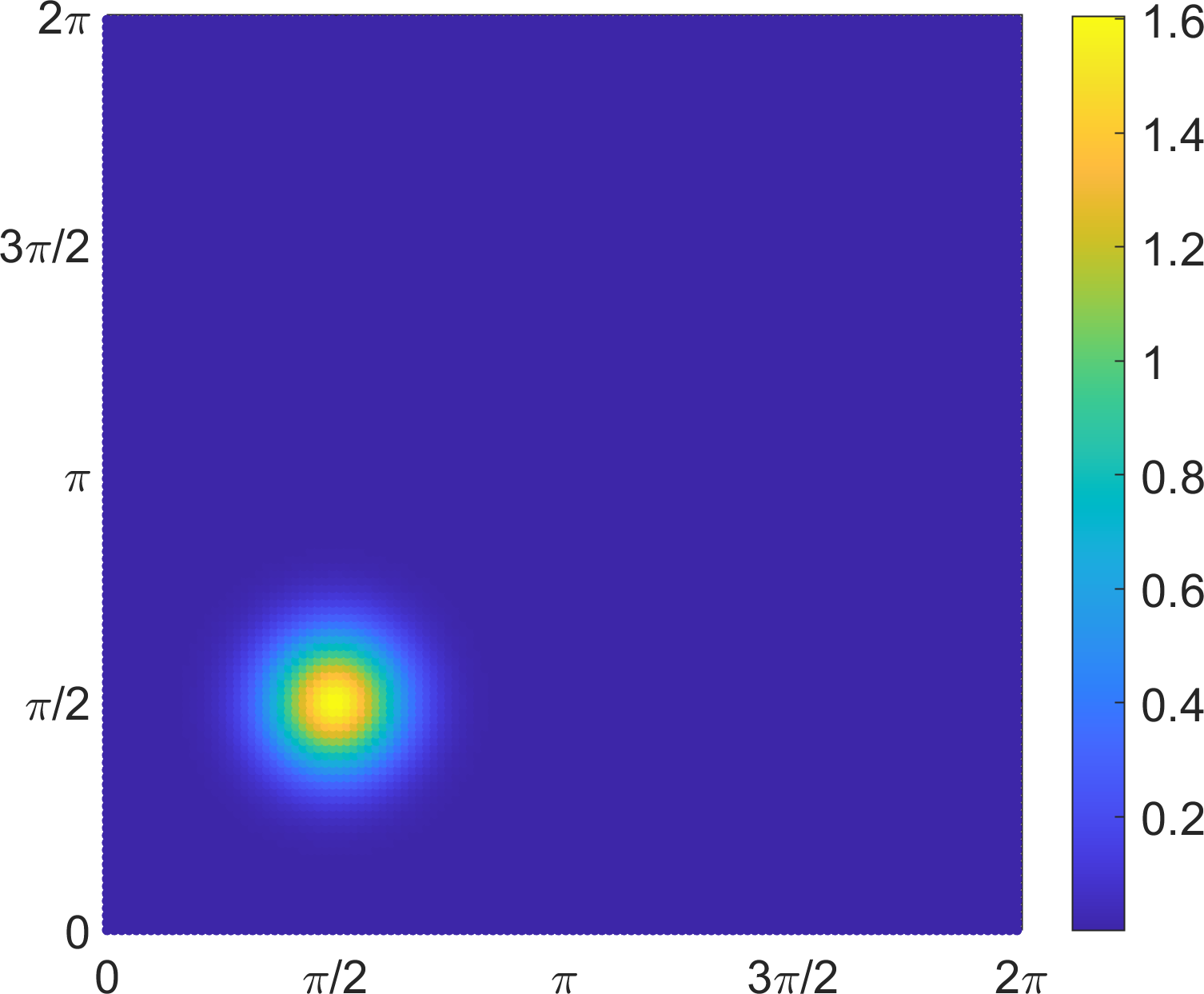} \\
{\scriptsize (a)} & {\scriptsize (b)} & {\scriptsize (c)}\\
\end{tabular}
\caption{ Velocity field of the cellular flow (a) and the two different initial conditions: (b) a Gaussian in the center of the cell and (c) a smaller Gaussian centered in left lower gyre.}
\label{fig:oc0}
\end{figure}

We start with a simple two-dimensional incompressible autonomous cellular flow, which is a variant of the Taylor-Green vortex \cite{Taylor1923}.  The flow is defined by the time-independent stream function
\begin{equation*}
\Psi(\bm{x})= A\sin(x)\sin(y),
\end{equation*}
where $A = \sqrt{2}$. We restrict to $M=[0, 2\pi]^2$ and thus to a single cell. The corresponding velocity field is shown in Figure \ref{fig:oc0}(a). For our mixing studies we consider two different initial conditions,  a Gaussian in the center of the cell (Figure \ref{fig:oc0}(b)) and a smaller Gaussian centered in left lower gyre to show the purely diffusive effect (Figure \ref{fig:oc0}(c)).

\begin{figure}[!htb]
\centering
\begin{tabular}{cccccc}
& {\scriptsize PDE } & {\scriptsize $h=0.05$} & {\scriptsize $h=0.1$} & {\scriptsize $h=0.2$}  \\[2mm]
{\scriptsize $D=0.001$} & 
\includegraphics[height=0.14\textheight]{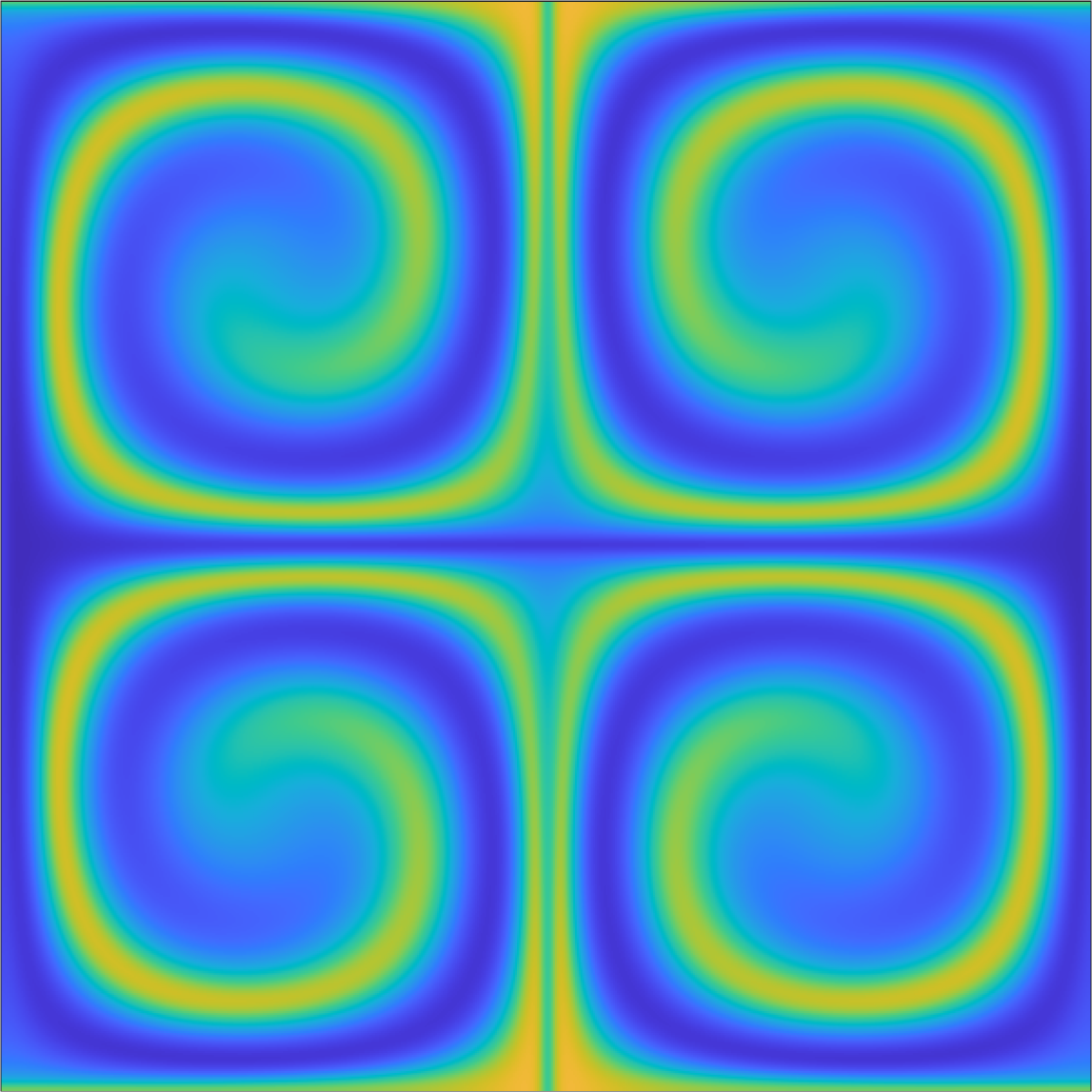} & 
\includegraphics[height=0.14\textheight]{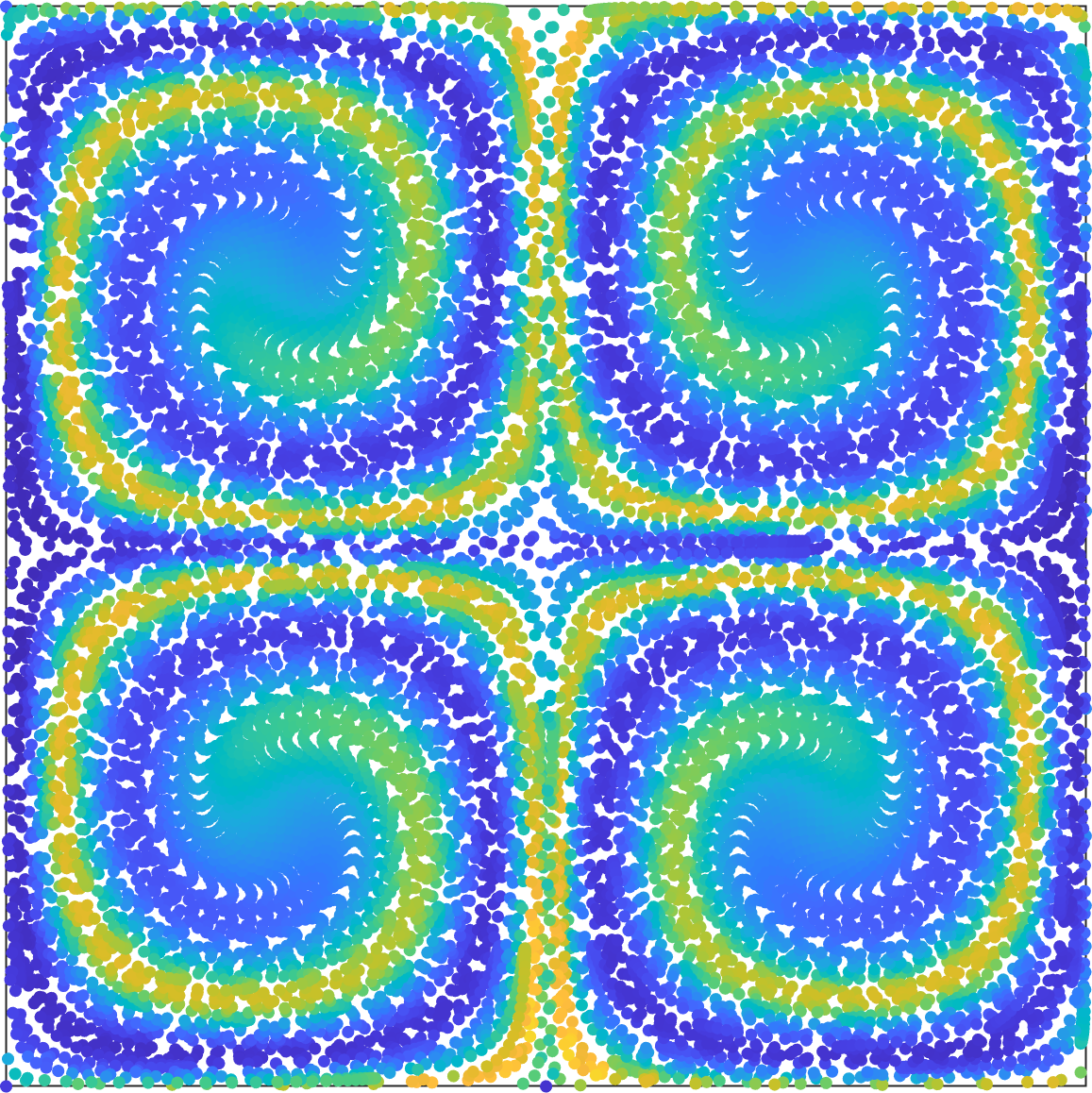} & 
\includegraphics[height=0.14\textheight]{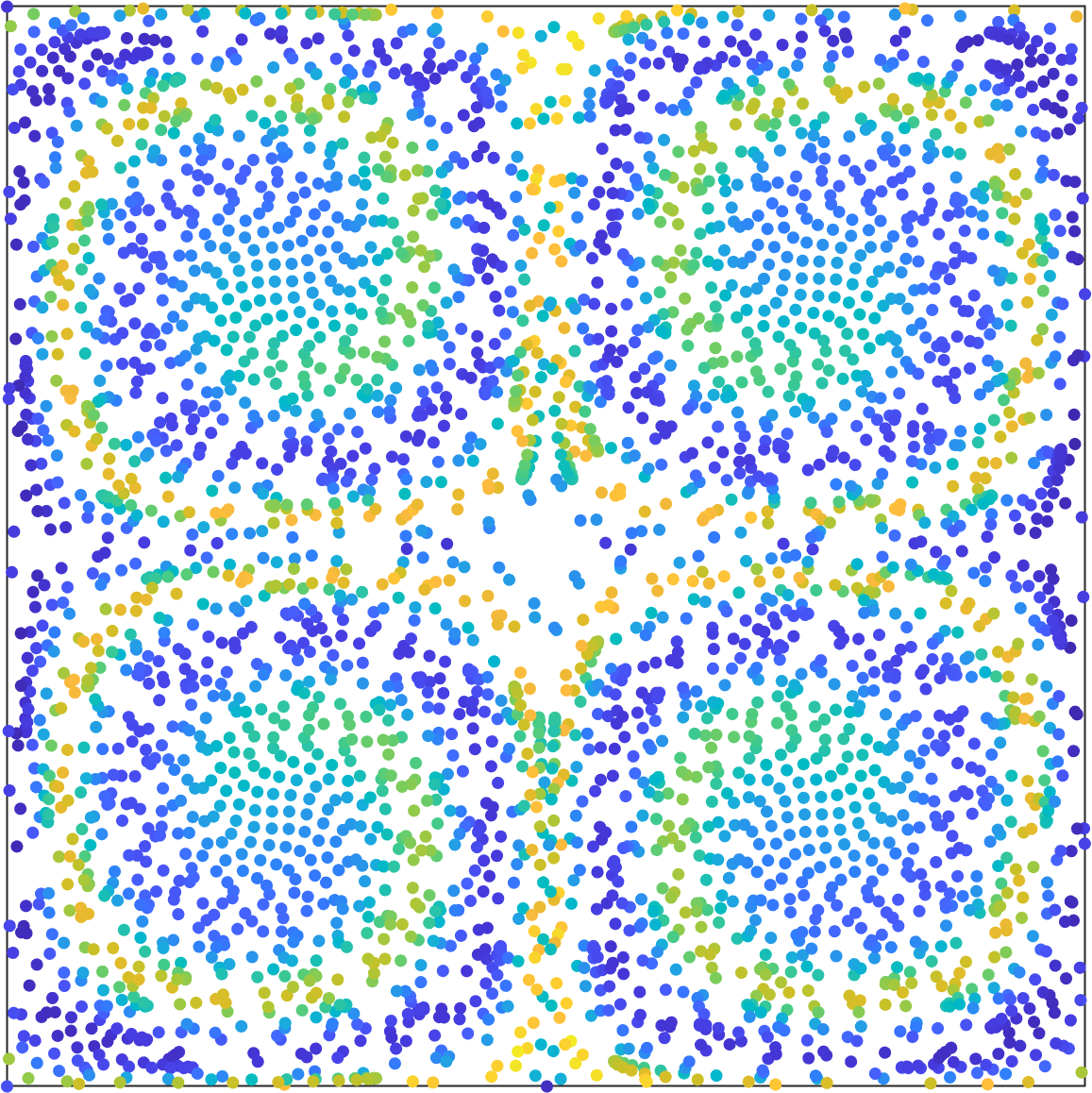} &
\includegraphics[height=0.14\textheight]{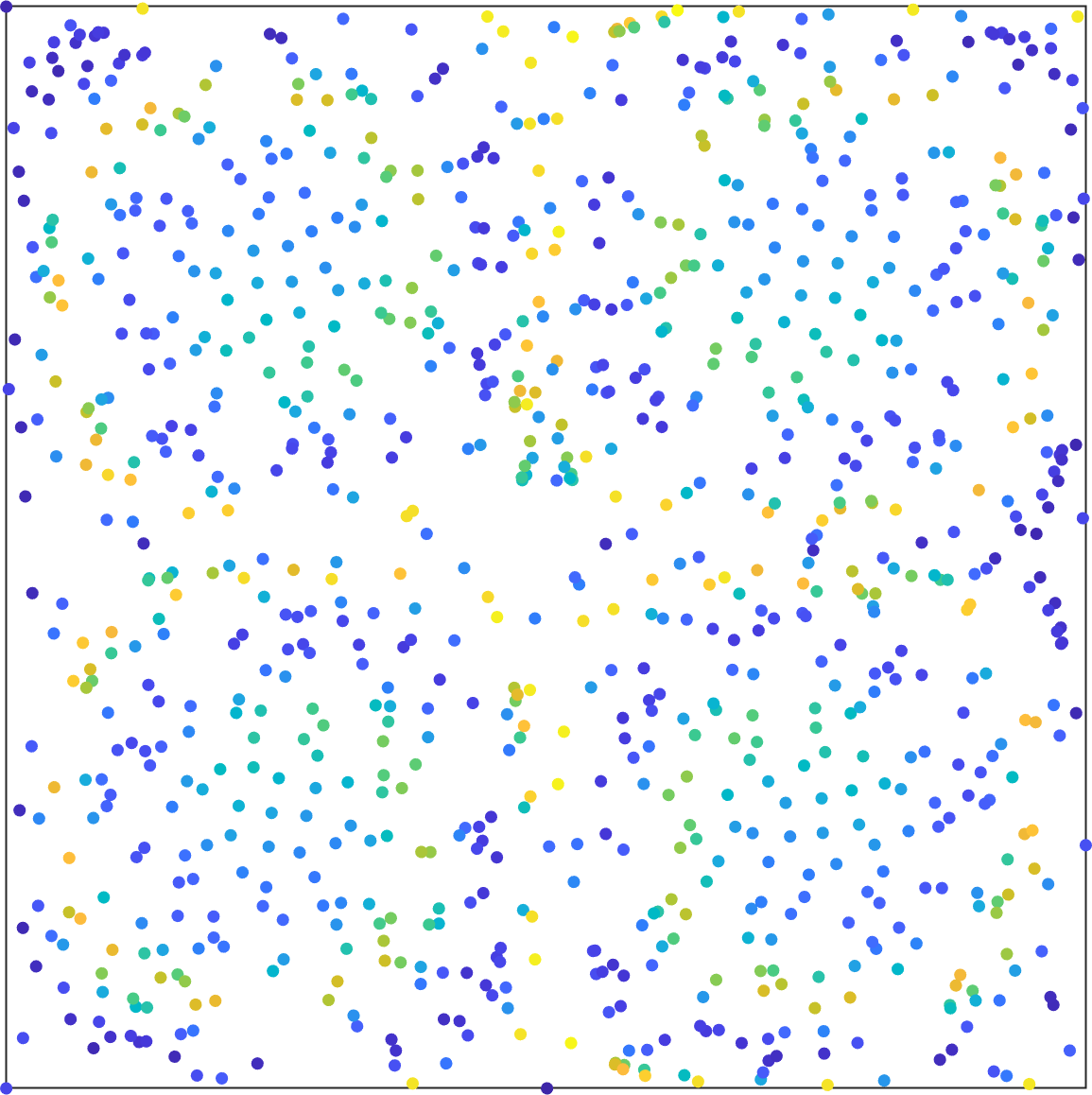} &  \includegraphics[height=0.14\textheight]{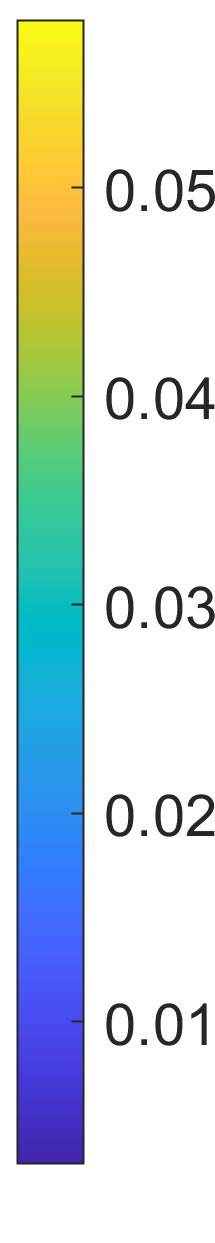} \\
{\scriptsize $D=0.01$} &
\includegraphics[height=0.14\textheight]{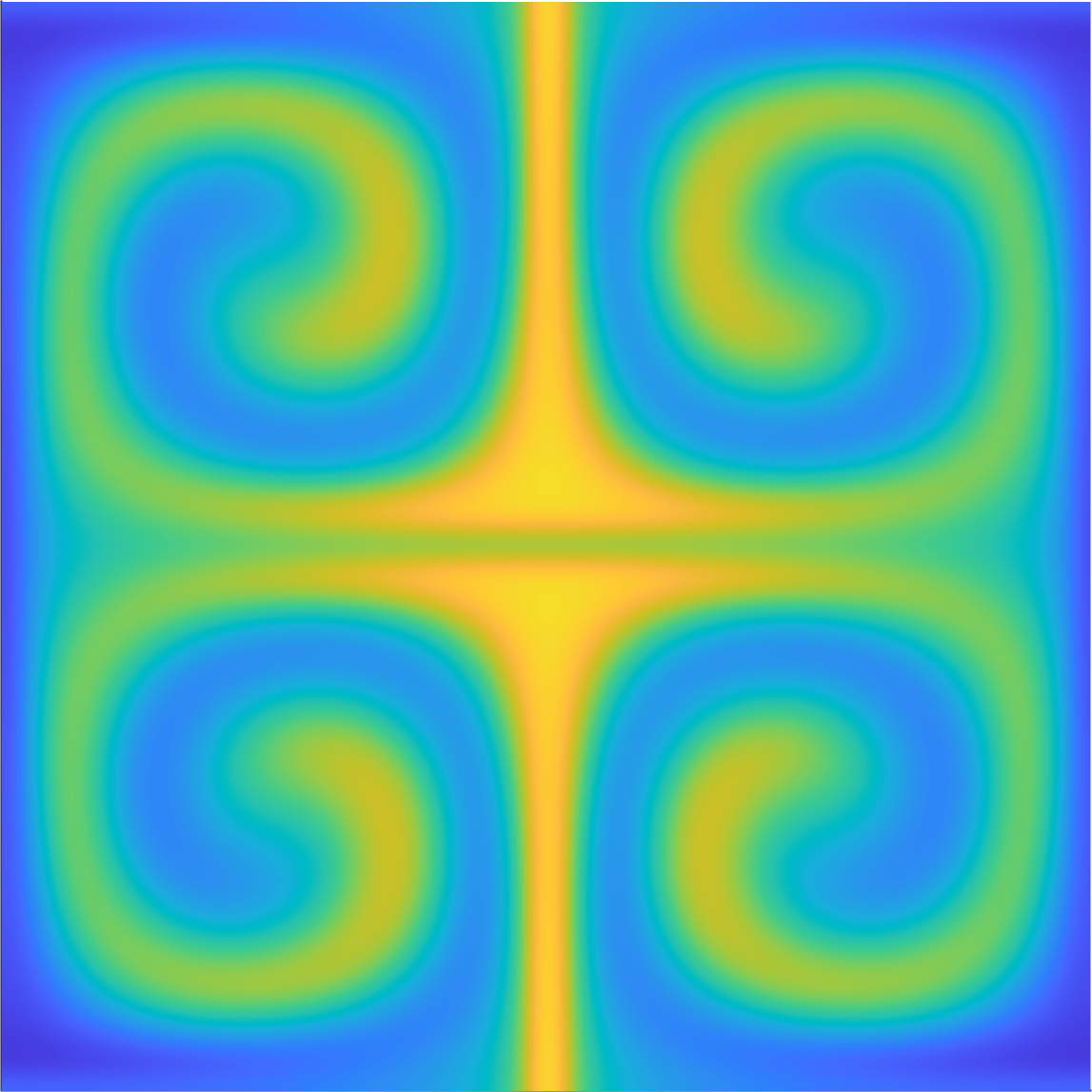} & 
\includegraphics[height=0.14\textheight]{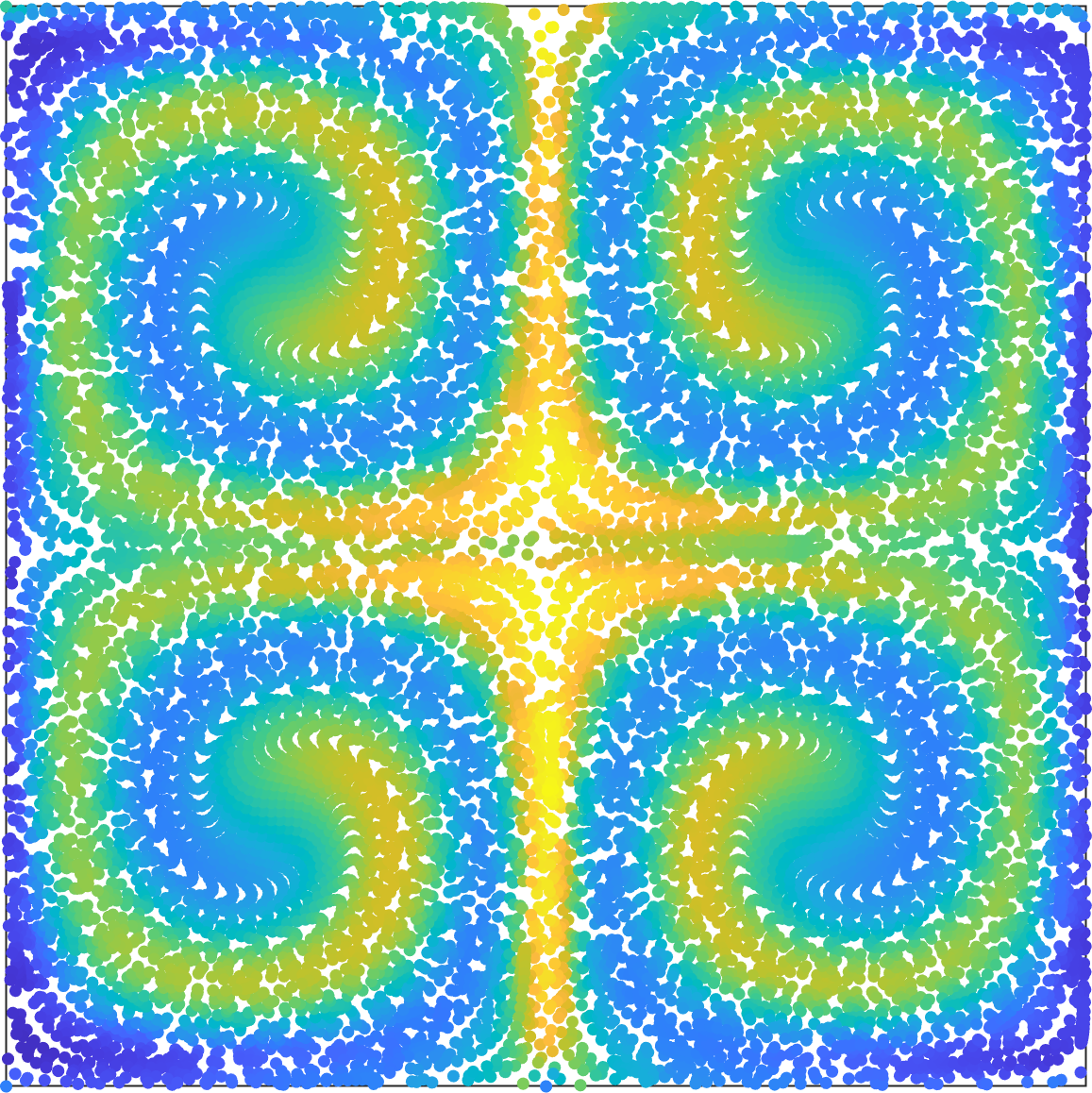} & 
\includegraphics[height=0.14\textheight]{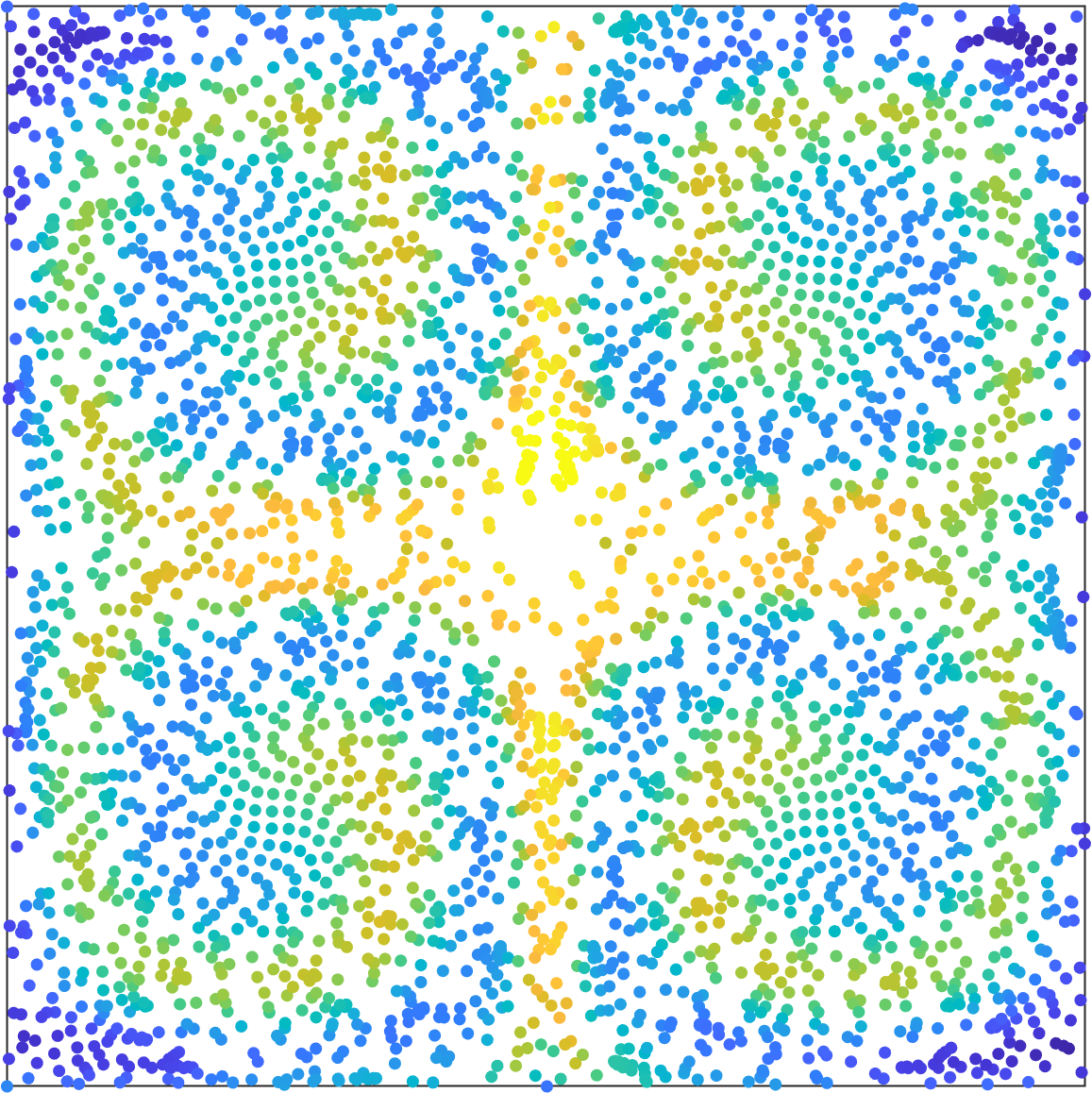} &
\includegraphics[height=0.14\textheight]{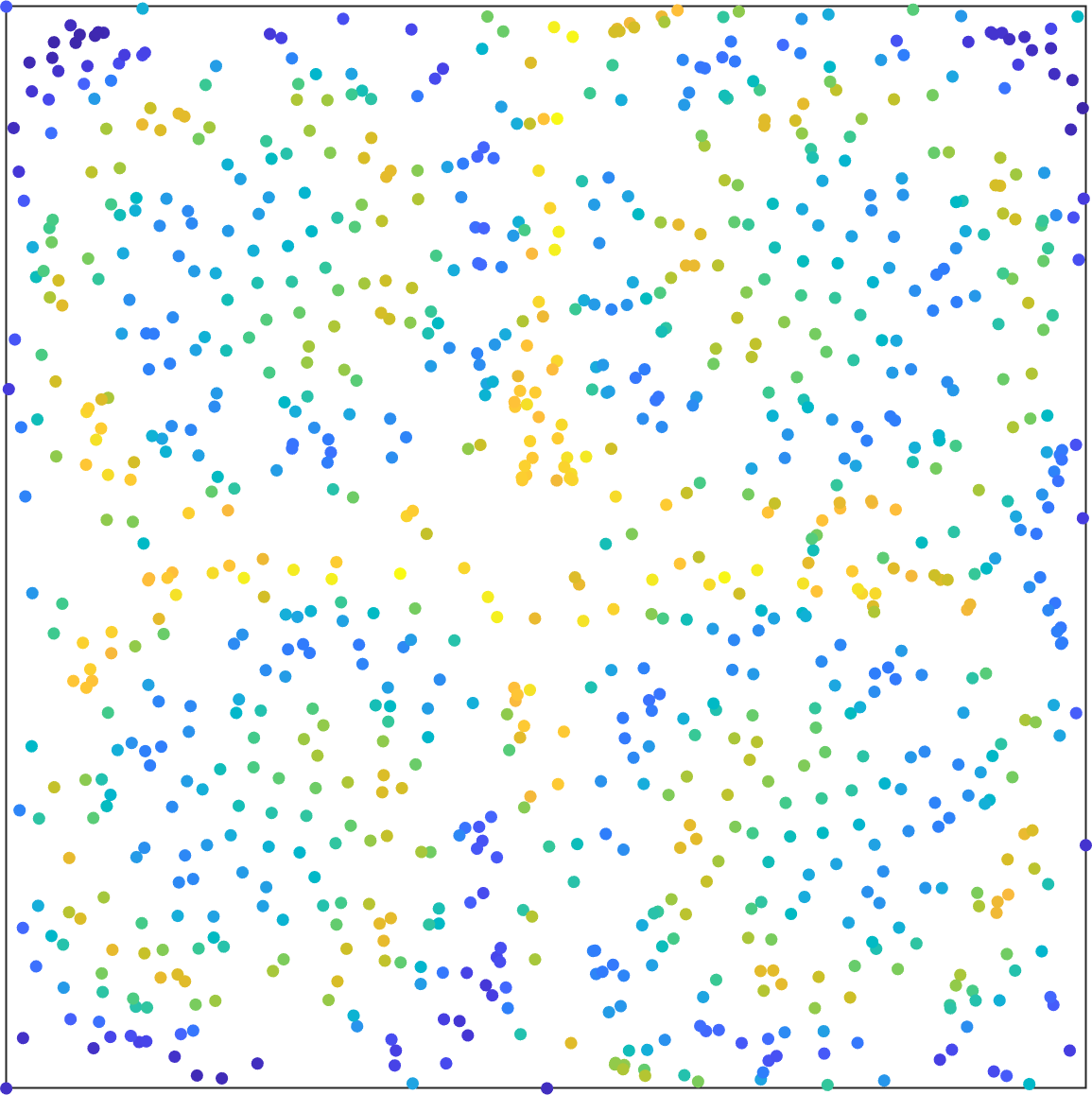} &
\includegraphics[height=0.14\textheight]{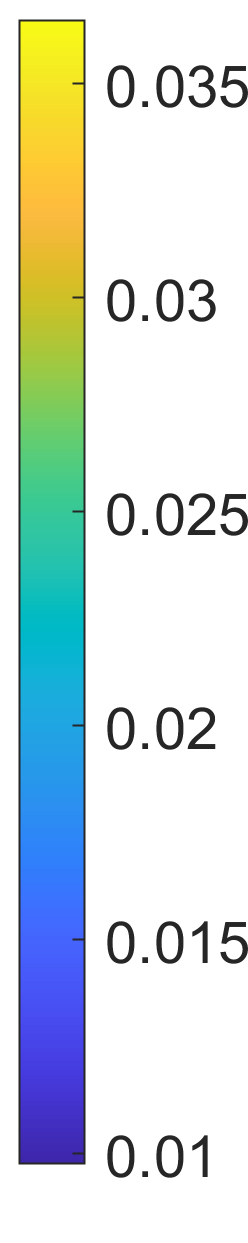}\\

\end{tabular}

\begin{tabular}{cc}
    \begin{minipage}[c]{0.65\textwidth}
        \includegraphics[width=\textwidth]{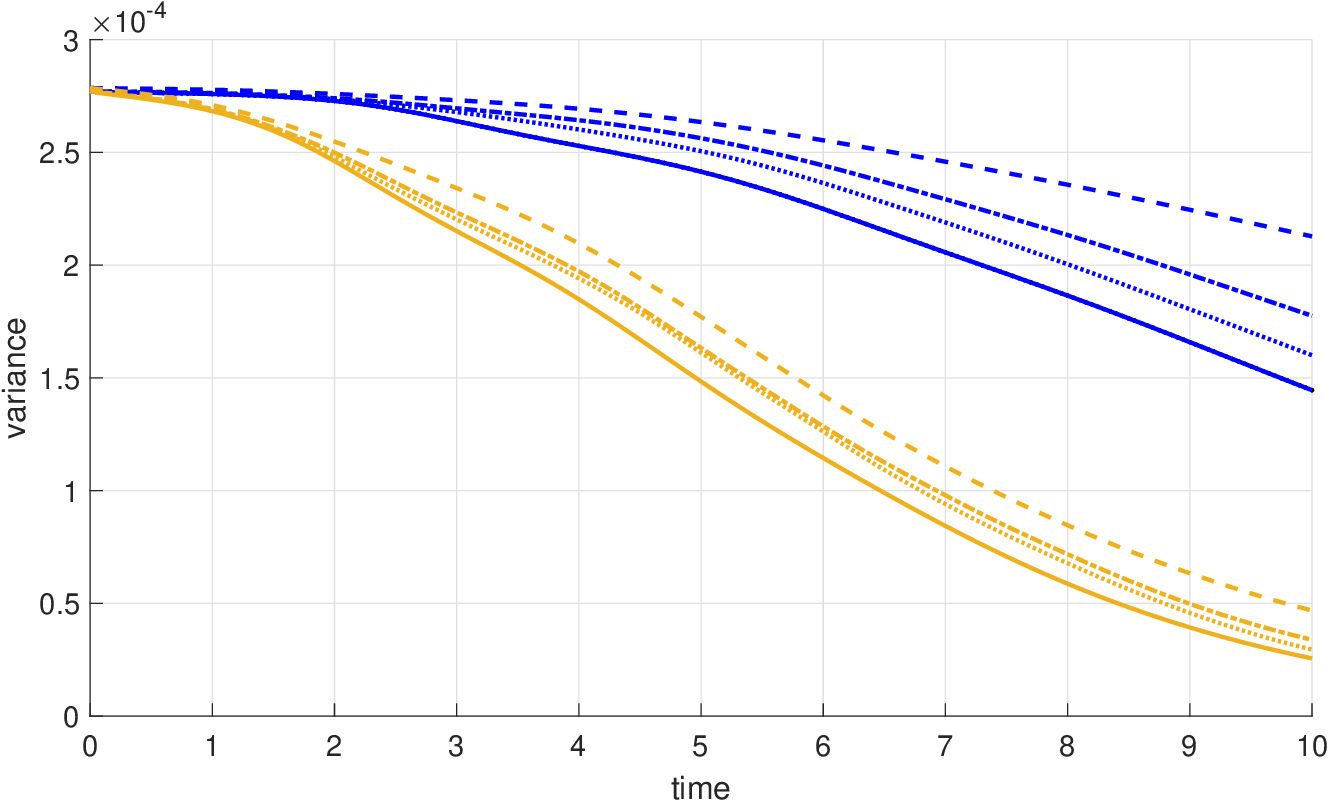}
    \end{minipage} 
    & 
    \begin{minipage}[c]{0.2\textwidth}
        \includegraphics[width=\textwidth]{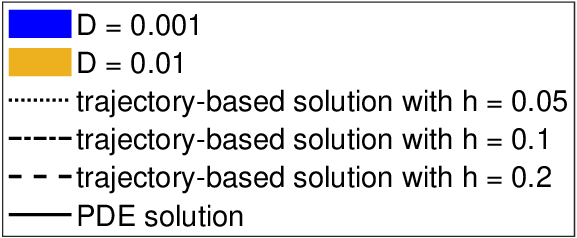}
    \end{minipage}
\end{tabular}

\caption{PDE solution (left column) at final time $t_{200}=10$ compared to the evolved trajectory-based density vectors $\bm{w}^{200}$ for the first initial condition (Figure \ref{fig:oc0}(b)) for the three different choices of grid spacing ($h=0.05, 0.1, 0.2$, columns 2--4) and two choices of the effective diffusion constant ($D=0.001, 0.01$) (rows 1--2). Bottom panel: For the quantification of mixing the respective variances over the time span $[0, 10]$ are plotted for $D=0.001$ (blue) and $D=0.01$ (orange) -- PDE solution (solid) and trajectory-based solutions with $h=0.05$ (dotted), $h=0.1$ (dash dotted), $h=0.2$ (dashed). }\label{fig:oc1}
\end{figure}

\begin{figure}[!htb]
\centering
\begin{tabular}{cccccc}
& {\scriptsize PDE } & {\scriptsize $h=0.05$} & {\scriptsize $h=0.1$} & {\scriptsize $h=0.2$}  \\[2mm]
{\scriptsize $D=0.001$} & 
\includegraphics[height=0.14\textheight]{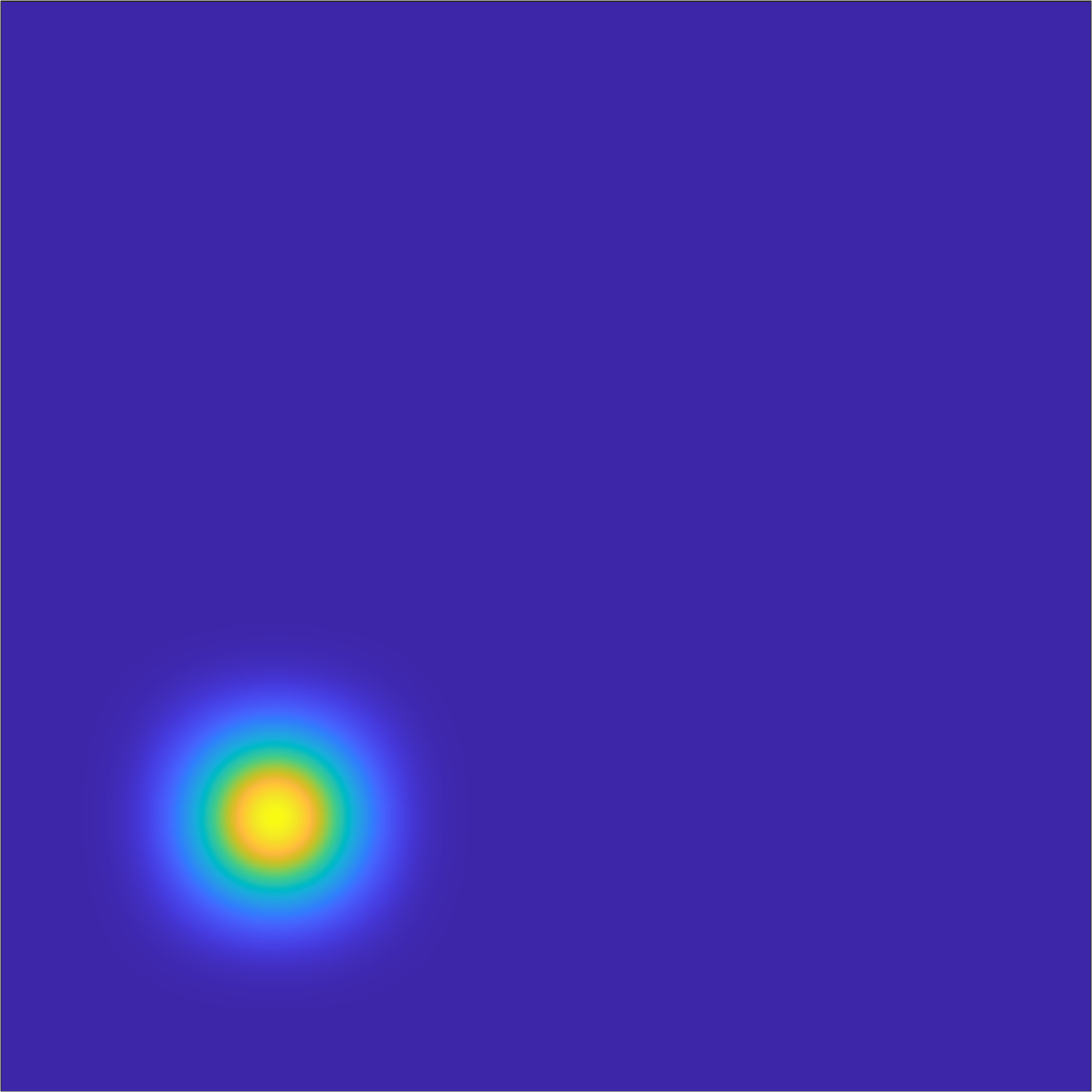} & 
\includegraphics[height=0.14\textheight]{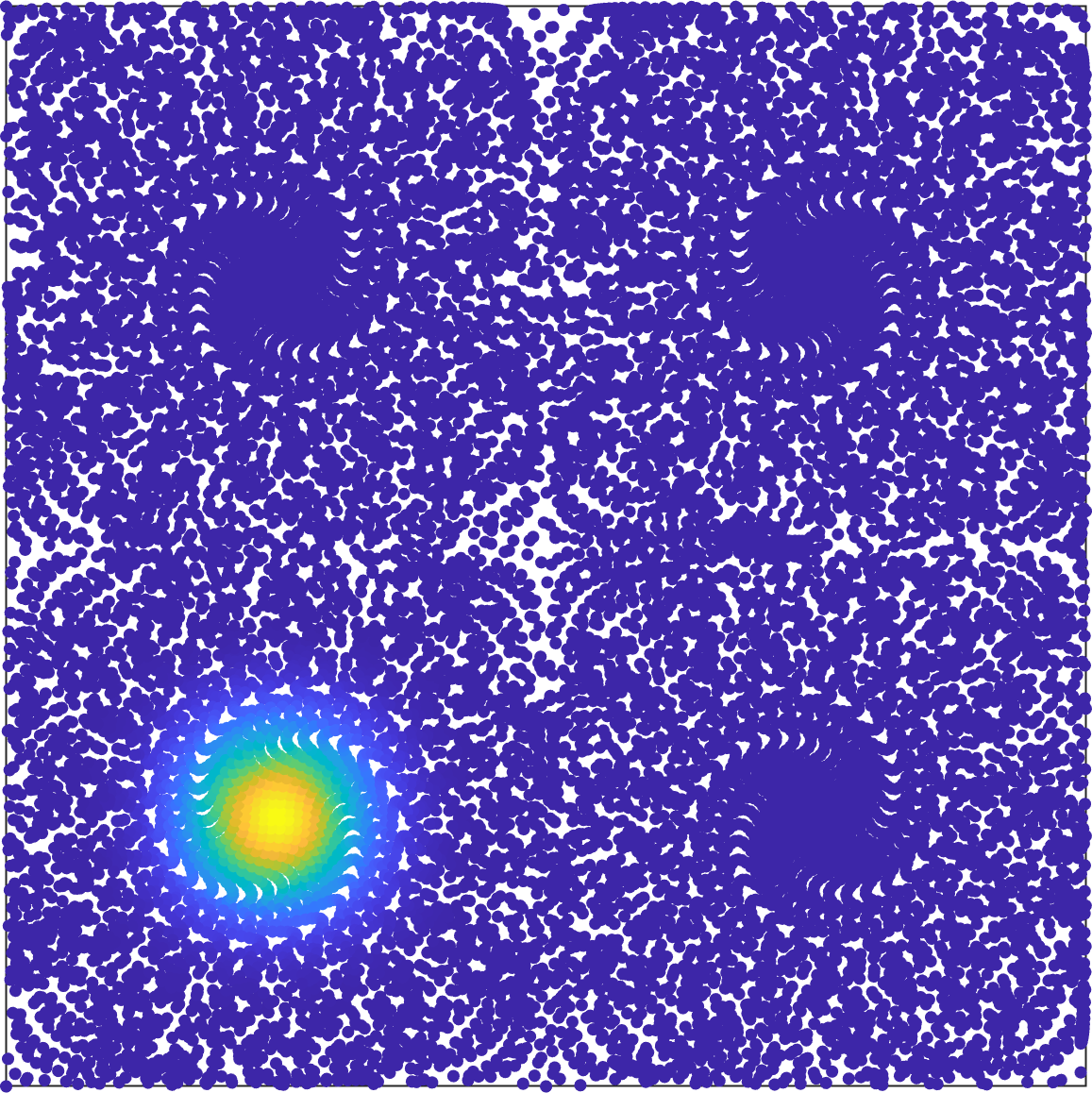} & 
\includegraphics[height=0.14\textheight]{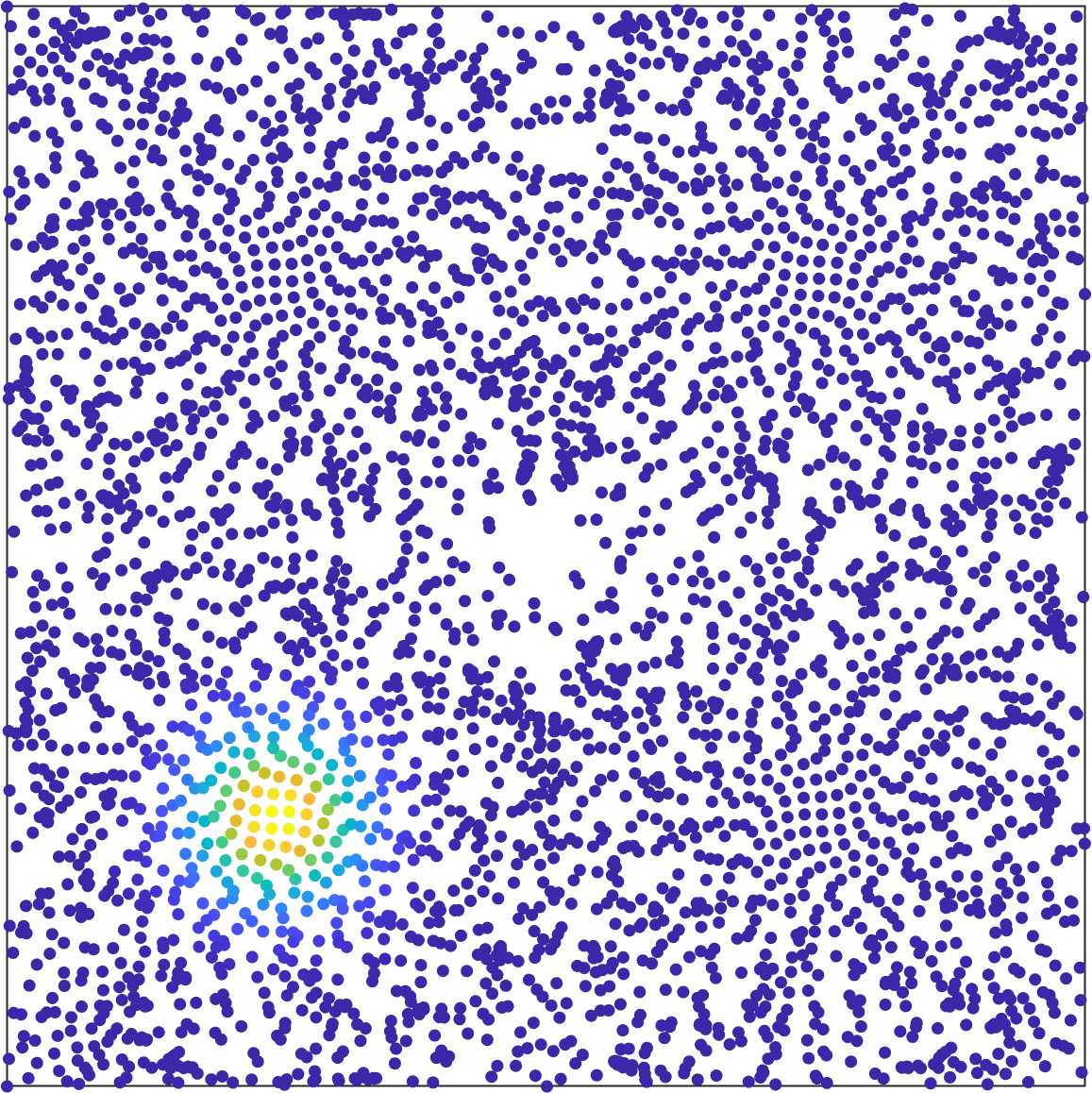} &
\includegraphics[height=0.14\textheight]{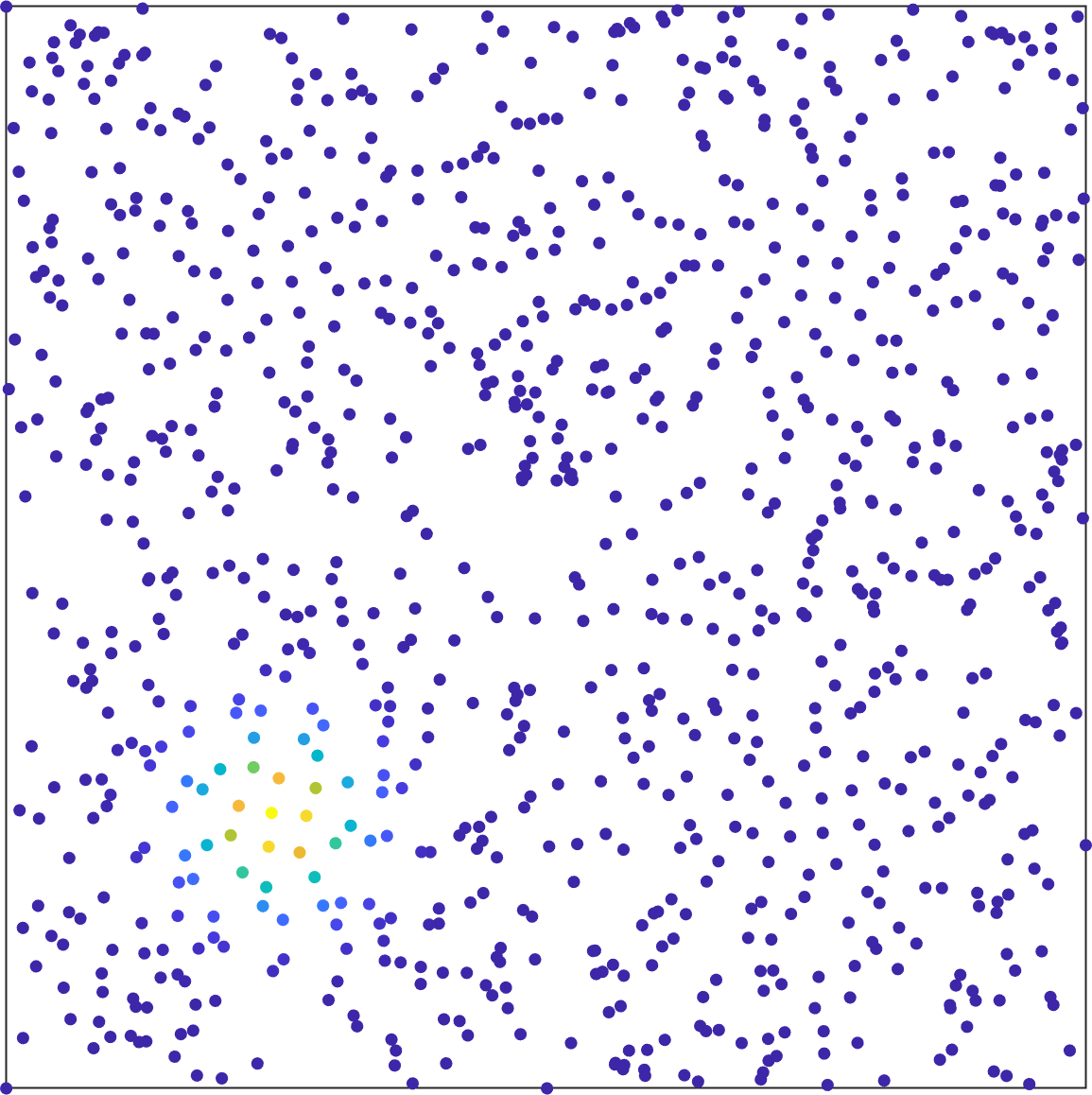} &
\includegraphics[height=0.14\textheight]{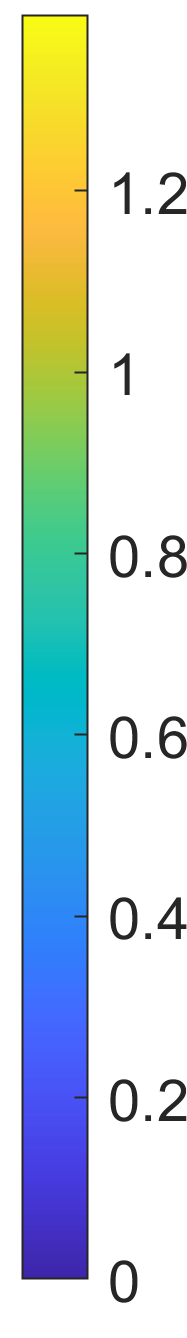} \\
{\scriptsize $D=0.01$} & 
\includegraphics[height=0.14\textheight]{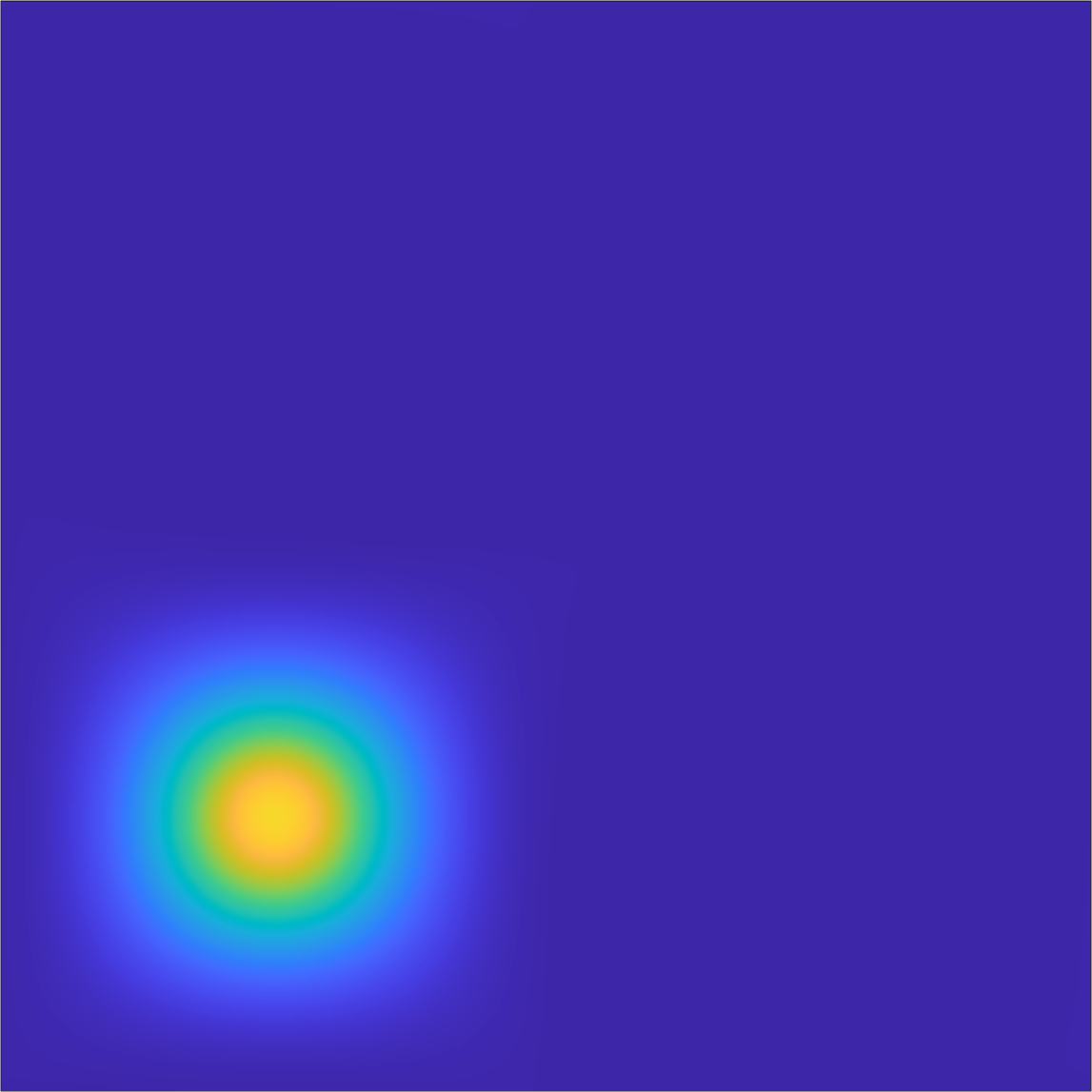} & 
\includegraphics[height=0.14\textheight]{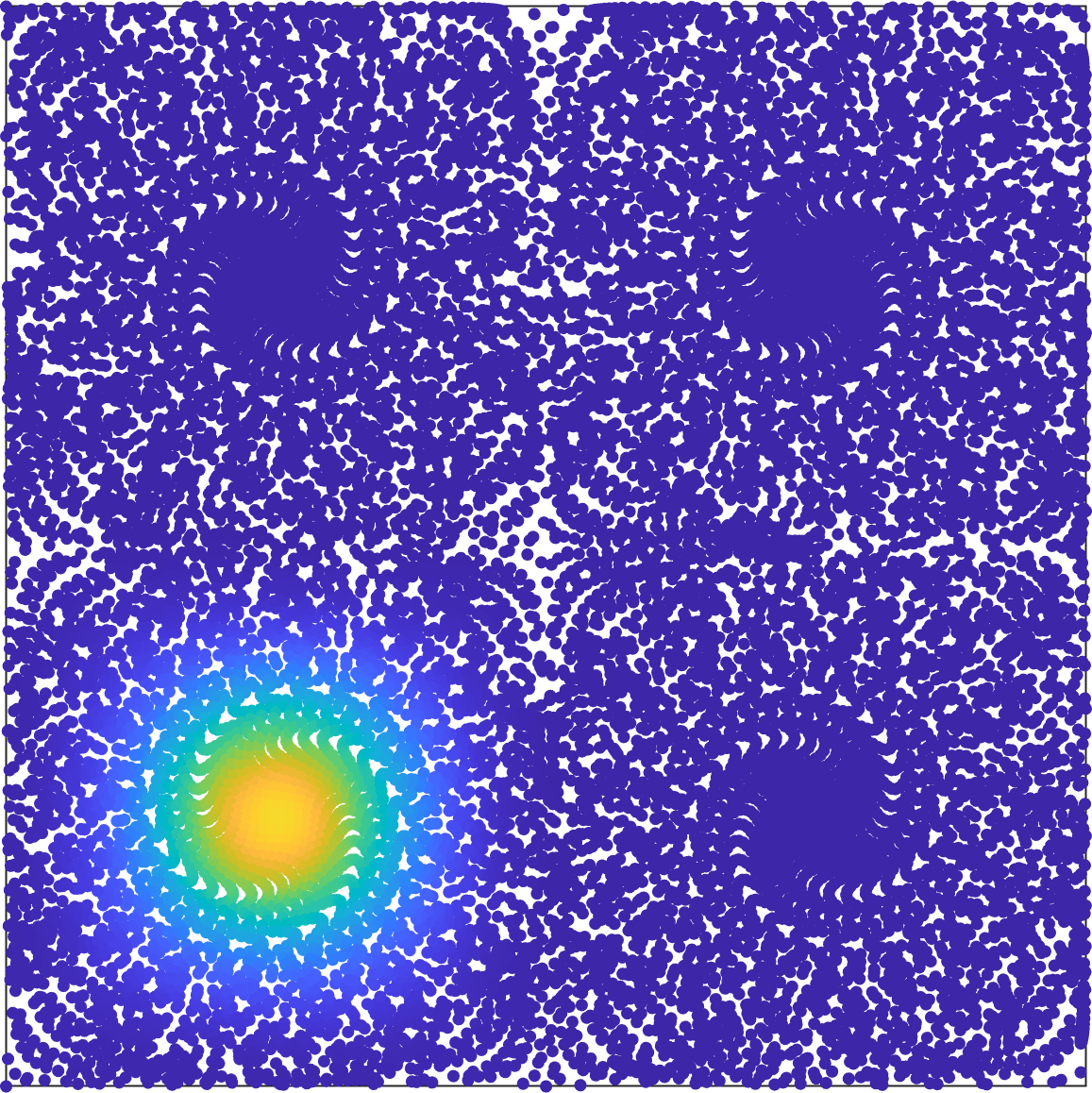} & 
\includegraphics[height=0.14\textheight]{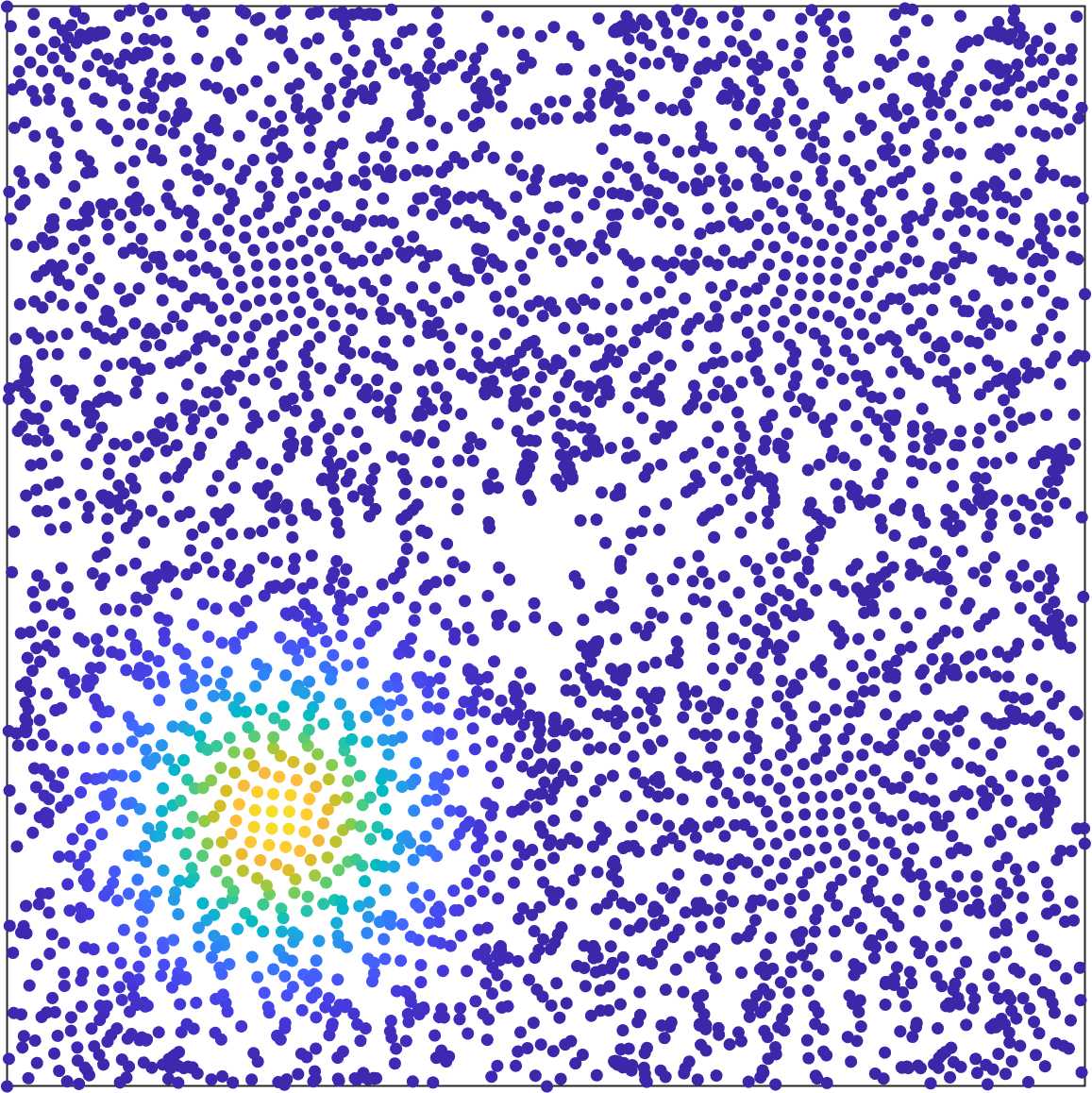} &
\includegraphics[height=0.14\textheight]{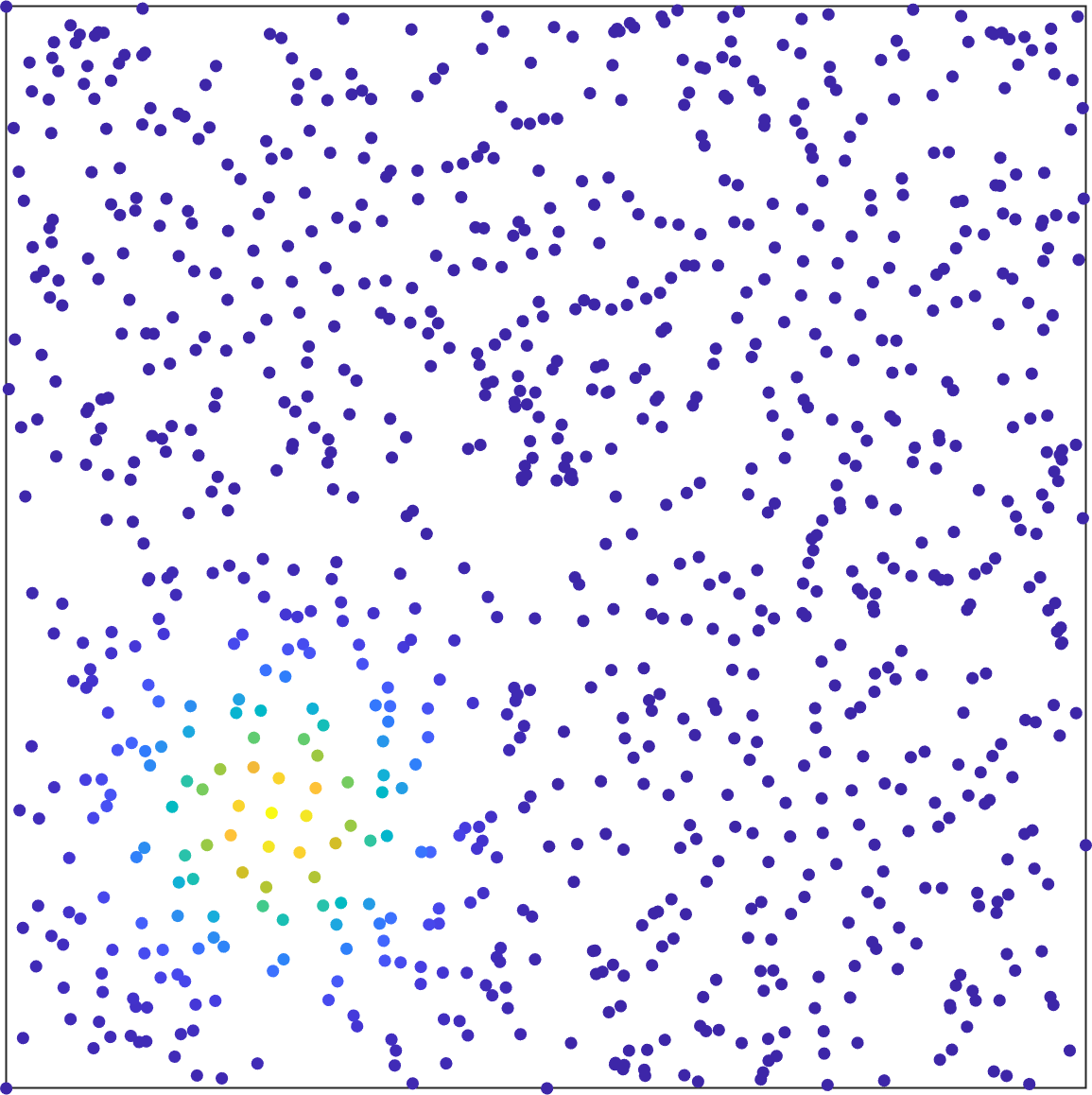}
&
\includegraphics[height=0.15\textheight]{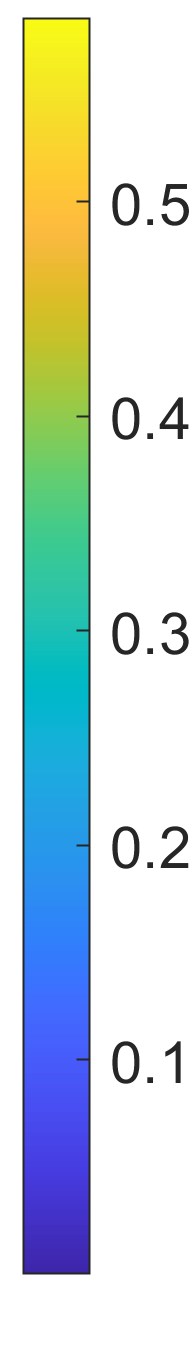}  \\
\end{tabular}

\begin{tabular}{cc}
    \begin{minipage}[c]{0.65\textwidth}
        \includegraphics[width=\textwidth]{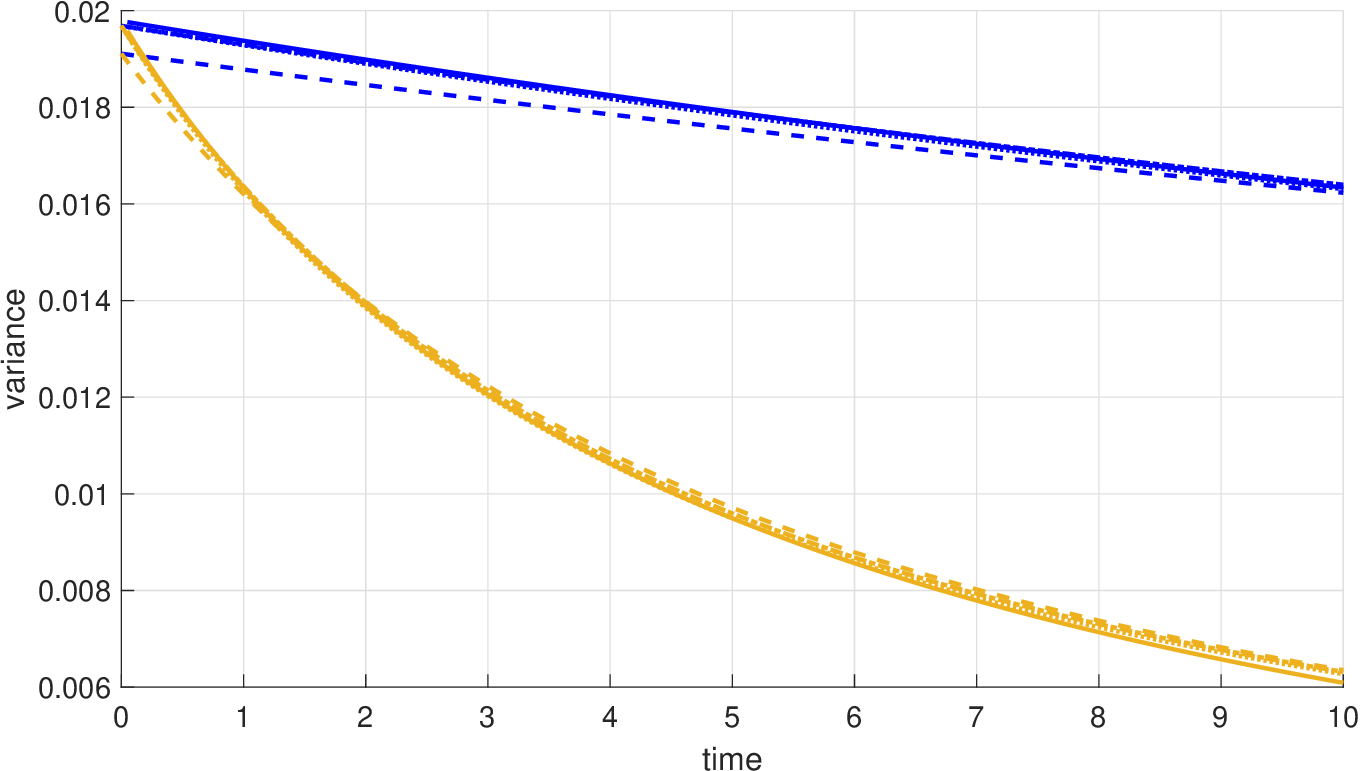}
    \end{minipage} 
    & 
    \begin{minipage}[c]{0.2\textwidth}
        \includegraphics[width=\textwidth]{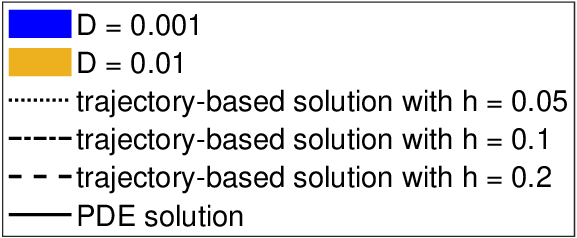}
    \end{minipage}
\end{tabular}

\caption{PDE solution (left column) at final time $t_{200}=10$ compared to the evolved trajectory-based density vectors $\bm{w}^{200}$ for the second initial condition (Figure \ref{fig:oc0}(c)) for the three different choices of grid spacing ($h=0.05, 0.1, 0.2$, columns 2--4) and two choices of the effective diffusion constant ($D=0.001, 0.01$) (rows 1--2). Bottom panel: For the quantification of mixing the respective variances over the time span $[0, 10]$ are plotted for $D=0.001$ (blue) and $D=0.01$ (orange) -- PDE solution (solid) and trajectory-based solutions with $h=0.05$ (dotted), $h=0.1$ (dash dotted), $h=0.2$ (dashed).}\label{fig:oc2}
\end{figure}

For our trajectory-based method, we initialize trajectories on a grid with spacing $h=0.05$ and on two coarser grids with $h=0.1, 0.2$, respectively. We form the respective density vector $\bm{w}^0$ by evaluating the initial condition field in the grid points. 
For each choice of $h$, we compute diffusion matrices $\bm{P}_{\epsilon}(t_k)$ for $\epsilon=\sqrt{2} h$ on the time span $[0, 10]$ using time step length $\tau=0.05$, i.e. we have 201 matrices, where $\mathbb{T}=\{0, 0.05, 0.1, \ldots, 10\}$.
We consider the advective-diffusive dynamics for two different effective diffusion constants $D=0.001, 0.01$ and resulting $\tilde{D}=4\tau D/\epsilon^2$. 

 Figures \ref{fig:oc1} and \ref{fig:oc2} show our evolved density vectors $\bm{w}^{200}$ for the trajectories at final time $t_{200}=10$ in comparison to the results of a highly resolved numerical solution of the corresponding advection-diffusion equation (using a spectral code provided by \cite{thiffeaultcode}) for the two different initial conditions. Visually there is good agreement throughout, even for the very coarse data set ($h=0.2$). 
 
 For the quantification of mixing we plot the sample variance over time for each setting, see the bottom panels of Figures \ref{fig:oc1} and \ref{fig:oc2}. For the first initial condition (as in Figure \ref{fig:oc0}(b)) all curves are very close to each other for the first couple of time steps, until about $t=2$ (Figure \ref{fig:oc1}). Afterwards, the trajectory-based variances start to increasingly deviate from the PDE results, although the overall shapes of the curves remains similar. This deviation is due to the fact that the particles are no longer uniformly distributed and gaps appear in particular in the vicinity of hyperbolic structures, which however should take larger values of the scalar field. There are two observations: the finer the data the closer are the results to that of the PDE. Moreover, a higher effective diffusion constant ($D=0.01$ compared to $D=0.001$) appears to have a regularizing effect (compare orange ($D=0.01$) and blue curves ($D=0.001$) in Figure \ref{fig:oc1}).
 For the second initial condition, a Gaussian centered on one of the gyres (as in Figure \ref{fig:oc0}(c)), all the variance curves nearly coincide, except for the coarse data set, where the grid spacing appears to be too large to capture the initial condition with sufficient accuracy (Figure \ref{fig:oc2}). Due to the elliptic dynamics close to the center point where the initial density is supported the particles do not move much and remain approximately uniformly distributed, so that we observe the results of effective diffusion rather than an advection-diffusion.

%\newpage

\subsection{Closed double gyre system} \label{sec:cdg}
\begin{figure}[!htb]
\begin{center}
\begin{tabular}{ccc}
\includegraphics[width=0.3\textwidth]{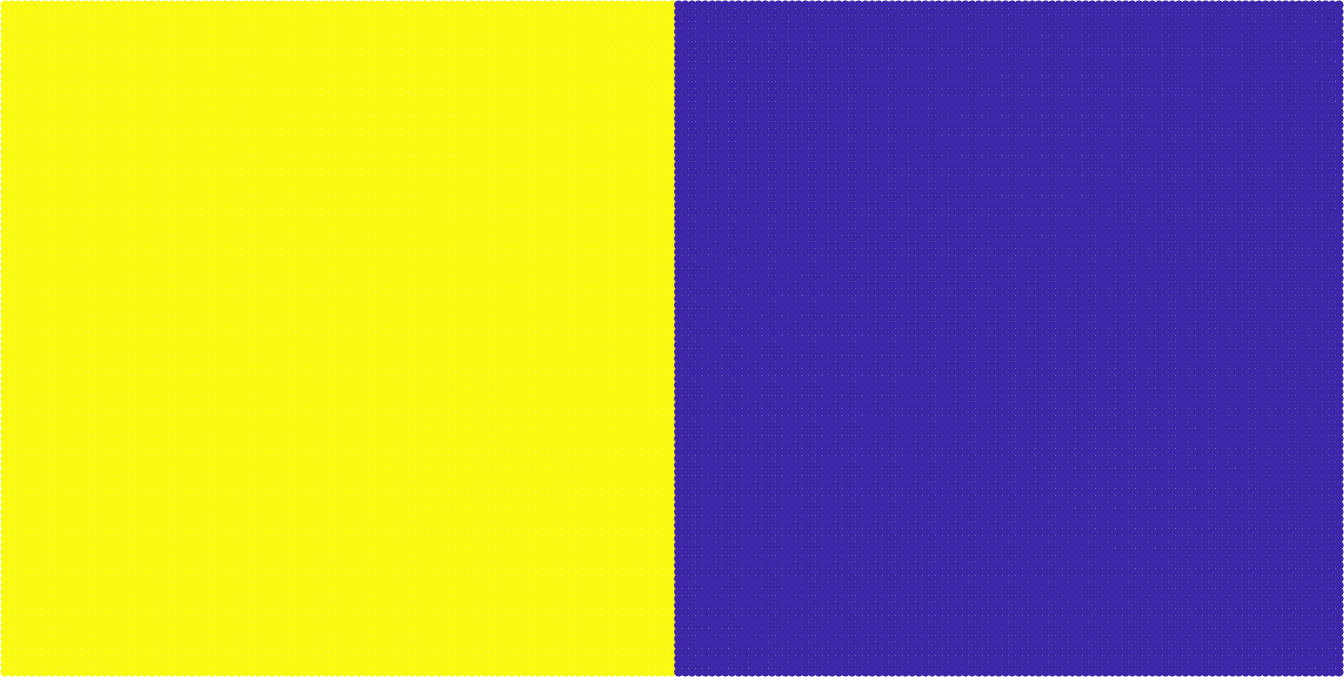} & \includegraphics[width=0.3\textwidth]{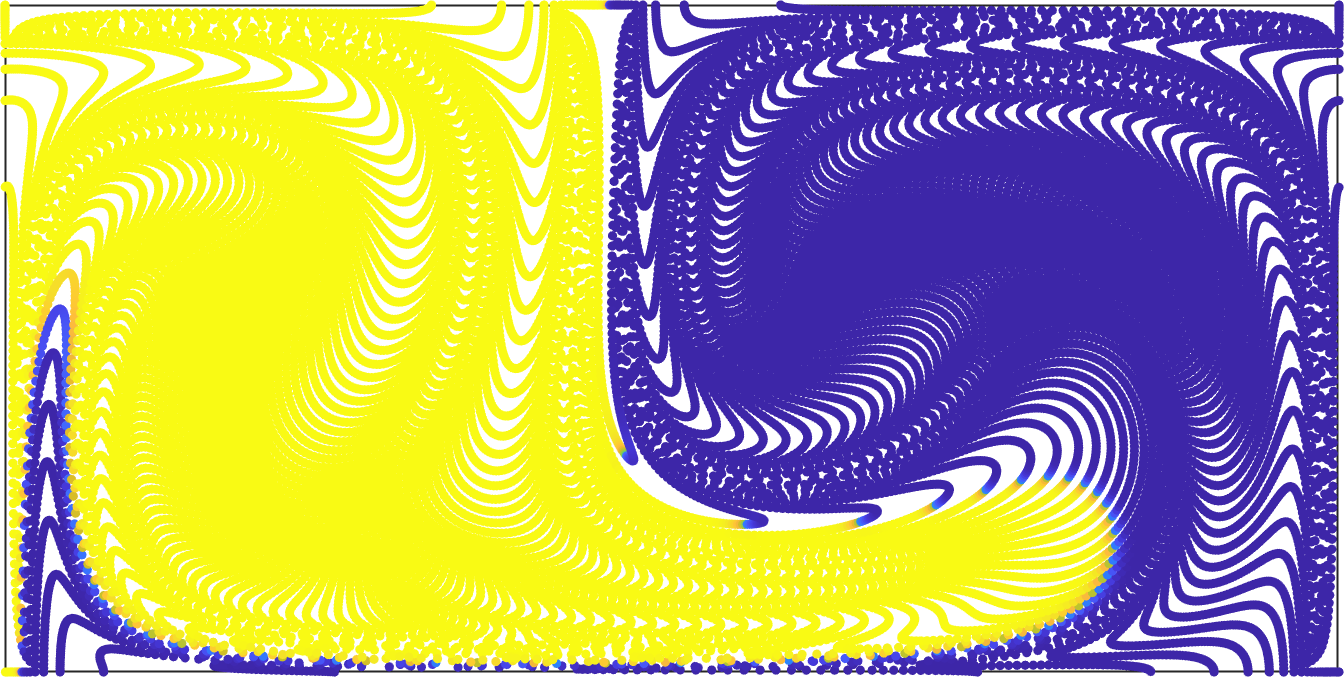}& \includegraphics[width=0.3\textwidth]{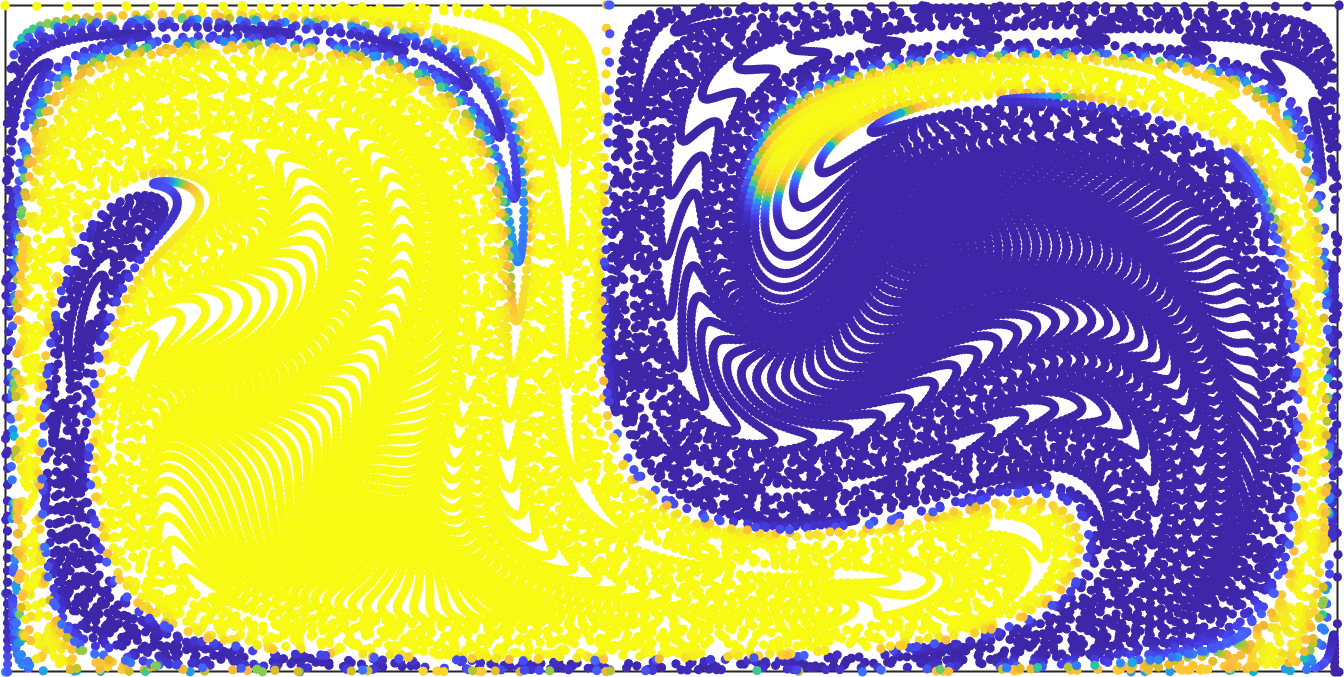} \\
{\scriptsize $\bm{w}^0$ (at time $t_0=0$) } & {\scriptsize $\bm{w}^{10}$ (at time $t_{10}=1$) } & {\scriptsize $\bm{w}^{20}$ (at time $t_{20}=2$) }  \\[2mm] 
 \includegraphics[width=0.3\textwidth]{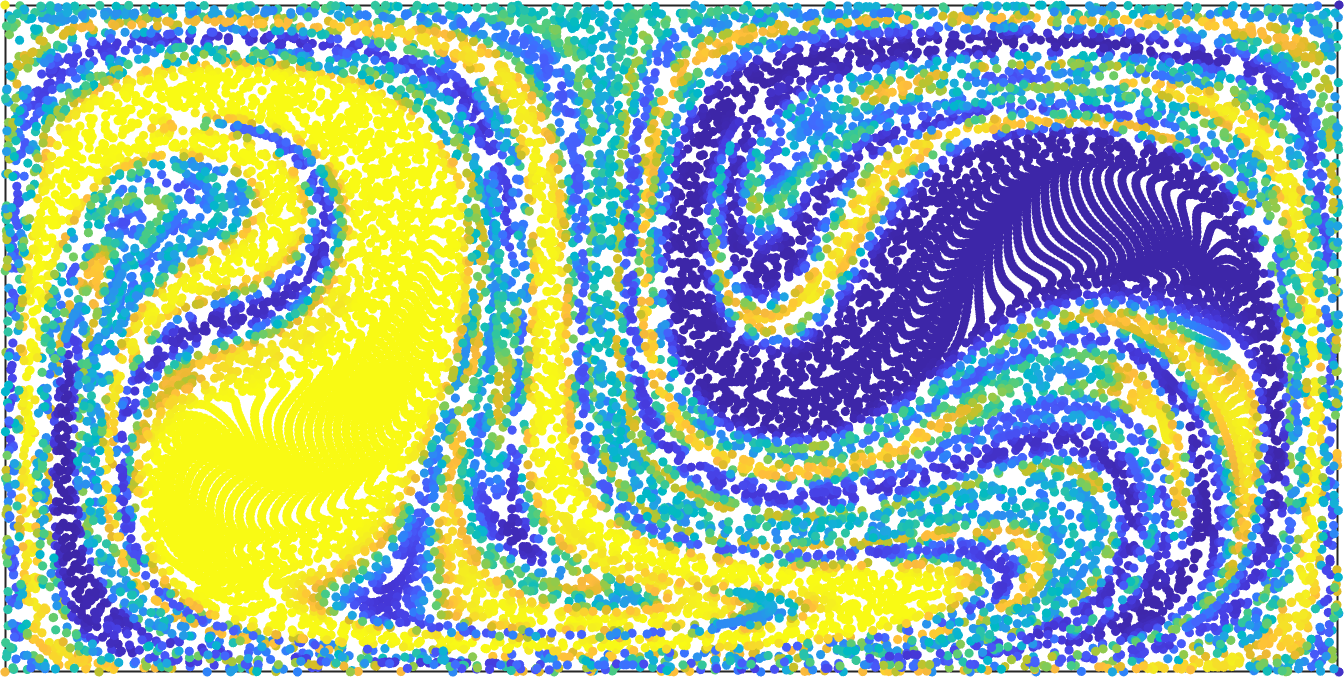} & \includegraphics[width=0.3\textwidth]{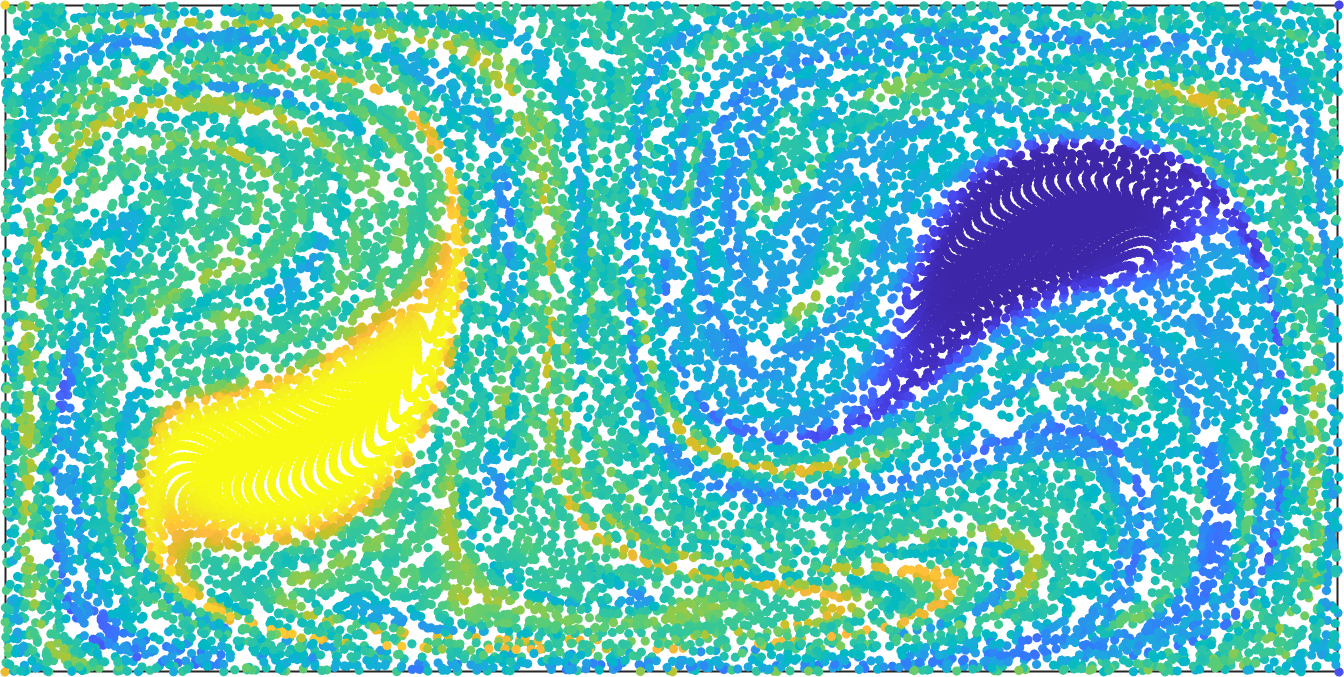}& \includegraphics[width=0.3\textwidth]{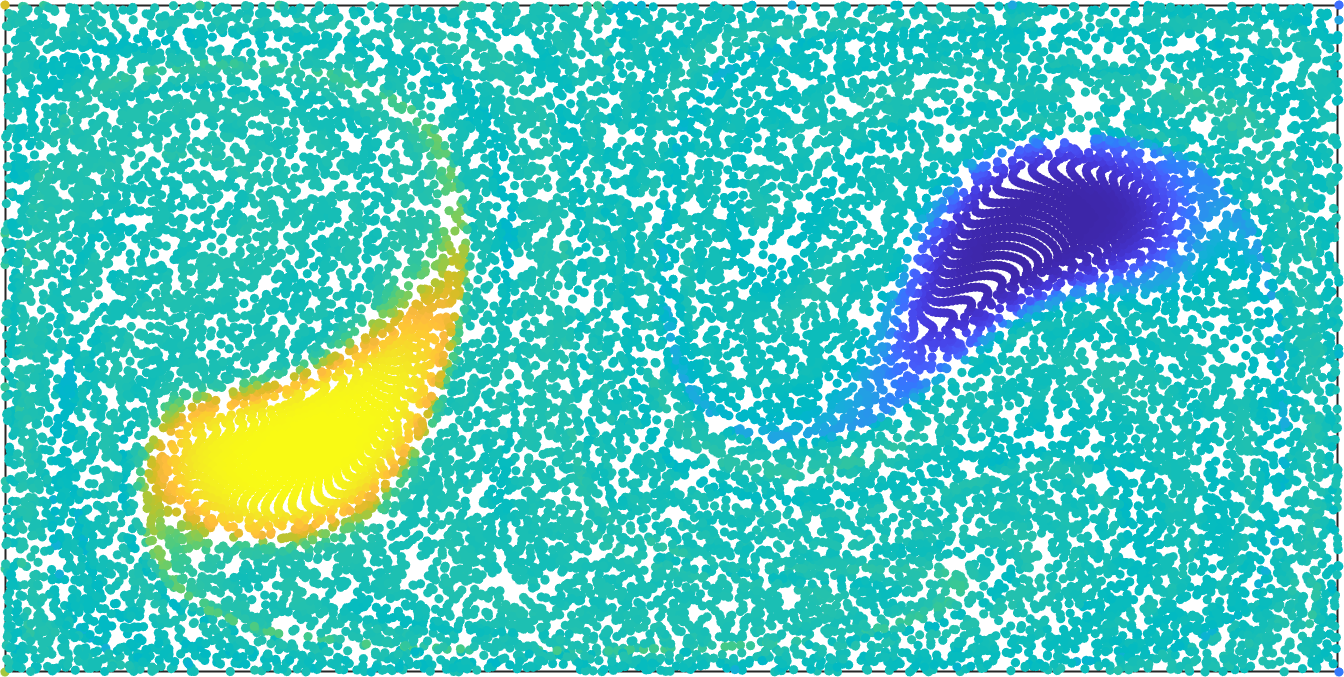} \\
{\scriptsize $\bm{w}^{50}$ (at time $t_{50}=5$)} & {\scriptsize $\bm{w}^{100}$ (at time $t_{100}=10$)} & {\scriptsize $\bm{w}^{200}$ (at time $t_{200}=20$)}  \\[2mm] 
 \includegraphics[width=0.3\textwidth]{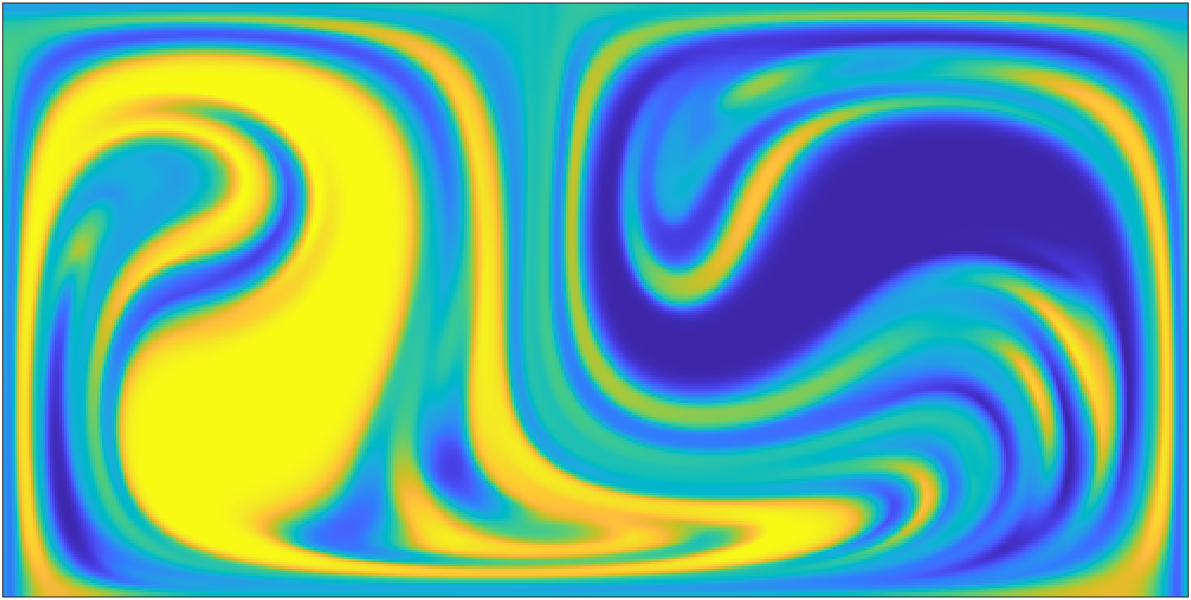} & \includegraphics[width=0.3\textwidth]{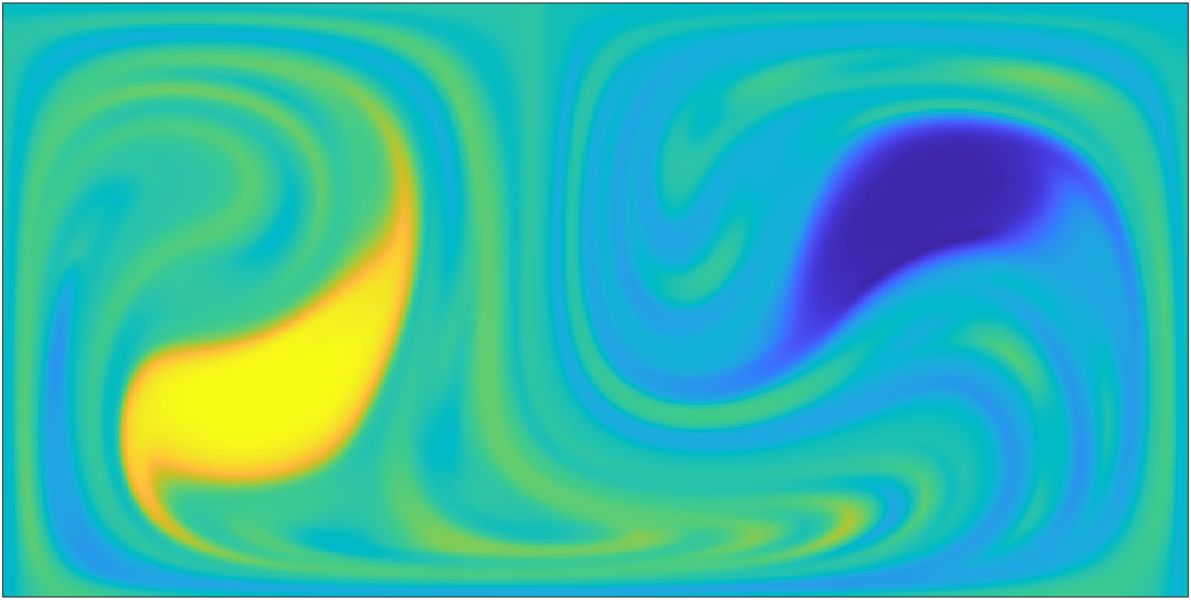}& \includegraphics[width=0.3\textwidth]{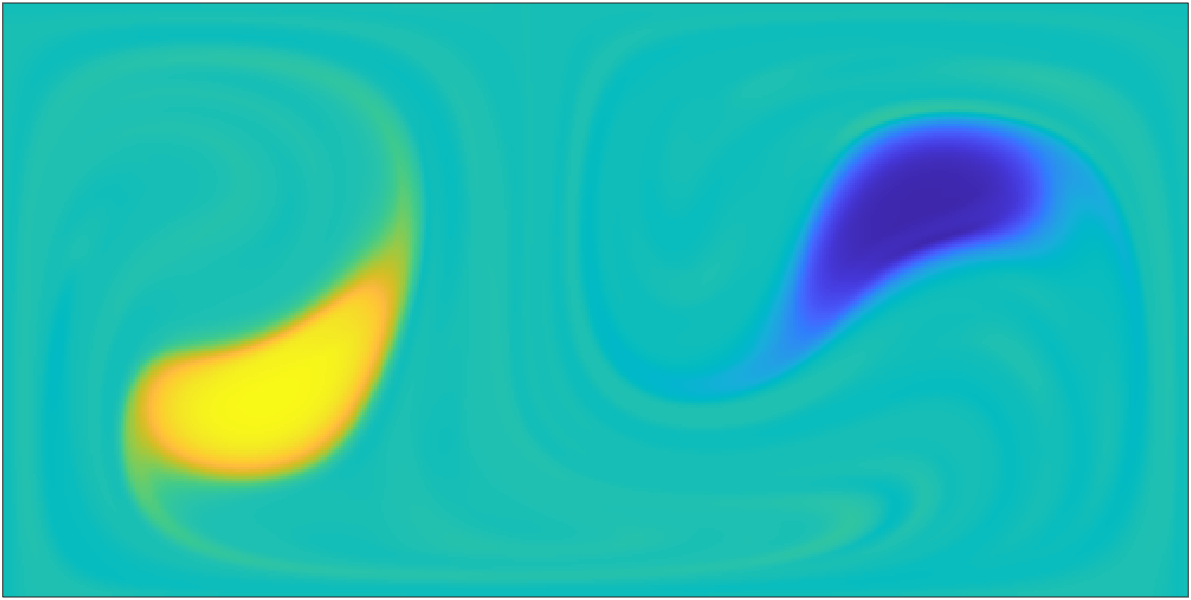} \\
{\scriptsize PDE solution (at time $t=5$)} & {\scriptsize  PDE solution (at time $t=10$)} & {\scriptsize  PDE solution (at time $t=20$)}  \\[2mm] 
& & \includegraphics[width=0.3\textwidth]{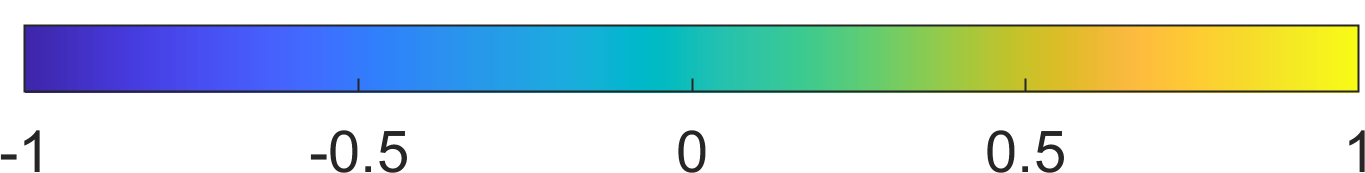}
\end{tabular}
\includegraphics[height=0.25\textheight]{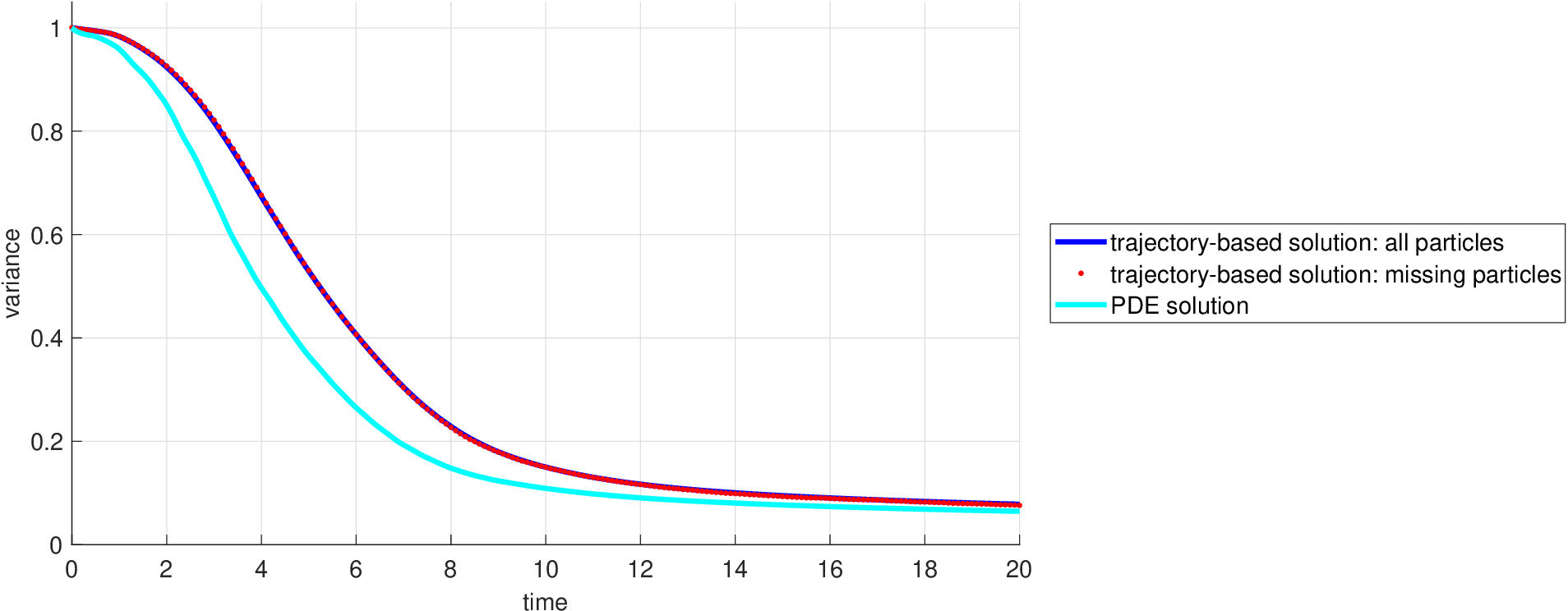} 
\end{center}
\caption{Particles with coevolved color vector $\bm{w}^k$ in the closed double gyre system with parameters $A=0.5$, $\delta=0.2$ (rows 1 \& 2). Mixing is observed outside the two coherent sets. 
The trajectory based results compare well to the numerical solution of the PDE using a finite-volume scheme (row 3). Variances over time for the complete and incomplete data set nearly coincide and only converge very slowly due to the two unmixed regions. This is also true for the variance of the PDE solution, which however decreases initially a bit faster, but with a very good alignment to the data-based curves for larger times.}\label{fig:cdg_zk2}
\end{figure}

We consider the well-known periodically perturbed double gyre flow \cite{shadden2005definition} 
with time-dependent stream function
\begin{equation}
\Psi_{\mathrm{m}}(t, \bm{x})= -A \sin(f(t,x)\pi)\sin(\pi y), \label{eq:dgstreamfunction}
\end{equation}
here $f(t,x)=\delta \sin(\omega t) x^2 +(1-2\delta \sin(\omega t)) x$ models the periodic perturbation with amplitude $\delta \geq 0$ and frequency $\omega$, and $A>0$ controls the amplitude of the rotation speed of the gyres. In the following, we fix $A=0.5$ and set $\omega=2 \pi$, so that the time period of the flow is $1$, and study the system for different $\delta$.

 First we demonstrate the mixing of two differently colored fluids on the invariant domain $M=[0,2]\times[0,1]$ with system parameter $\delta= 0.2$. For this, we initialize 20,301 particles on a grid with width $h=0.01$. Particles on the left half of the domain belong to the first fluid and get the value 1 in $\bm{w}^0$ (presented by yellow); and particles on the right of the domain belong to the second fluid and get the value -1 in $\bm{w}^0$ (presented by blue). We obtain the trajectory data by the classical Runge Kutta method for a time span of length $20$ and evaluate the trajectories at 201 time steps (time step length $\tau=0.1$).

We choose $ \epsilon=\sqrt{2}h$ and compute the 201 diffusion matrices $\bm{P}_{\epsilon}(t_k)$ for $t_k\in \mathbb{T}=\{0, 0.1,  \ldots, 20\}$. The vector $\bm{w}^0$ is evolved over these time steps to $\bm{w}^{200}$ using \eqref{eq:coevolution} (see section \ref{sec:trajdiffmap}), where we choose a small diffusion constant of $D=0.00005$.

Figure \ref{fig:cdg_zk2} (upper panel) shows the particles colored according to the values of the coevolved vectors $\bm{w}^k$ for a selection of times $t_k \in \mathbb{T}$. As expected from the initial condition and the dynamics of the double gyre, mixing happens predominantly along the unstable manifold of the hyperbolic periodic orbit that oscillates in the center of the upper boundary of the domain. Inside the gyre cores the particles are maintaining their color -- this is not surprising as these regions correspond to well-studied coherent sets. Therefore, the variance (blue curve in Figure \ref{fig:cdg_zk2}, lower panel) converges only very slowly.
The trajectory-based results compare very well to those of a numerical solution of the corresponding advection-diffusion equations, obtained using a high-order finite-volume scheme. This especially applies to the time-evolved fields (compare rows 2 and 3 in Figure \ref{fig:cdg_zk2}).  The variance observed over time on $[0,20]$ (cyan curve in Figure \ref{fig:cdg_zk2}, lower panel) decreases initially a bit faster for the PDE case, which is to be expected given the very small effective diffusion constant and in good accordance with the results of the previous example in section \ref{sec:cellflow}. For larger times the trajectory-based and PDE results align very well, which demonstrates the correctness of the data-based construction. 

We now consider the same set-up but with gaps in observation. For this, we delete particle positions: The time span of missing for each particle is geometrically distributed ($p=0.02$). The centers of the missing time spans are then uniformly distributed over $\mathbb{T}$, resulting in about 10-25\% of the particles missing in each time step. We set the vector entries for missing particles to NaN and update the value for (re-)appearing particles using a weighted average based on the neighboring particles' values (approach (ii) in section \ref{sec:opensystem}). Mixing over time as measured by the sample variance appears to be unaffected by the present gaps in observation as demonstrated in Figure \ref{fig:cdg_zk2} (lower panel).

%%%%%%%%%%%%%%%%%%%%%%%%%%%%%%%%%%%%%%%%%%%%%%%%%%%%%%%%%%%%%%%%%%%%%%%%%%%%%%%%%%%%%%%%%%%%%%%%

\subsection{Open double gyre mixer}\label{sec:odg}

\begin{figure}[htb]
\centering
\includegraphics[width=1\textwidth, trim={4cm 4cm 4cm 1cm},clip]{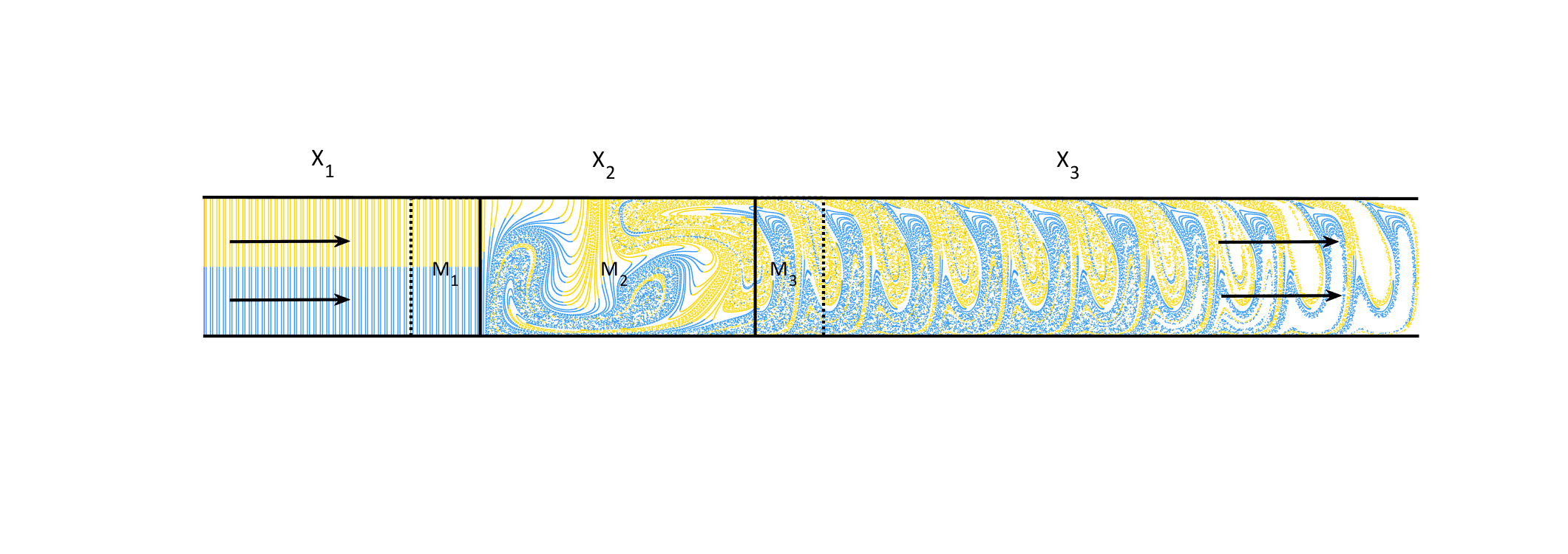} 
\caption{Set-up of the open double gyre mixer \eqref{eq:dgmixer}, where fluid of two different colors in inserted from the left, gets advected through a stirring region ($X_2$) to an outlet region. The sketch shows the stirring of the two differently colored fluids without additional effective diffusion.} \label{odg:shema}
\end{figure}

We now model an open system with an in- and outflow region based on the double gyre system as proposed in \cite{klunker2022open}. By adding a stationary background flow, the domain $M$ of the double gyre flow becomes the bounded stirring region $X_2$ on an infinite strip $X=(-\infty, \infty)\times[0,1]$, which consists further of an unbounded  unmixed region $X_1$ and an unbounded mixed region $X_3$ (Figure \ref{odg:shema}). 

For the background flow that advects fluid from $X_1$ through the stirring region and finally into $X_3$, we choose the constant velocity field $\bm{u}_{\mathrm{b}}$ with stream function
\begin{equation*}
  \Psi_{\mathrm{b}}(\bm{x})= \beta y, \qquad
  \mbox{ with } \beta >0 \mbox{ and } \bm{x}=(x,y).
\end{equation*}
The velocity field on $X$ then has the form
\begin{equation}\label{eq:dgmixer}
  \bm{u}(t, \bm{x})=\bm{u}_{\mathrm{b}}(\bm{x})+\bm{u}_{\mathrm{m}}(t, \bm{x}) \mathbf{1}_{[0,2]}(x),
\end{equation}
where $\bm{u}_{\mathrm{m}}$ is derived from the stream function $\Psi_{\mathrm{m}}$ of the double gyre flow \eqref{eq:dgstreamfunction}. 

In the following, we study the dynamics of the open subsystem restricted to $M=[-0.5,2.5]\times [0,1]$ with inlet region $M_1=[-0.5,0]\times [0,1]\subset X_1$, stirring region $M_2=X_2$, outlet region $
M_3=[2,2.5]\times [0,1] \subset X_3$. We assume that the particles that start (or later get inserted) on the upper half of $X_1$ are of a different color (yellow, modeled by $+1$ in the coevolved vector $\bm{w}^k$) than particles that start on the lower half (blue,  $-1$).

We focus on the mixing patterns resulting from the simulated advective dynamics combined with our data-based diffusion process on the outlet region $M_3$ after a time span of length $8$. We will consider different choices of the system parameter $\delta$ of the double gyre flow \eqref{eq:dgstreamfunction}, when the two types of fluids are sent through the stirring region by means of the flow field \eqref{eq:dgmixer}. 

\begin{figure}[ht]
\centering
\begin{tabular}{cc}
\includegraphics[width=0.35\textwidth]{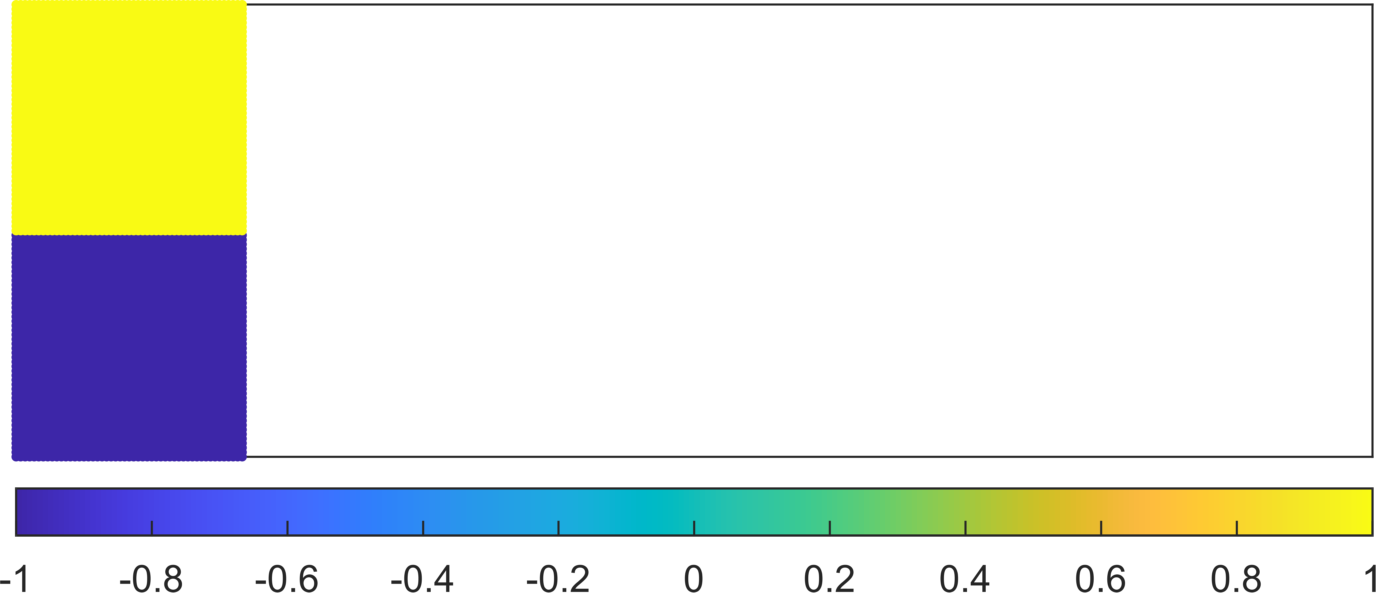} 
&\includegraphics[width=0.35\textwidth]{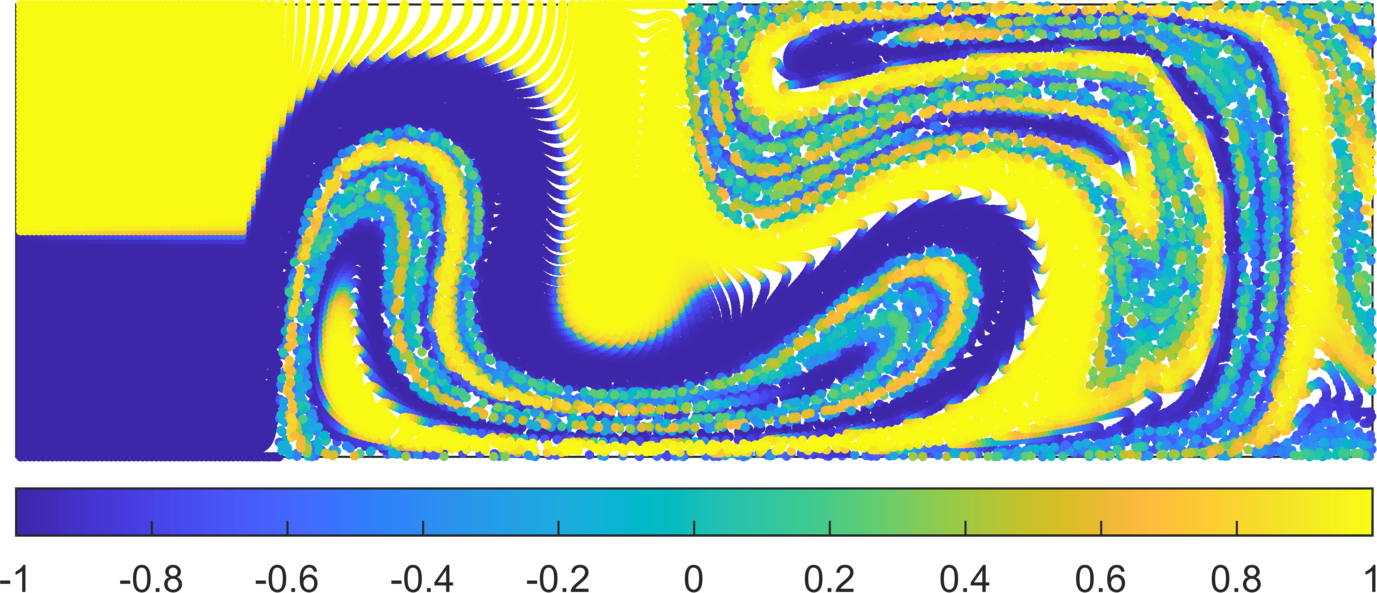}\\
\includegraphics[width=0.35\textwidth]{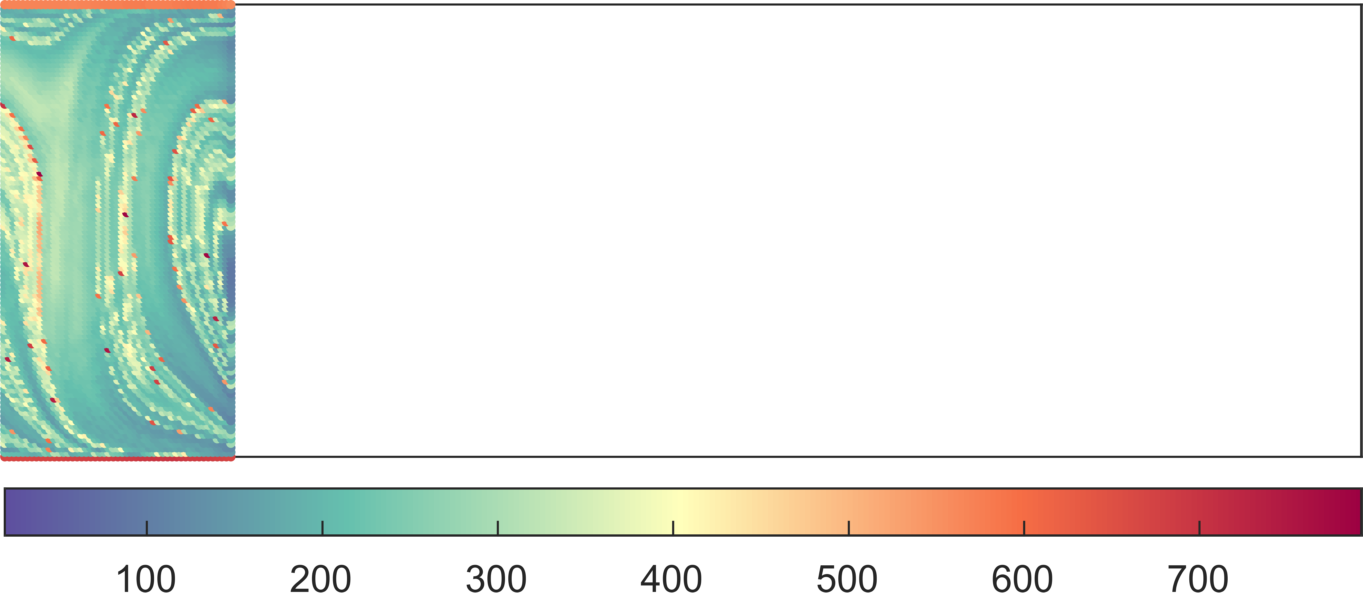} 
&\includegraphics[width=0.35\textwidth]{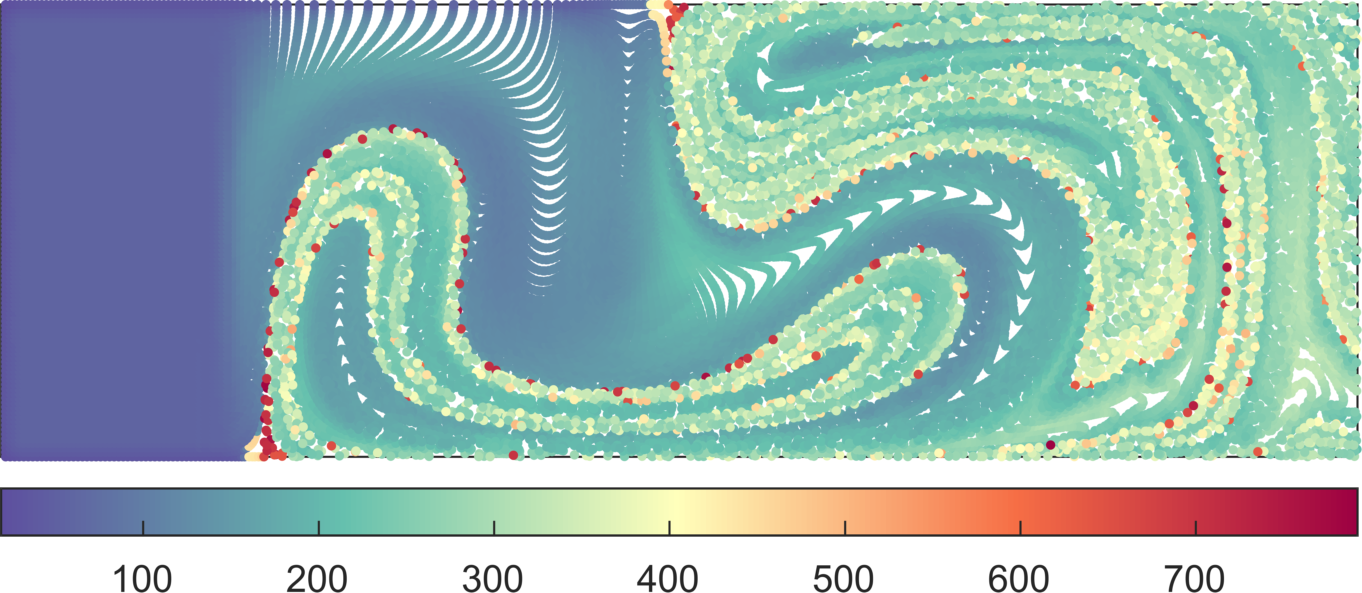}\\
 \includegraphics[width=0.35\textwidth]{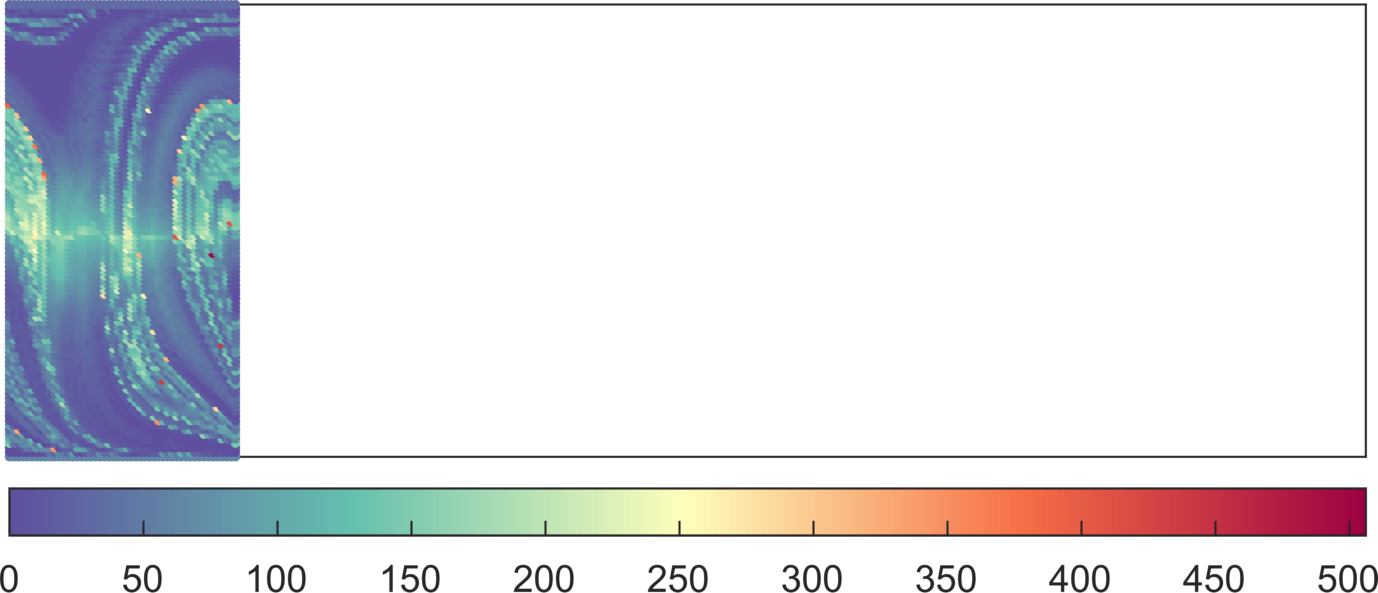} 
&\includegraphics[width=0.35\textwidth]{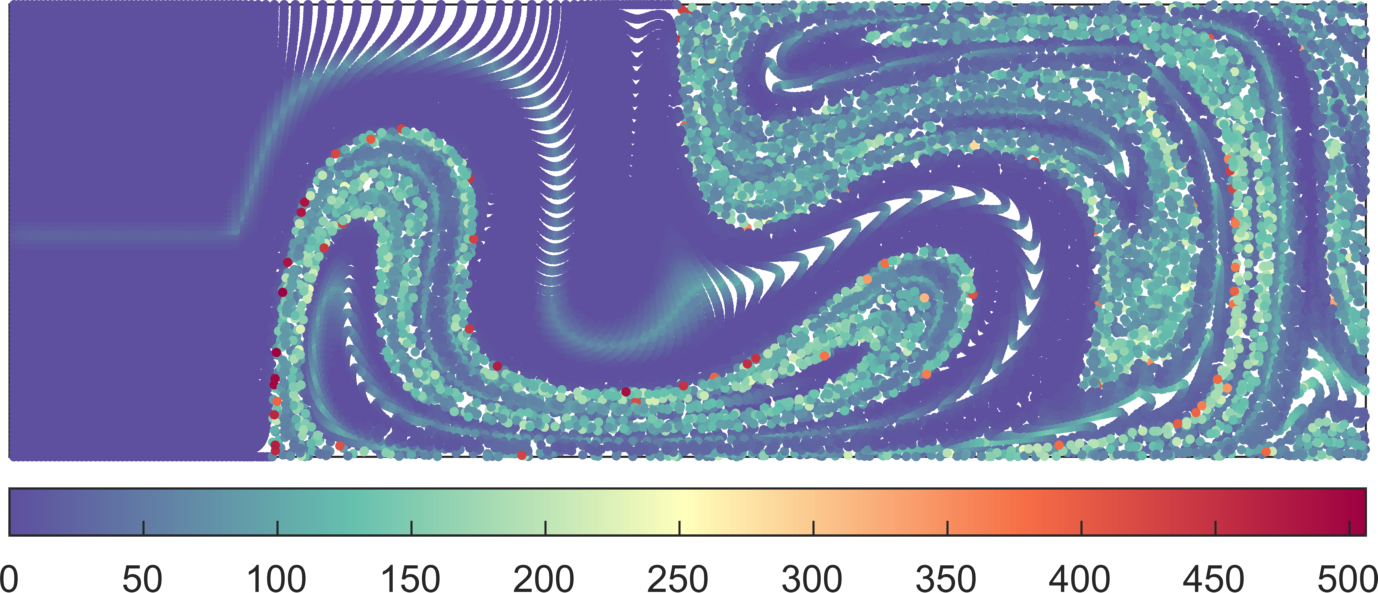}
\end{tabular}
\caption{Open double gyre mixer \eqref{eq:dgmixer} for the parameter choice $\delta=0.3$. Top row: coevolved vector $\bm{w}^0$ (initial condition) and $\bm{w}^{800}$. Second row: accumulated node degrees plotted at initial (so we see here the particles that will meet many others over the time span) and final time. Third row: accumulated sign-based node degree plotted at initial and final time.}\label{fig:odg_zksix}
\end{figure}

We initialize particles on a grid with width $h=0.01$ in the unmixed region $X_1$ and compute the trajectories on the time span $[0,8]$. 
This grid is chosen such that in the numerical simulation new particles will enter the system on the grid nodes and no particles are in the mixing and outlet region at $t_0=0$. In total, 45,450 particle trajectories are computed and we discard trajectory data that is outside of $M$. We compute diffusion matrices $\bm{P}_{\epsilon}(t_k)$ using the time step length $\tau=0.01$ (801 matrices), which is also the step size of the classical Runge Kutta scheme for simulating the particle trajectories, and $\epsilon=\sqrt{2} h$. As in the previous section we choose a small effective diffusion constant $D=0.00005$ for the evolution of the color vector.

 Figure \ref{fig:odg_zksix} (top row) shows the initial condition $\bm{w}^0$ and the resulting mixing pattern for the choice $\delta =0.3$ at time $t=8$ (corresponding to $\bm{w}^{800}$), the second row highlights particles with a high node degree (particles plotted at their initial and final positions, respectively), in the third row the sign-based node degree is considered. These plots show organizing structures as stable manifolds and unstable manifolds of chaotic saddles (which could also be detected by eigenvectors of time-averaged transition matrices in analogy to \cite{klunker2022open}).
 
\begin{figure}[htb]
\centering \includegraphics[width=0.85\textwidth]{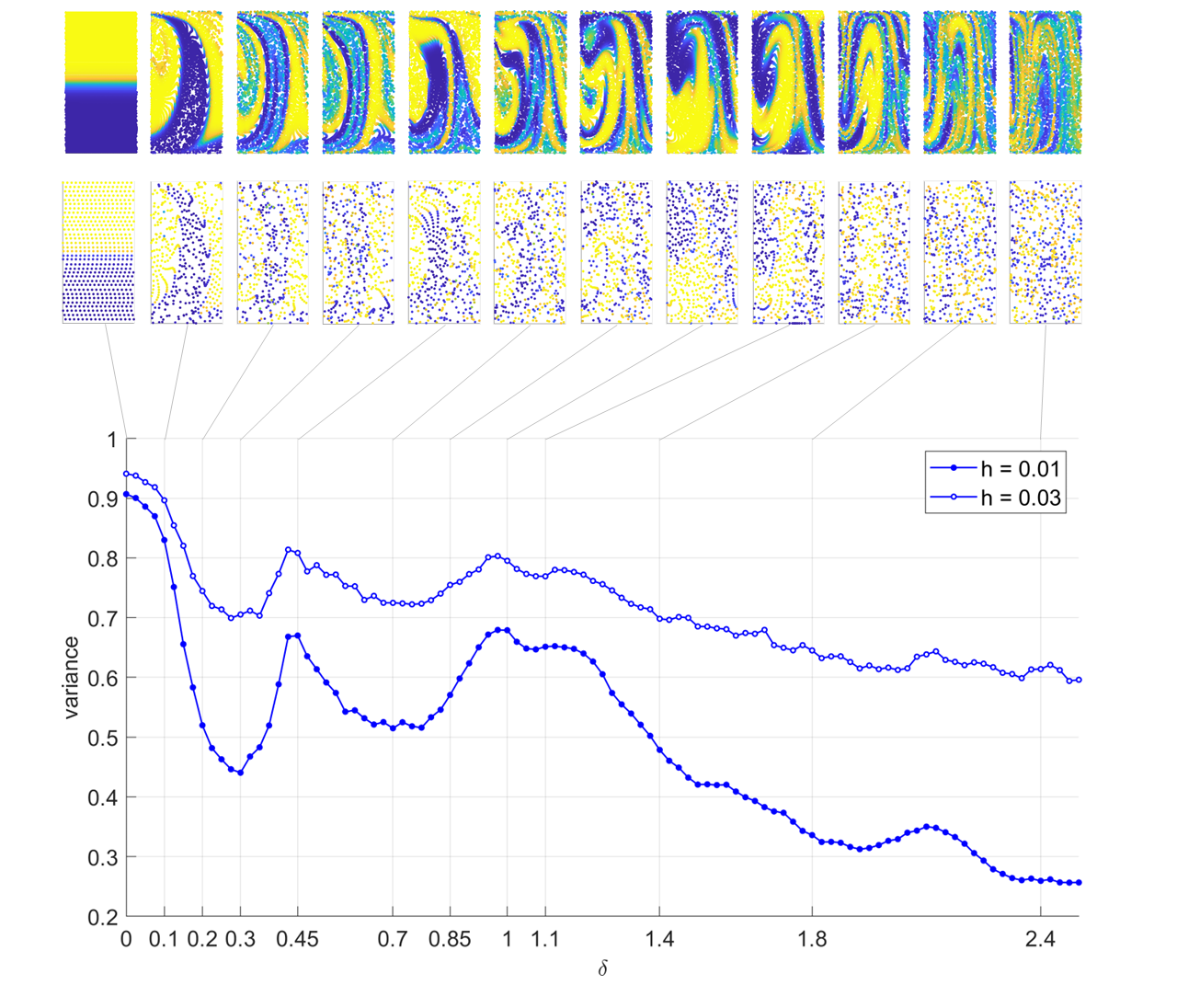}
\caption{Top rows: Color vectors $\bm{w}^{800}$ on $M_3$ for the open double gyre mixer calculated with many and a coarser number of particles for different $\delta$. Sample variance of the color vector $\bm{w}^{800}$ for the open double gyre mixer in the outlet region $M_3$ for different numbers of particles: grid width $h=0.01$ and $h=0.03$.}\label{fig:odg_var}
\end{figure}

We carry out parameter studies by varying $\delta$ from $0$ to $2.5$ in steps of  $0.025$ (resulting in 101 different systems) and consider two different numbers of particles, based on grids with $h=0.01$ and $h=0.03$, where the latter is obtained by discarding respective particles from the $h=0.01$ simulation. For both settings we plot the coevolved vectors $\bm{w}^{800}$ restricted to the outlet region $M_3$ and compute the sample variance (shown in Figure \ref{fig:odg_var}). The results for the fine and the coarse data resolution are qualitatively similar, but differ quantitatively, as there is less mixing for the coarse setting due to the diffusion being suppressed by the larger distances between particles. Overall, the results  are very much in agreement with those obtained by the transfer operator method in \cite{klunker2022open}, including the non-monotone dependence of the mixing results on $\delta$.  

\begin{figure}[htb]
\centering \includegraphics[width=0.85\textwidth]{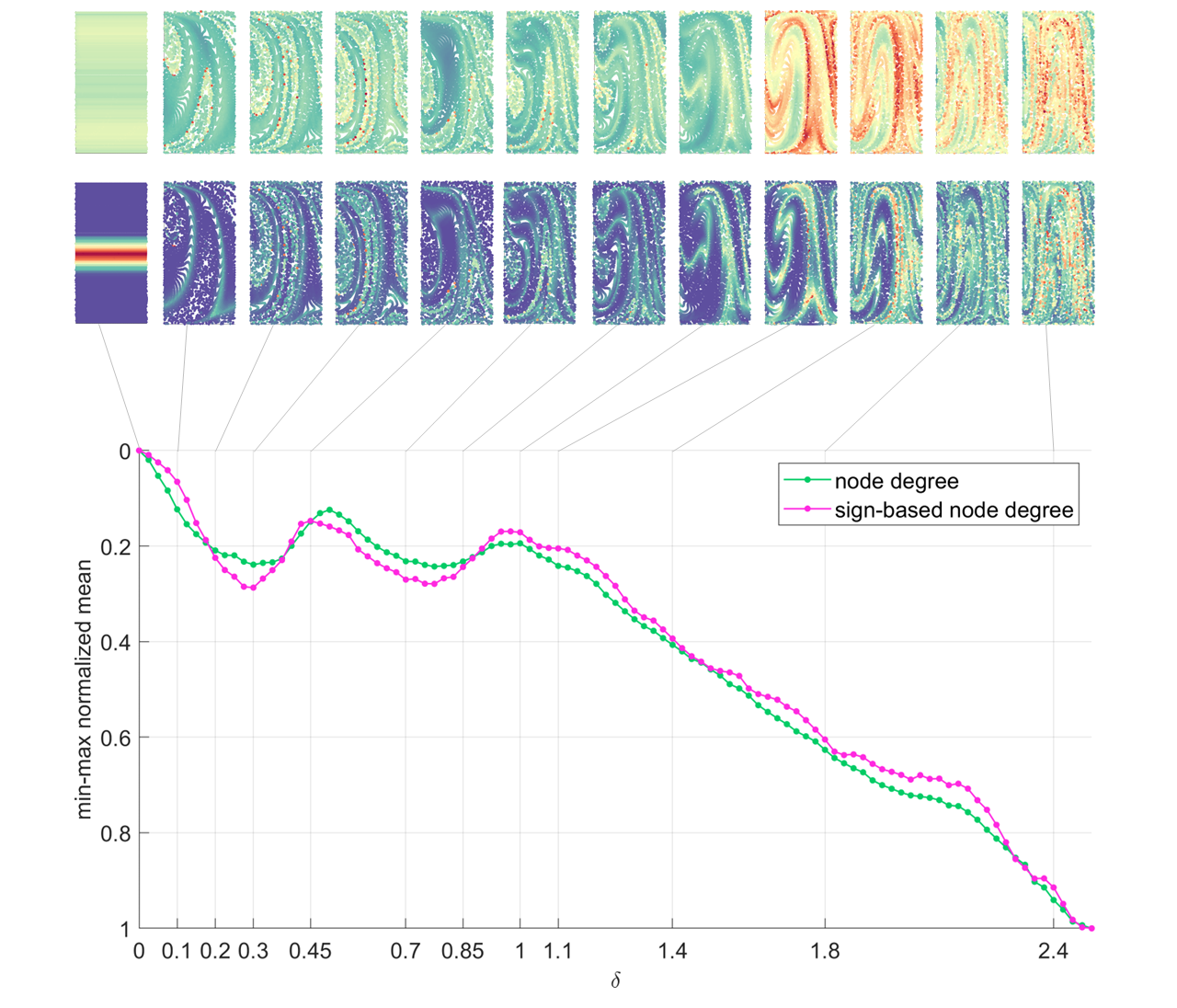}
\caption{Top rows: accumulated node and sign-based node degrees in the outlet region $M_3$. 
The graphs show the min-max-normalized mean node degree and sign-based node degree (flipped y-axis). }\label{fig:odg_nodedegree}
\end{figure}

Finally, we compute the node degree and the sign-based node degree depending on $\delta$ and plot the respective results with respect to the outlet region (Figure \ref{fig:odg_nodedegree}, upper rows). The respective mean node and sign-based node degrees serve as mixing measures. Here they are normalized to that they take values between $[0,1]$, where $0$ corresponds to the smallest and $1$ to the largest mean (sign-based) node degrees, see Figure \ref{fig:odg_nodedegree} (bottom panel). For simpler comparison with the sample variances the $y$-axis has been flipped. We observe a similar non-monotone behavior in the graphs to those of the sample variance in Figure \ref{fig:odg_var} with local maxima and minima at the same positions, i.e.\ high node degrees correspond to stronger mixing. Interestingly, the usual and the sign-based node degrees give very similar results.  

\subsection{Stirred tank reactor}\label{sec:str}
\begin{figure}[ht]
\centering \includegraphics[width=0.355\textwidth]{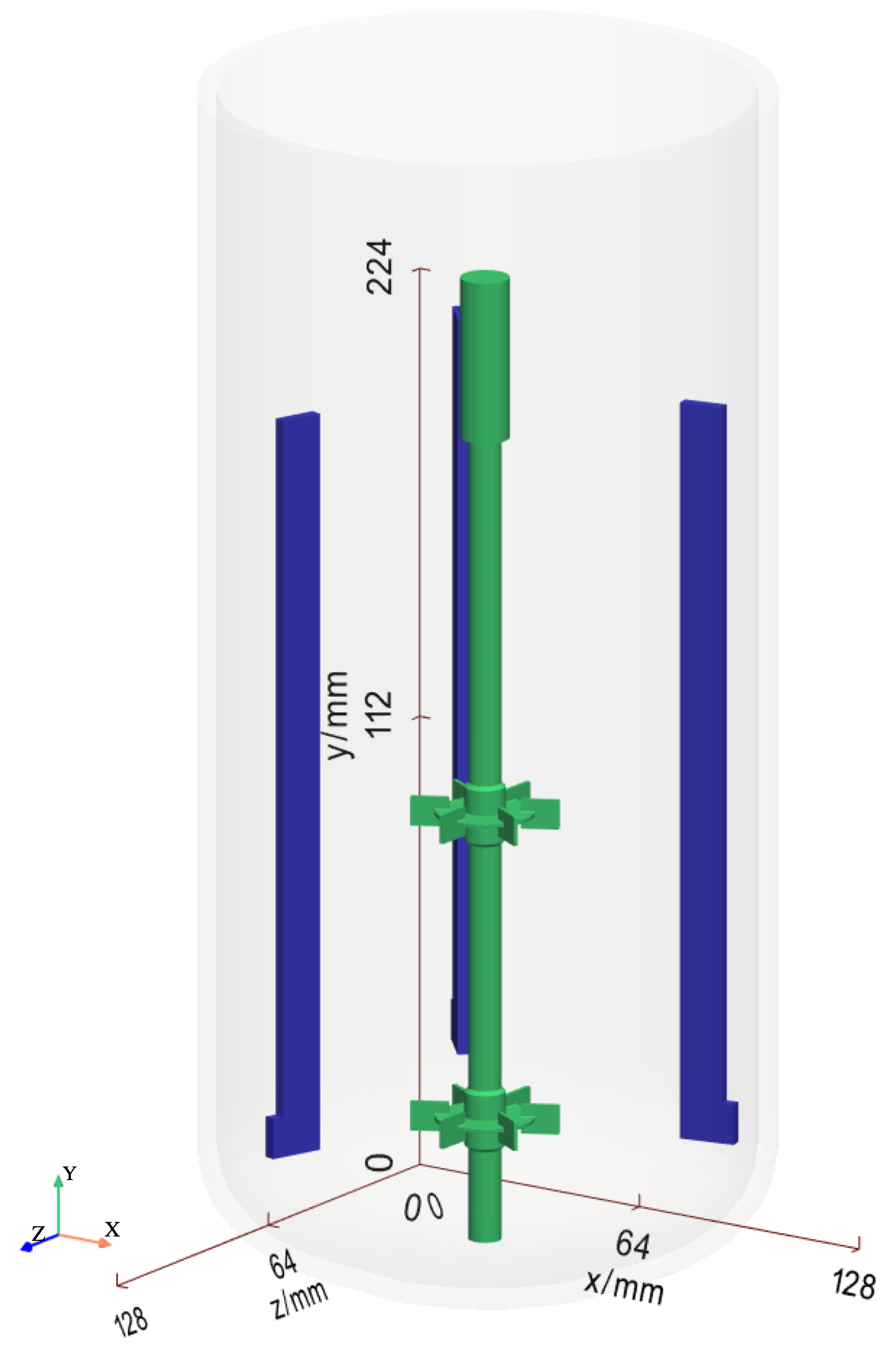}
\caption{Geometry of the stirred tank reactor with three baffles and two Rushton turbines, see \cite{weiland2023}. }\label{fig:reactorsetup}
\end{figure}

Finally, we study a model of stirred tank reactor. In such chemical engineering systems mixing processes are crucial as they determine the outcome of chemical reactions.
We consider a lab-scale stirred tank reactor of 2.8L water as in \cite{weiland2023} with stirrer speed 252 rpm. The reactor has a korbbogen head bottom, three baffles and two Rushton turbines, see Figure \ref{fig:reactorsetup} for the geometry and set-up.

Trajectories are computed for ten stirrer revolutions corresponding to a time span of $\SI{2.38}{s}$ from a three-dimensional Lattice-Boltzmann simulation, see \cite{weiland2023} for details on the numerical model that has been validated against experimental data \cite{Hofmann2022,Kuschel2021}.
Particles are initialized on a regular grid of mesh width $h=\SI{2}{mm}$, yielding 360,978 trajectories  which are evaluated at 239 time steps, where $\tau=\SI{0.01}{s}$. As in the previous section we choose $\epsilon=\sqrt{2}h$ and compute the 239 diffusion matrices $\bm{P}_{\epsilon}(t_k)$ for $t_k \in \mathbb{T}=\{0, 0.01, 0.02, \ldots, 2.38\}$.
We consider two different effective diffusion constants $D=2 \cdot 10^{-9} \text{m}^2/\text{s}$, which is similar to the molecular diffusion constant of water, as well as a considerably larger one $D=1\cdot 10^{-5} \text{m}^2/\text{s}$, which could be understood as a turbulent diffusion on unresolved scales of the large eddy simulation.

\begin{figure}[!ht]
\centering
\begin{tabular}{cccc}
\includegraphics[height=0.22\textheight]{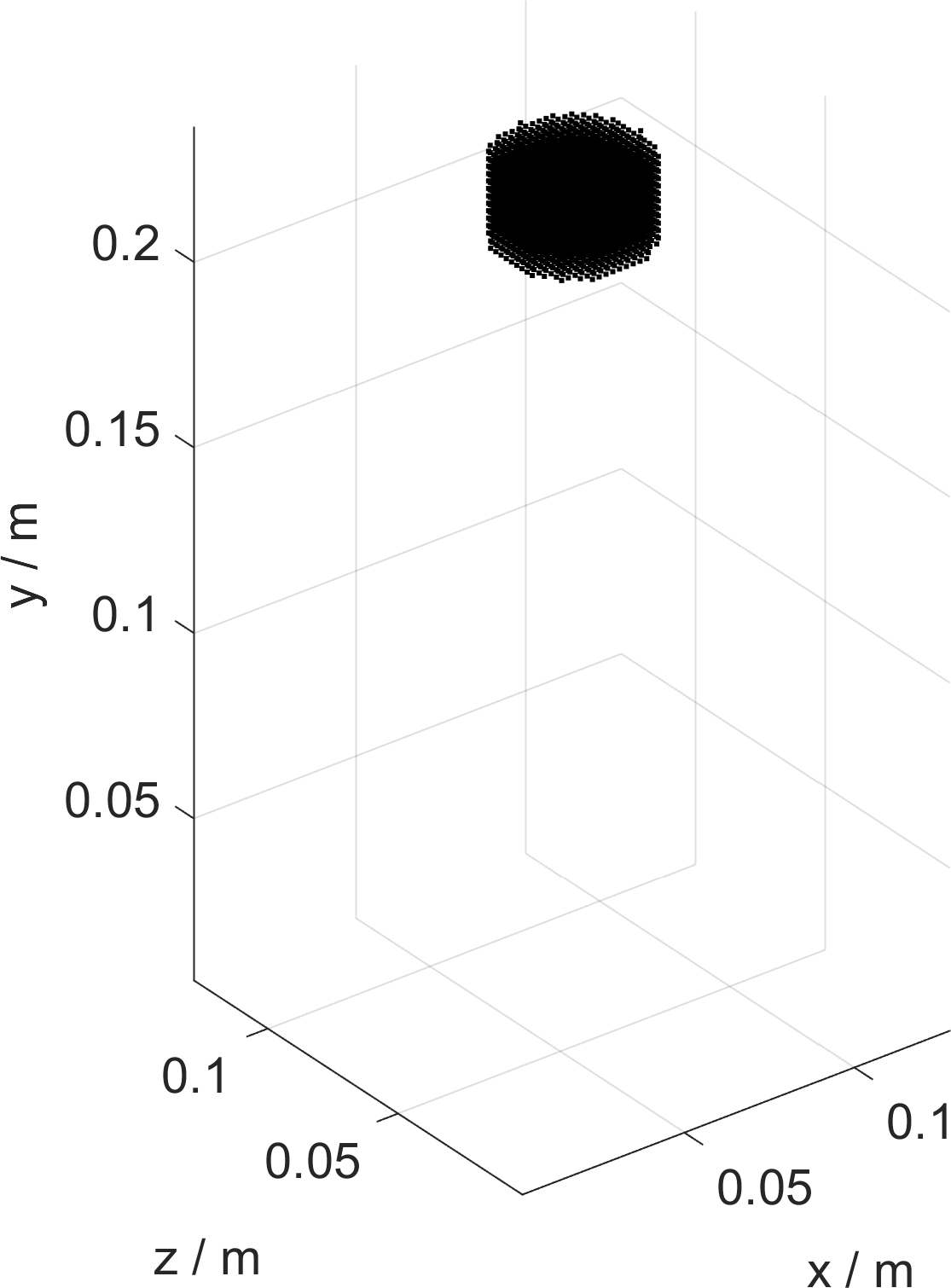} & 
\includegraphics[height=0.22\textheight]{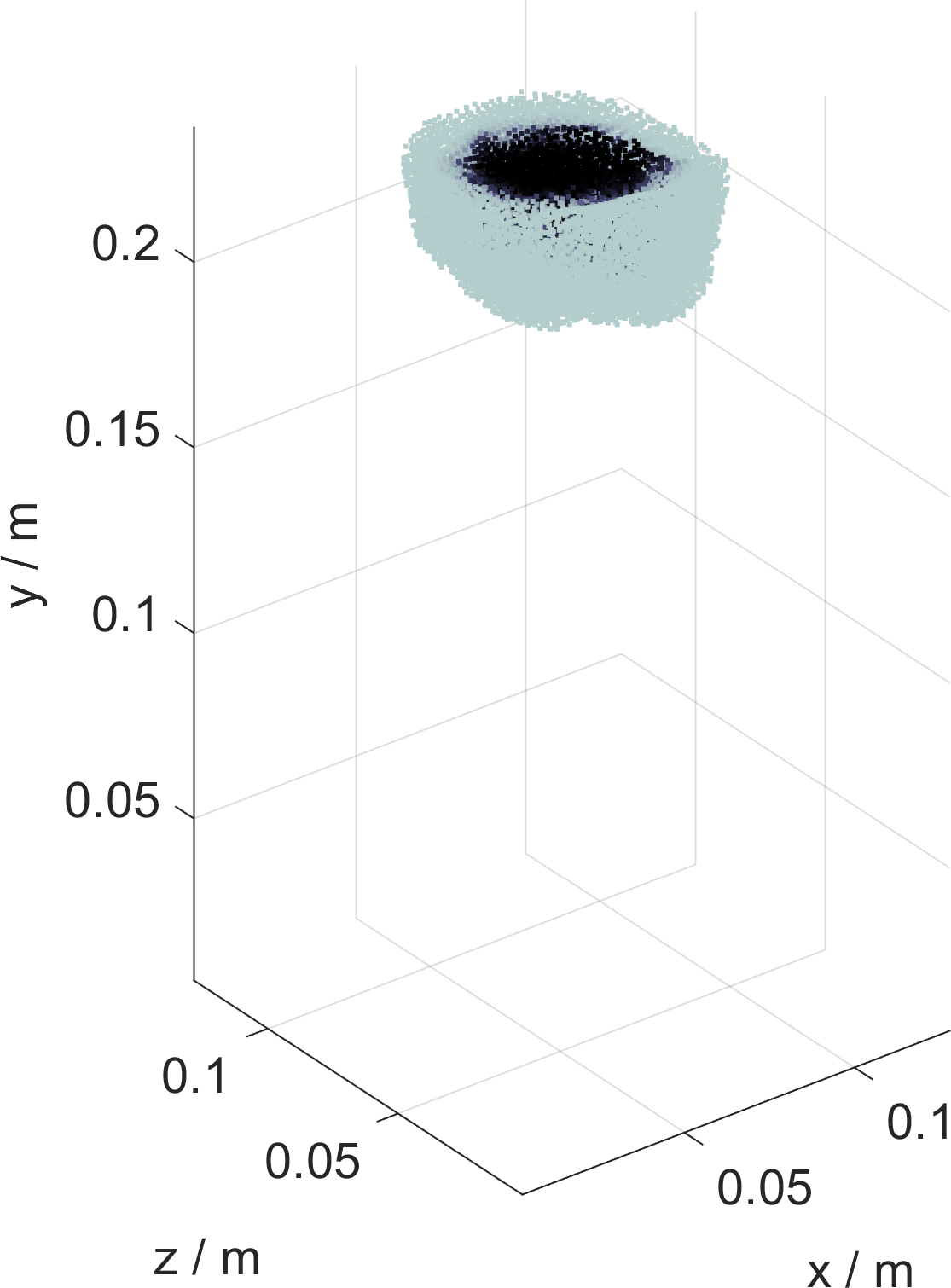} & 
\includegraphics[height=0.22\textheight]{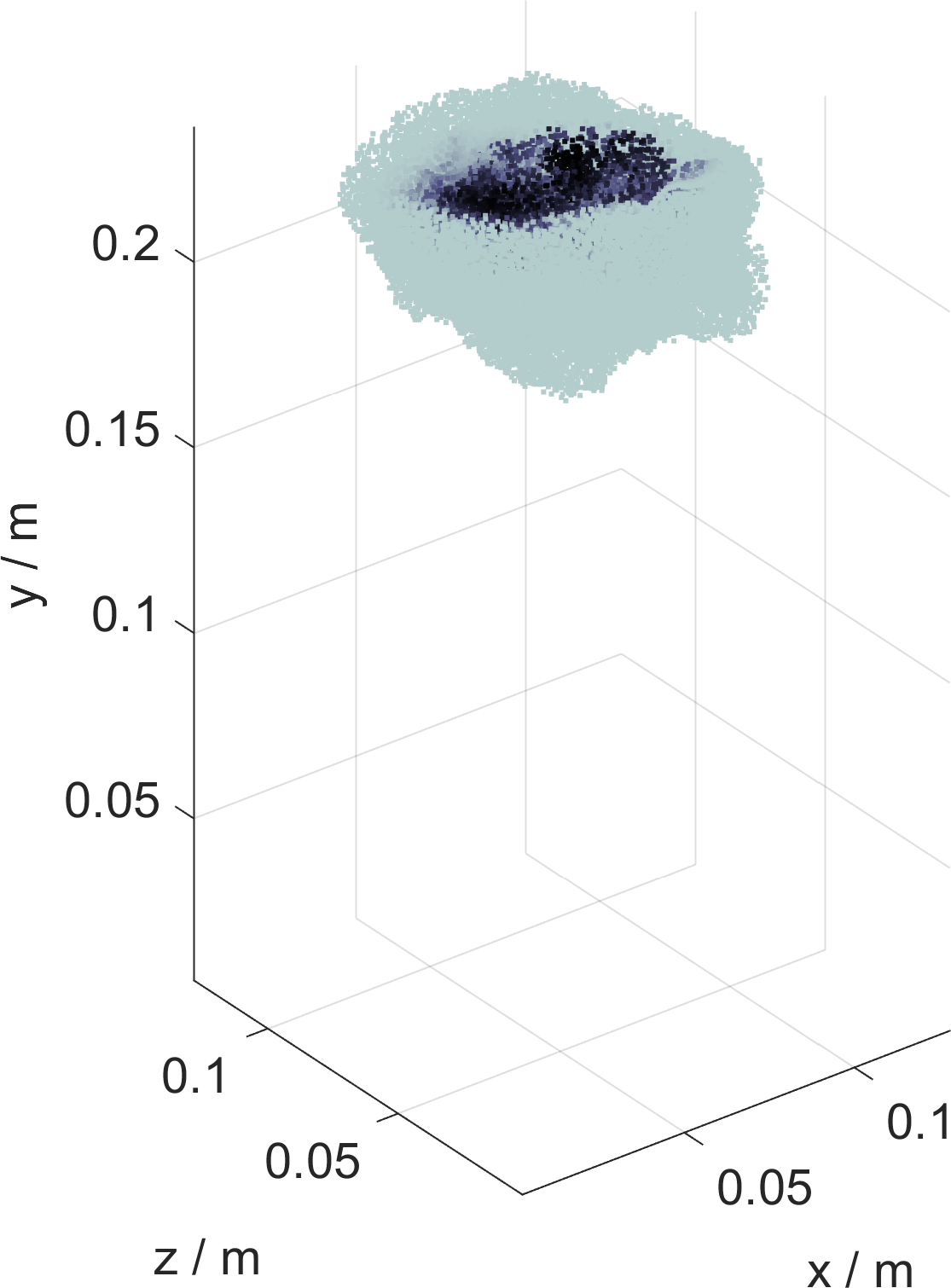} 
& \includegraphics[height=0.22\textheight]{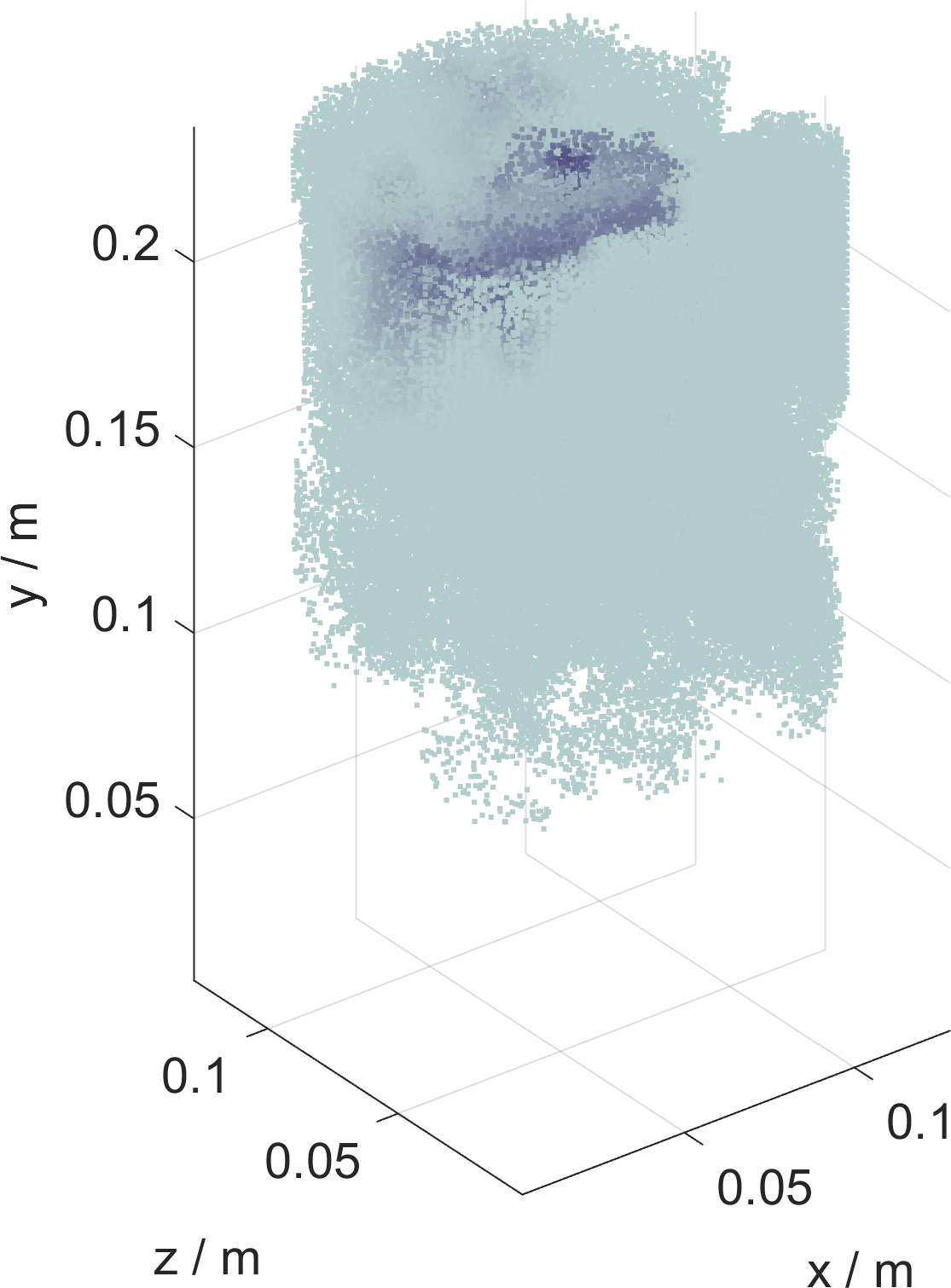}\\ 

\includegraphics[height=0.22\textheight]{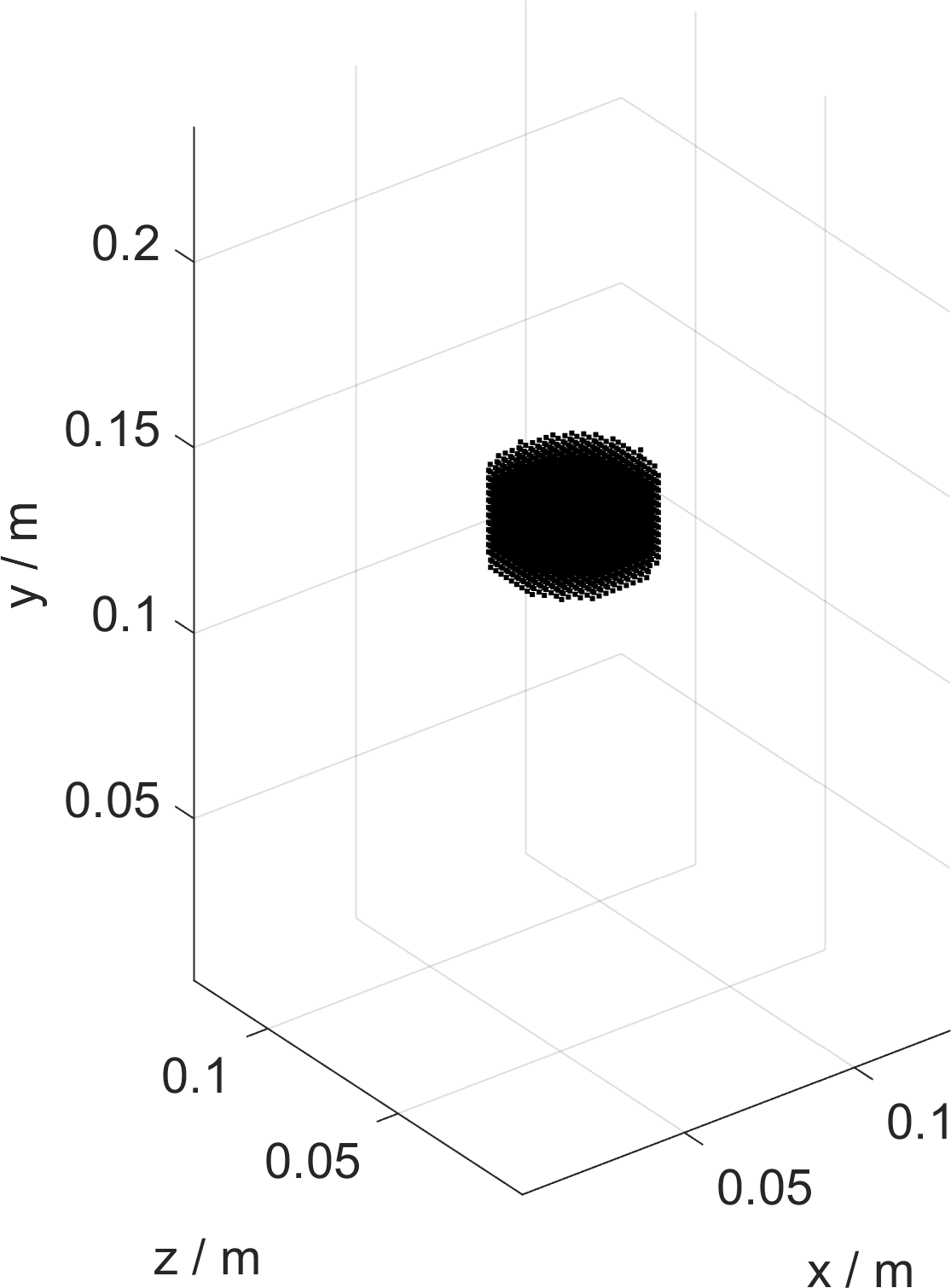} & 
\includegraphics[height=0.22\textheight]{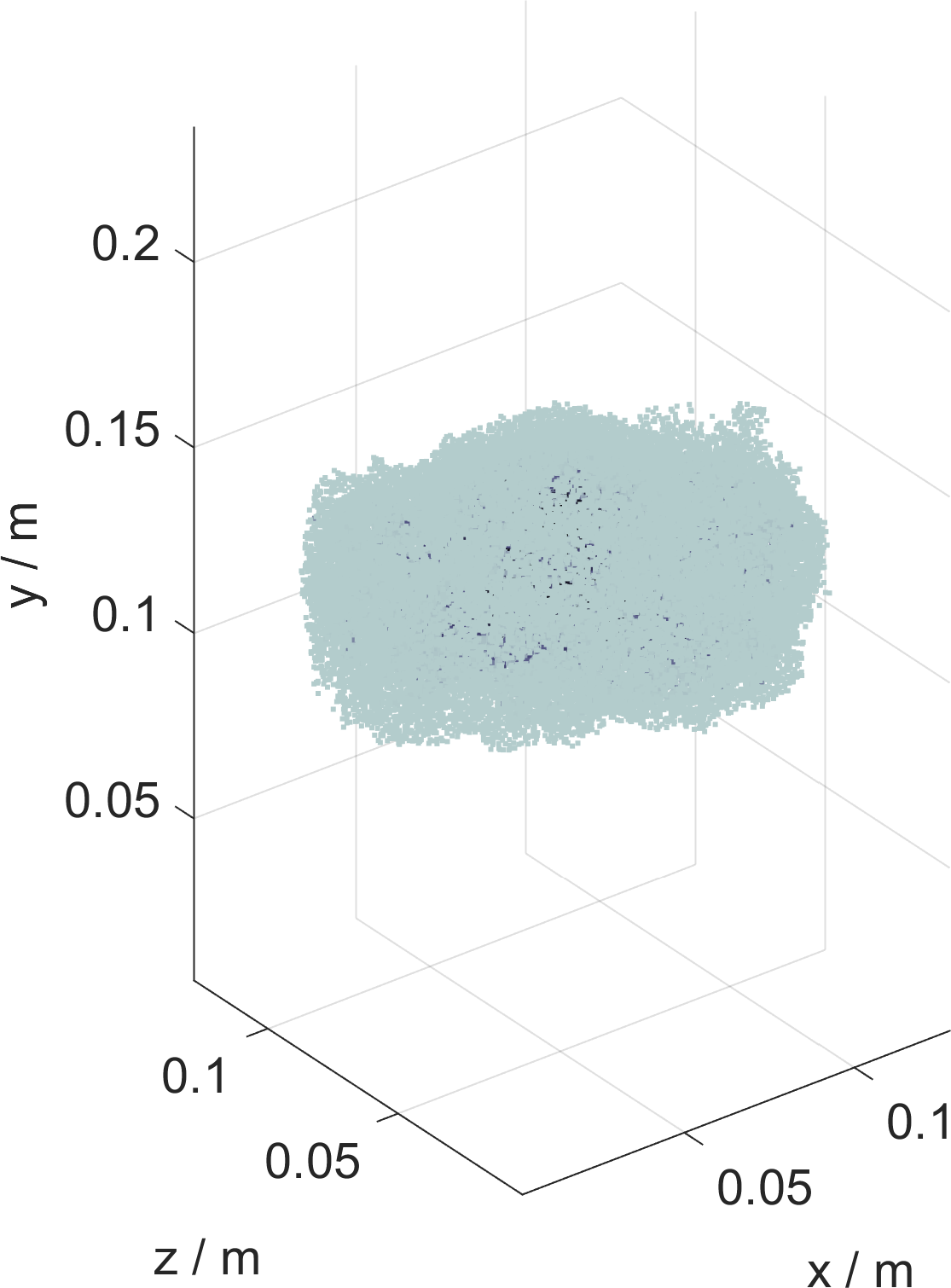} & 
\includegraphics[height=0.22\textheight]{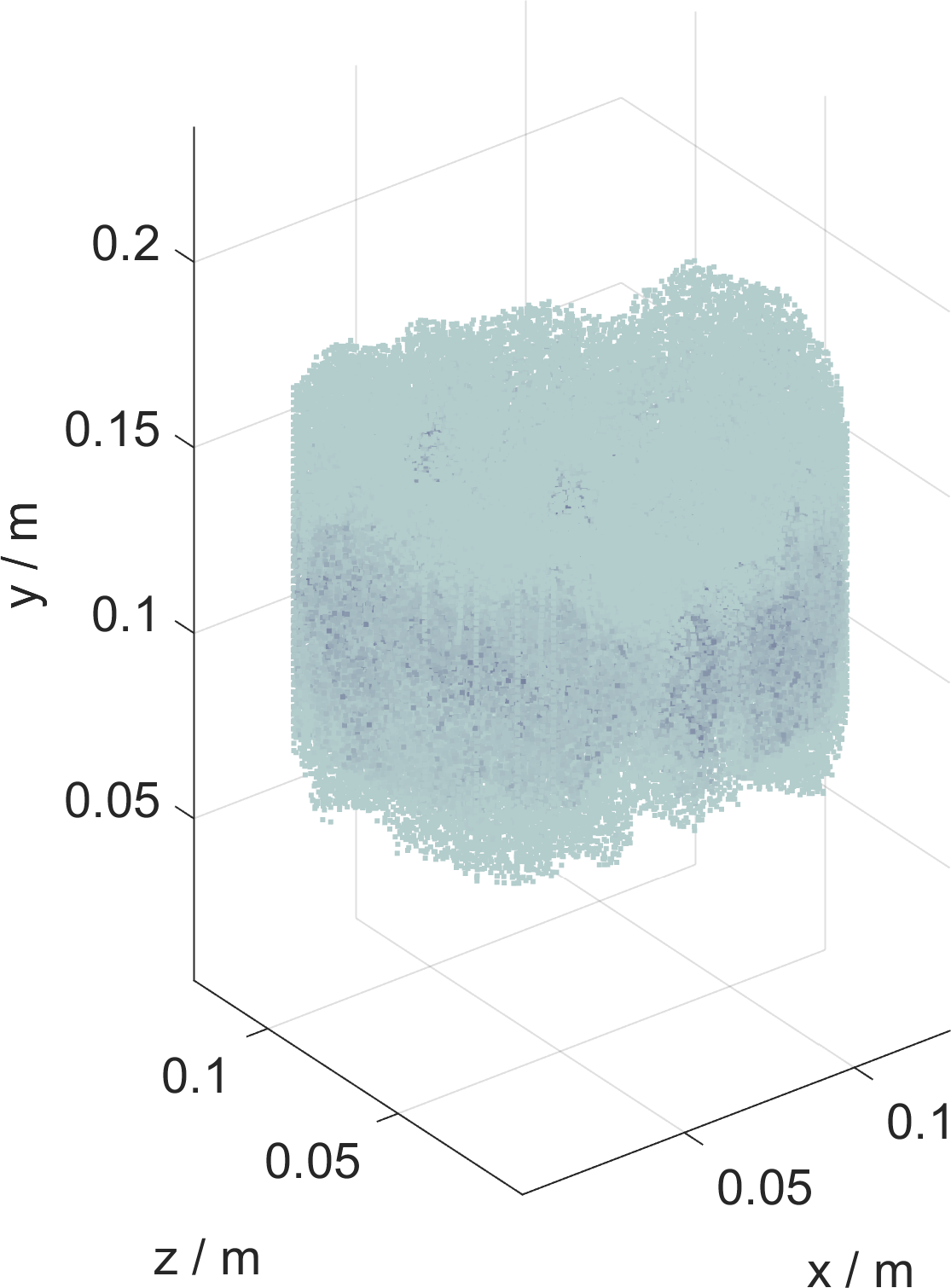} &
\includegraphics[height=0.22\textheight]{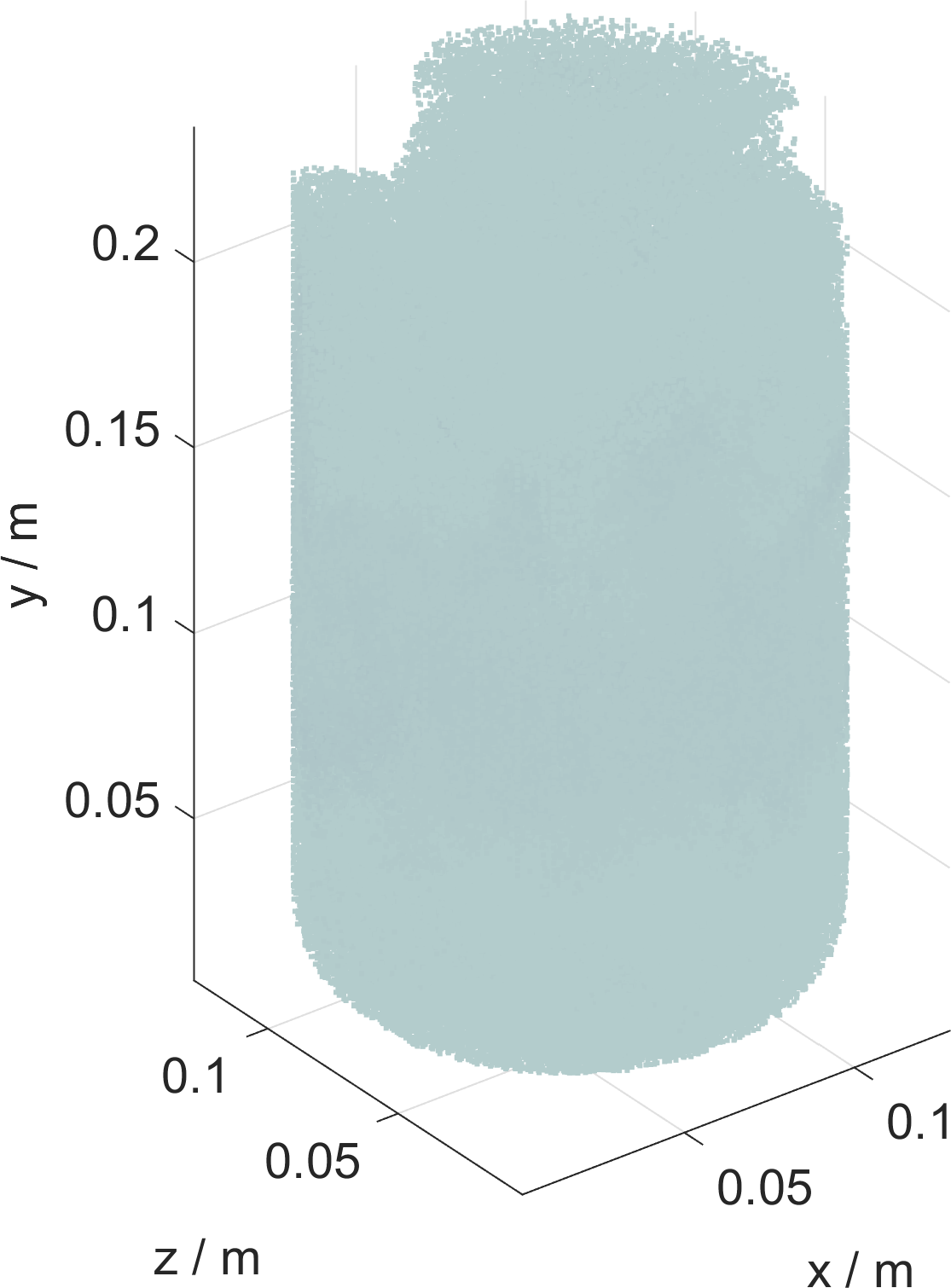} \\ 

\includegraphics[height=0.22\textheight]{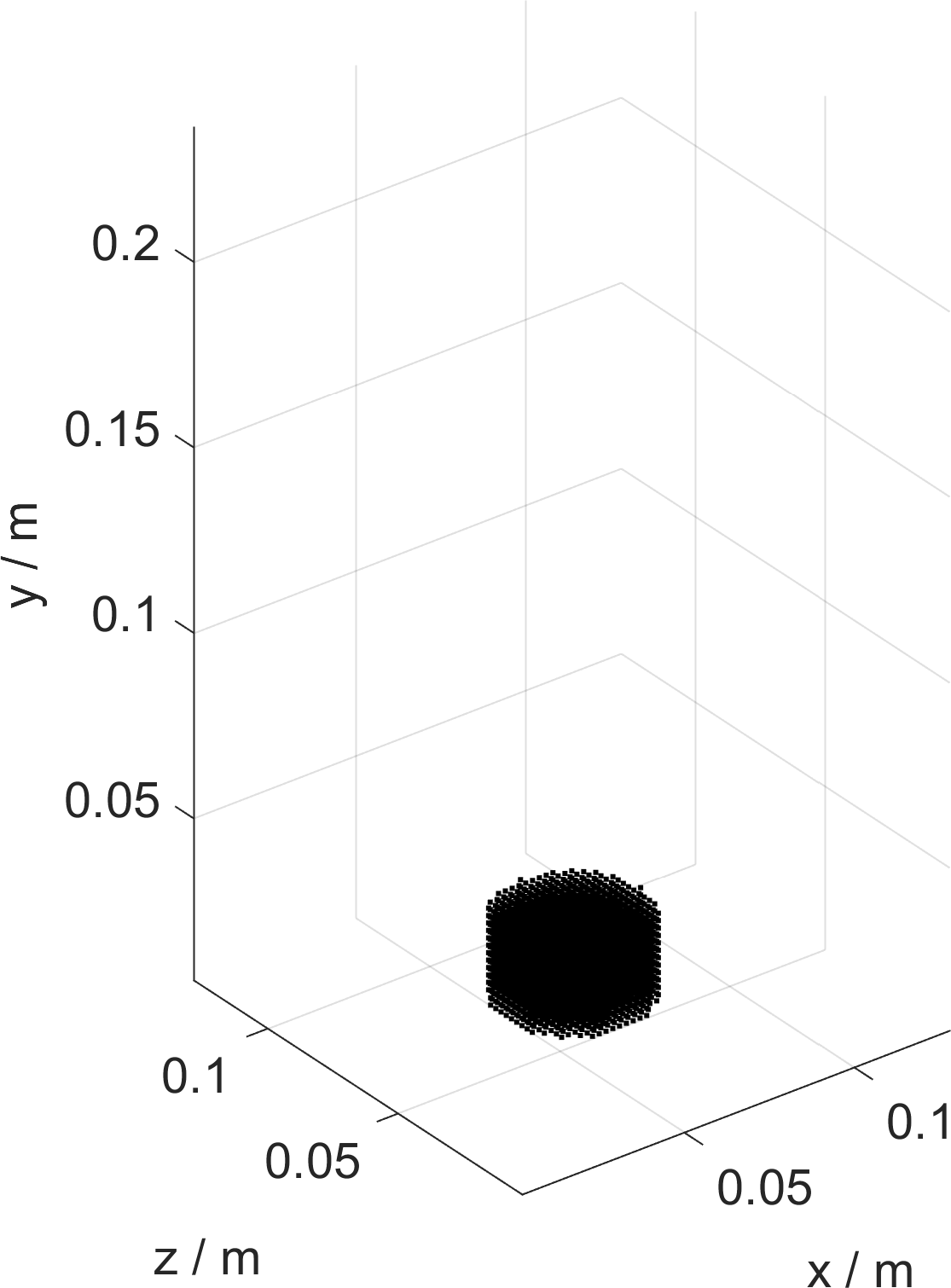} & 
\includegraphics[height=0.22\textheight]{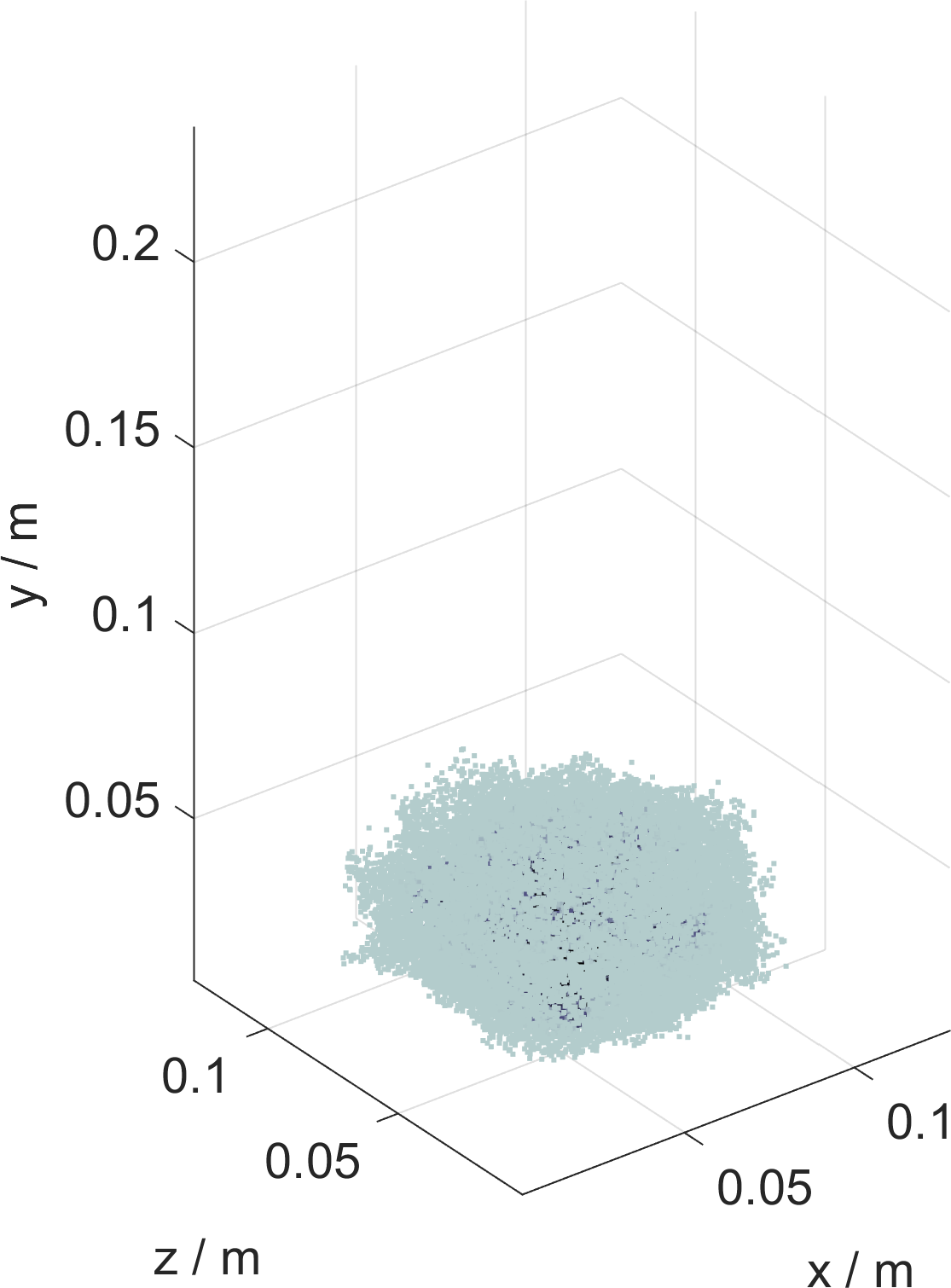} & 
\includegraphics[height=0.22\textheight]{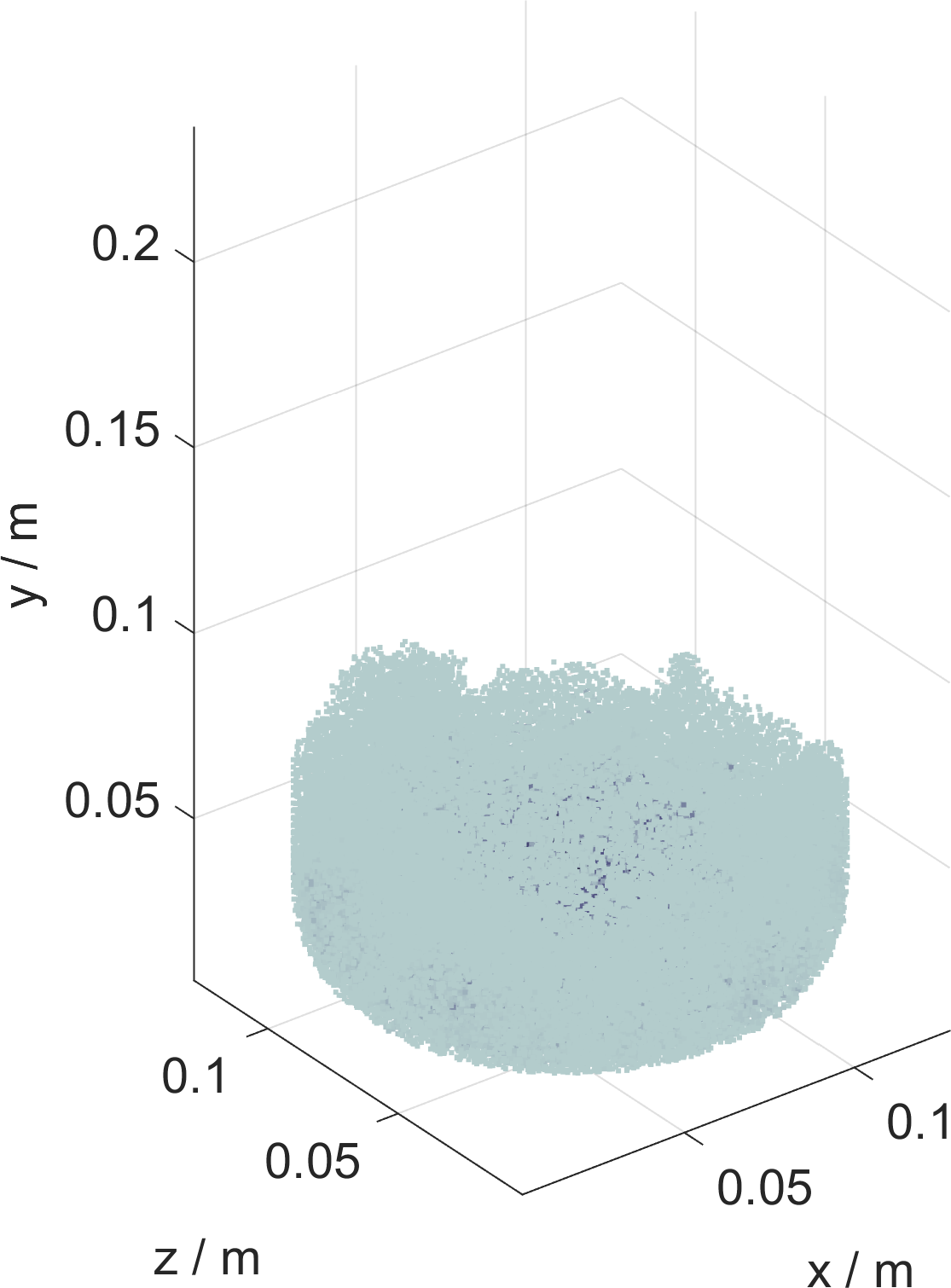} &
\includegraphics[height=0.22\textheight]{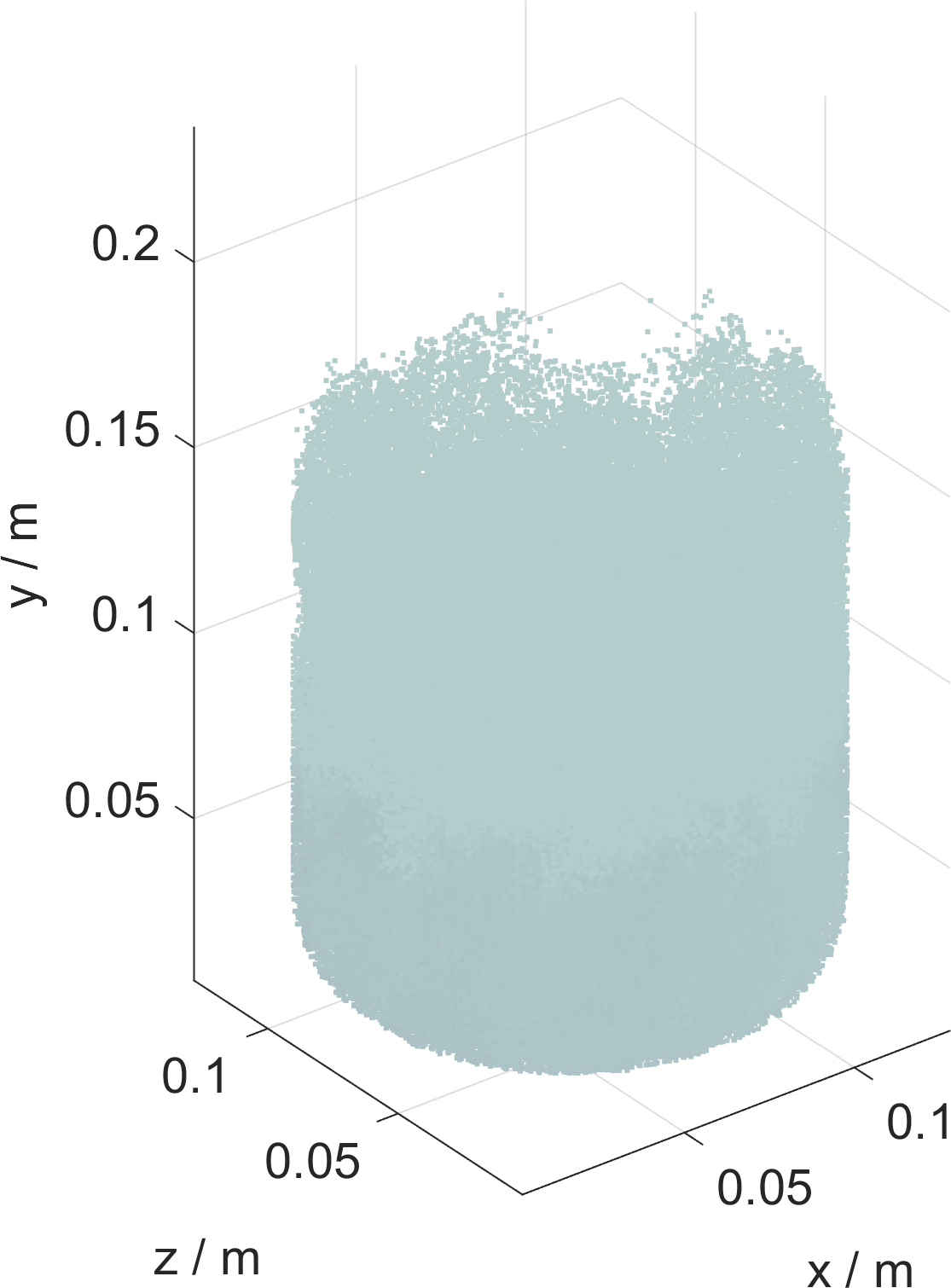} \\ 

&\multicolumn{3}{c}{\includegraphics[width=0.5\textwidth]{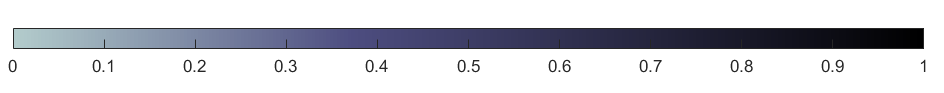} }
\end{tabular}
\caption{Evolution of blobs initialized at three different positions in the reactor for the diffusion constant $D=1\cdot 10^{-5} \text{m}^2/\text{s}$. Shown are the respective coevolved vectors at initial time ($\bm{w}^0$, first column) and at time slices closest to one ($\bm{w}^{24}$), three ($\bm{w}^{71}$) and 10 stirrer rotations  ($\bm{w}^{238}$). For better visibility, only particle positions $\bm{x}_i(t_k)$ for which $w_i^{k}>0.00001$ are plotted.}\label{fig:str_b1}
\end{figure}

In Figure \ref{fig:str_b1} we follow three blobs that have been initialized at different positions in the tank reactor. For this, particles in the blob have been given the value 1 at initial time in $\bm{w}^0$ and particles outside of the blob the value 0 (left column). The respective vector is evolved using the larger effective diffusion constant $D=1\cdot 10^{-5} \text{m}^2/\text{s}$, and the results after around one, three and ten stirrer rotations are plotted (columns 2--4), where we only show particles with color value $>0$. Clearly, particles are much more dispersed when starting in the middle of the reactor than when starting at the top or bottom. The top appears to be the worst location of for good mixing, which is an important finding, considering that substances are usually fed in at the top in bioreactors to avoid contamination.

\begin{figure}[!ht]
\centering
\begin{tabular}{ccc}
&\includegraphics[height=0.2\textheight]{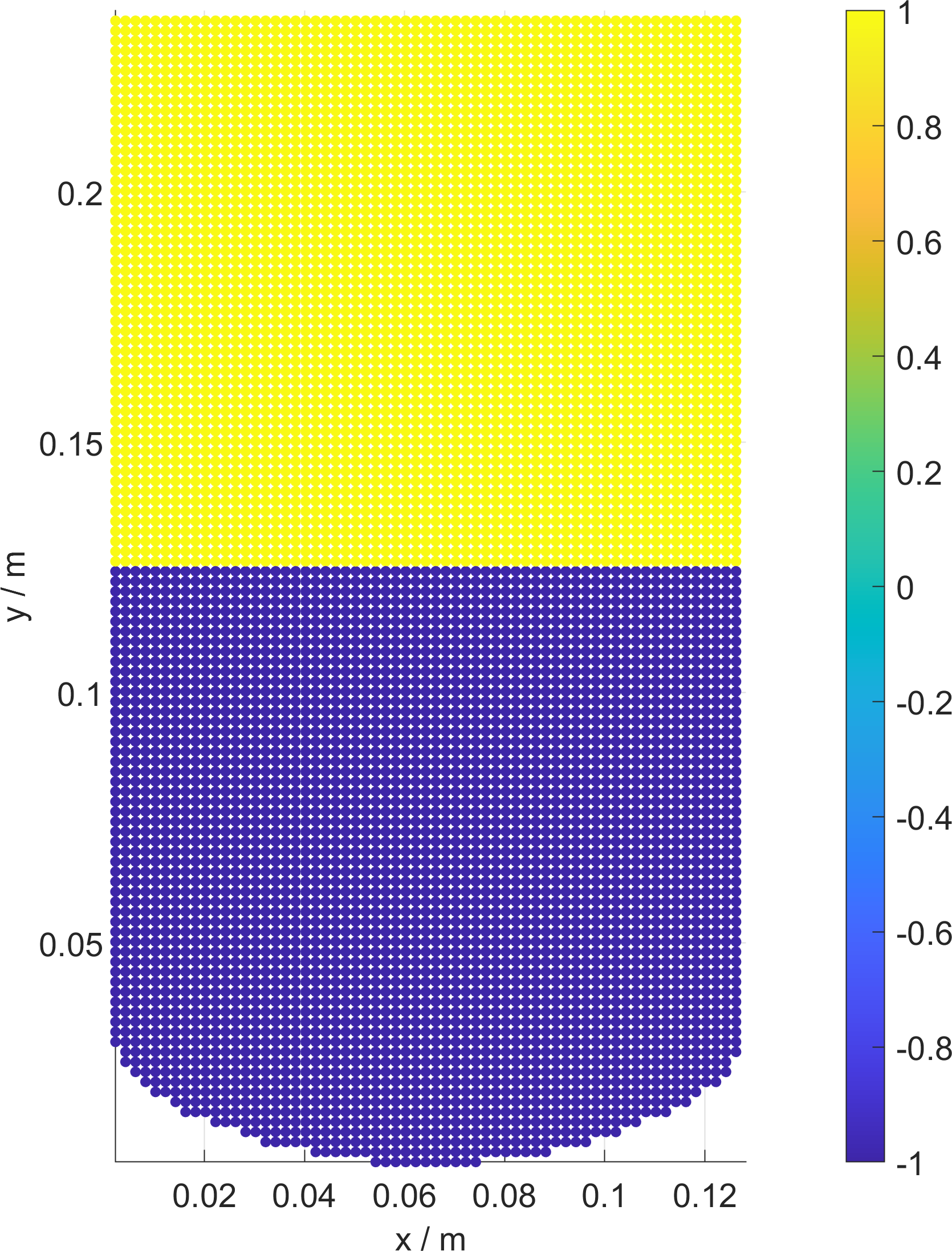}&\\

  \includegraphics[height=0.2\textheight]{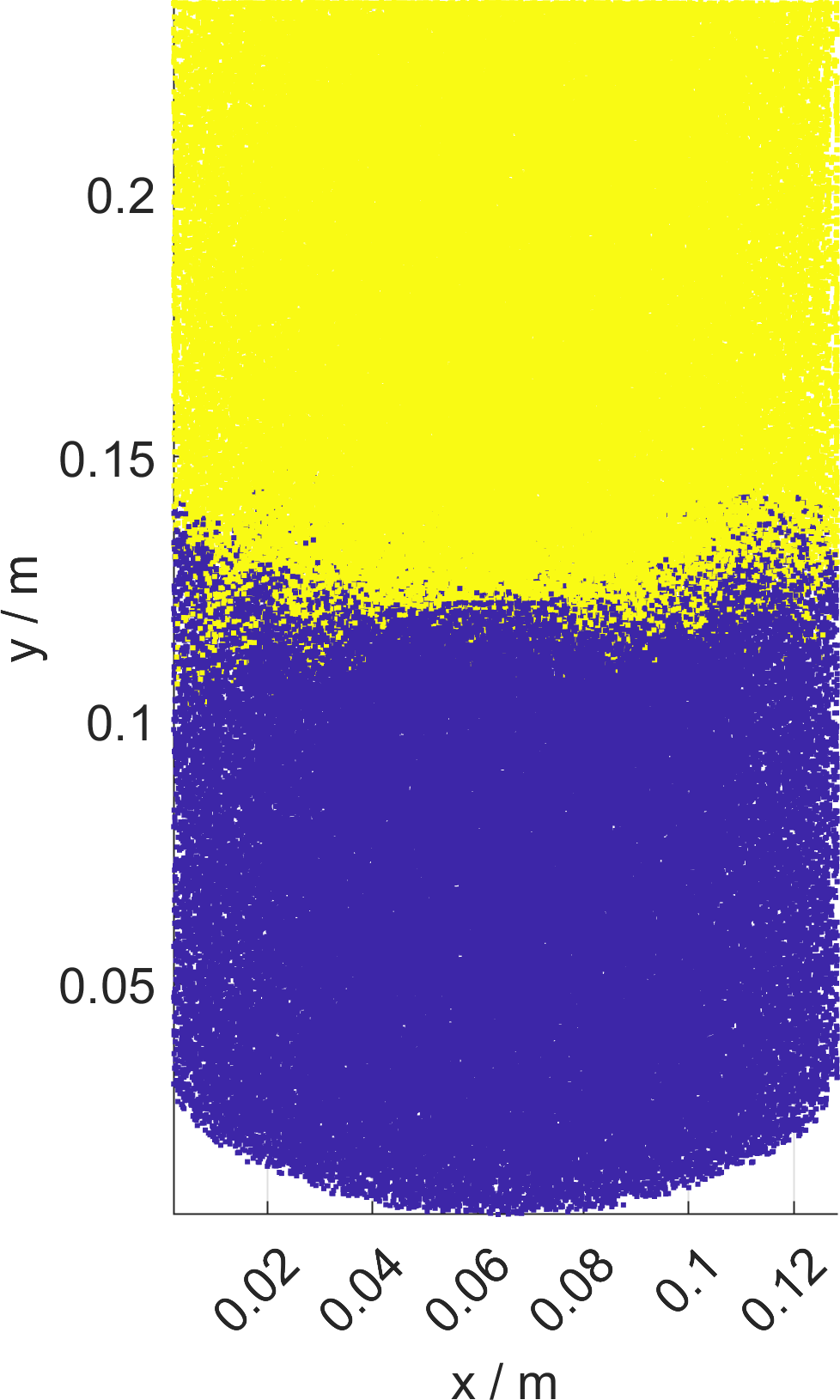}& 
  \includegraphics[height=0.2\textheight]{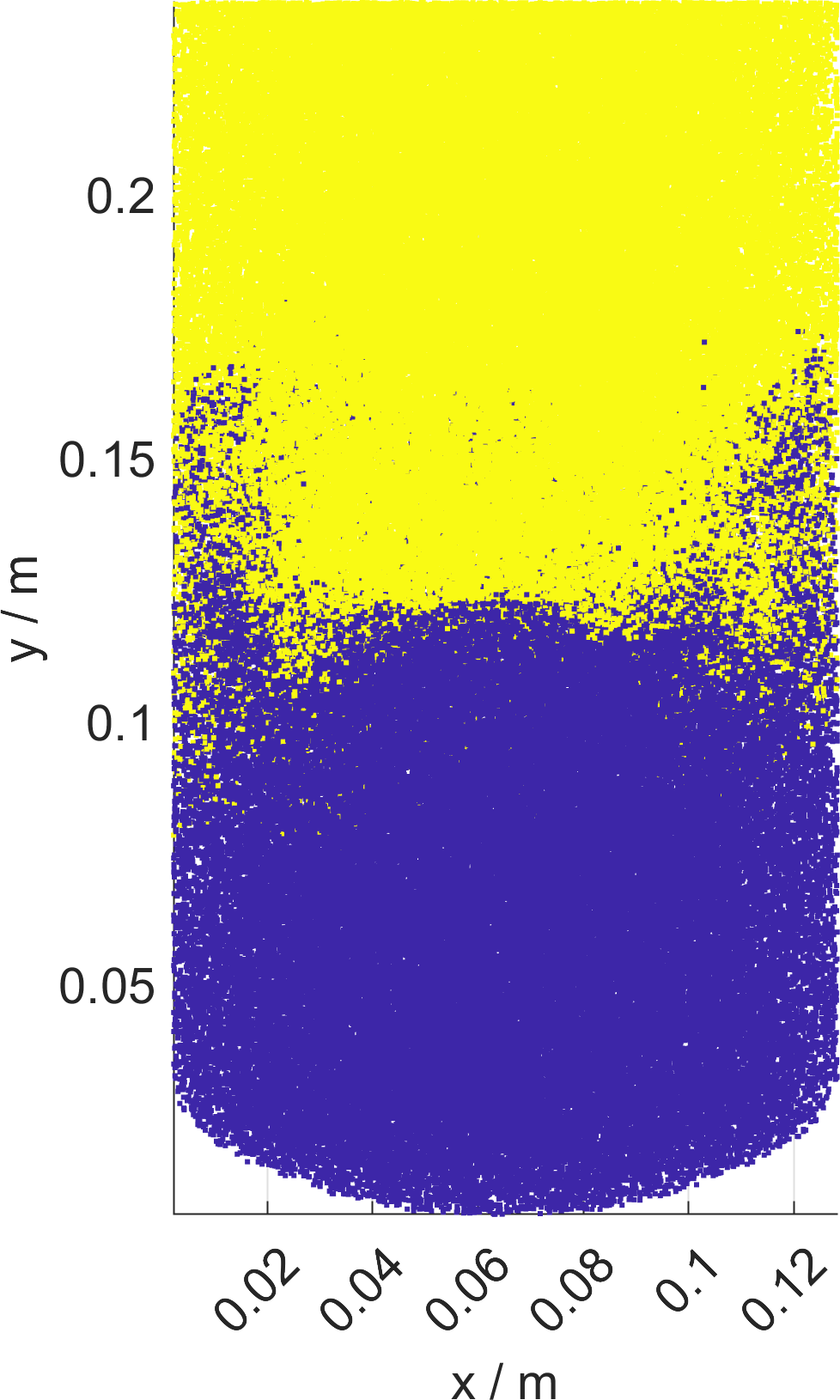}&
\includegraphics[height=0.2\textheight]{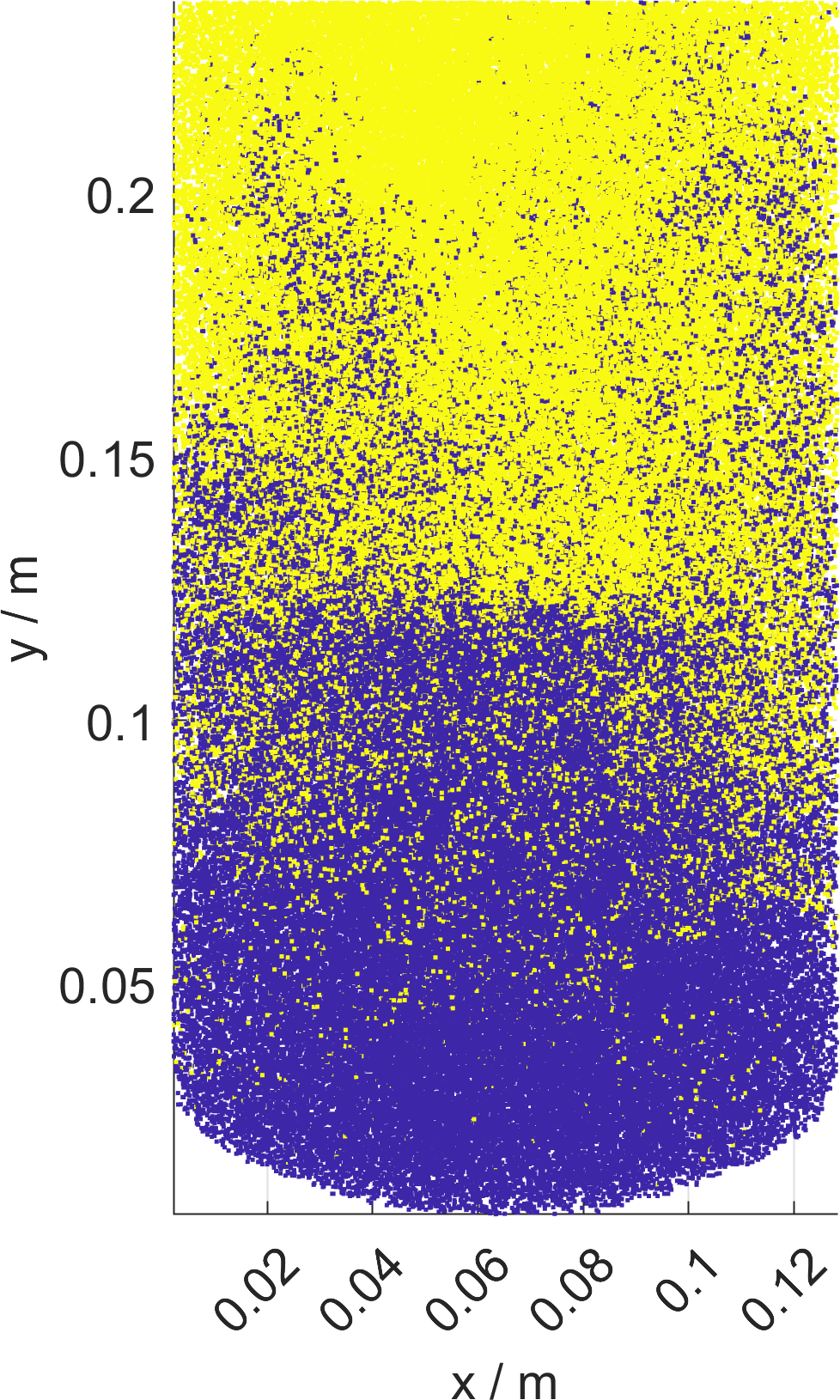}\\
\includegraphics[height=0.2\textheight]{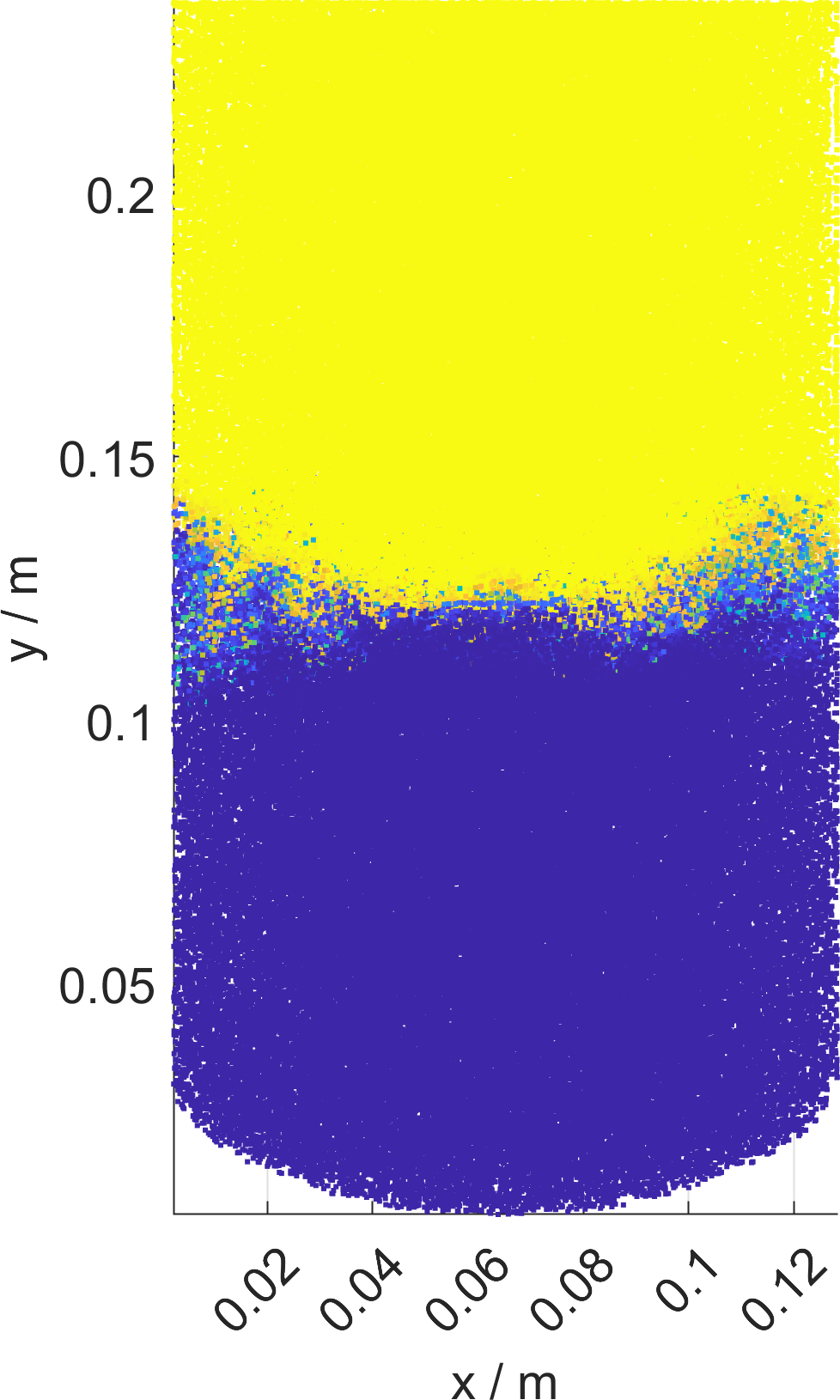}& 
\includegraphics[height=0.2\textheight]{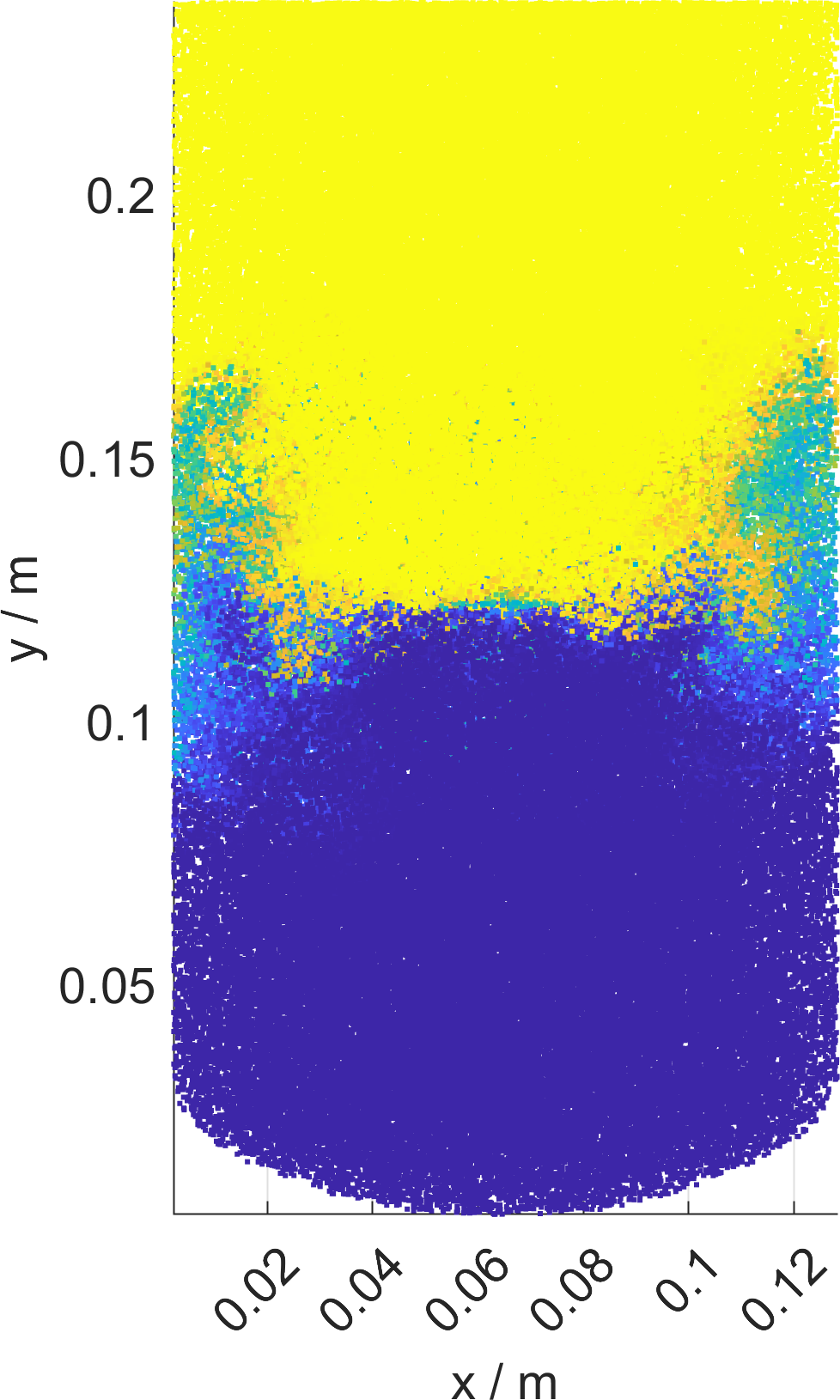}&
\includegraphics[height=0.2\textheight]{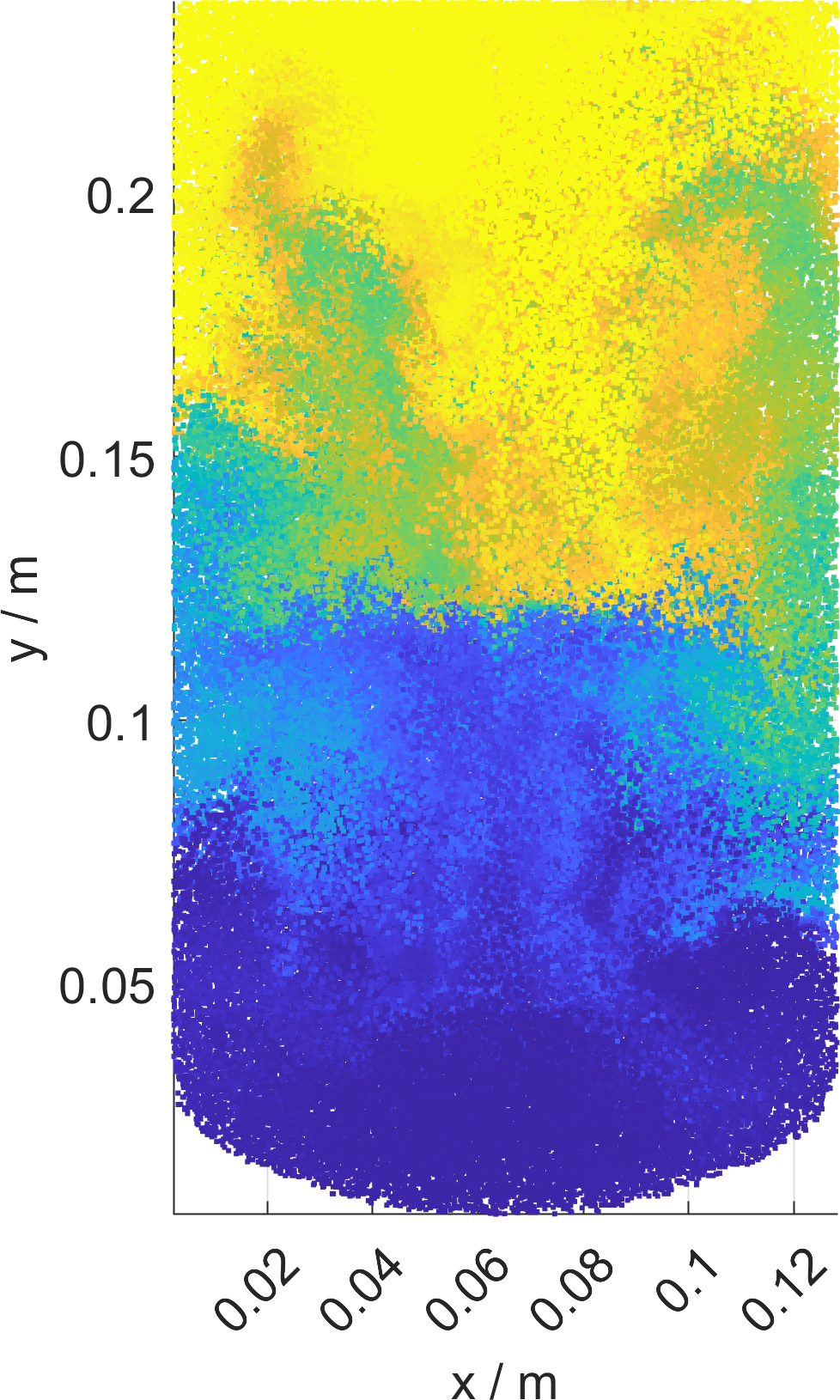}\\
\includegraphics[height=0.2\textheight]{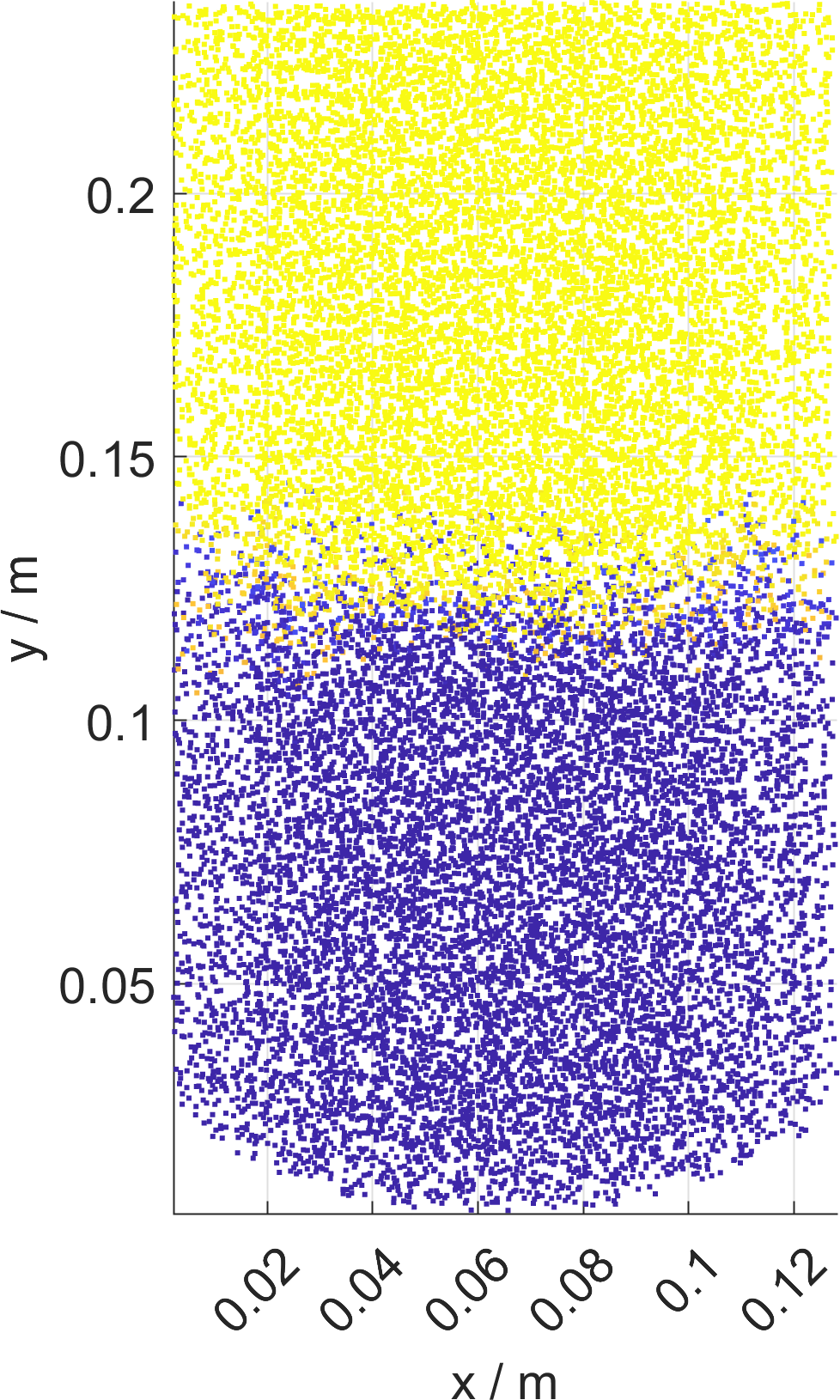}& 
\includegraphics[height=0.2\textheight]{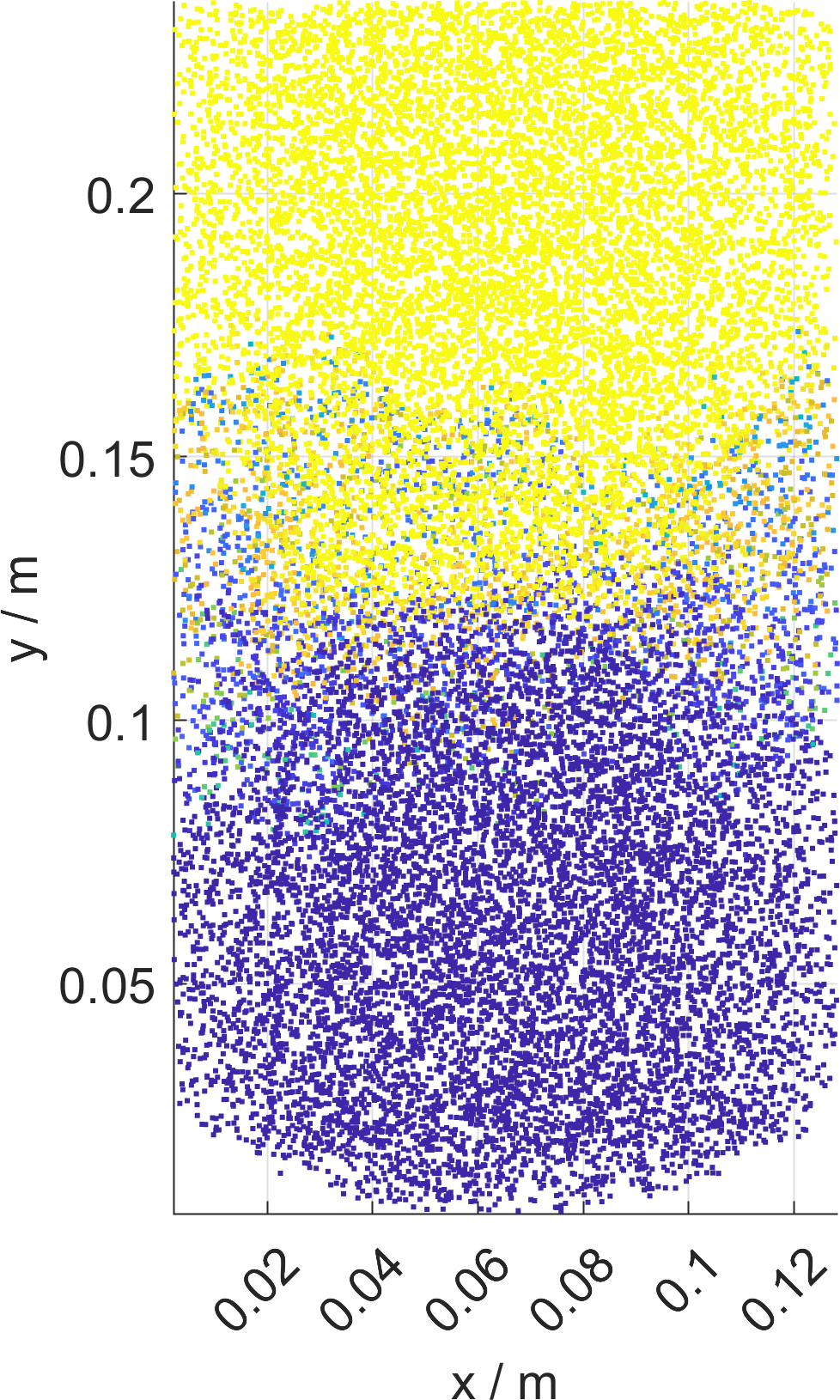}&
\includegraphics[height=0.2\textheight]{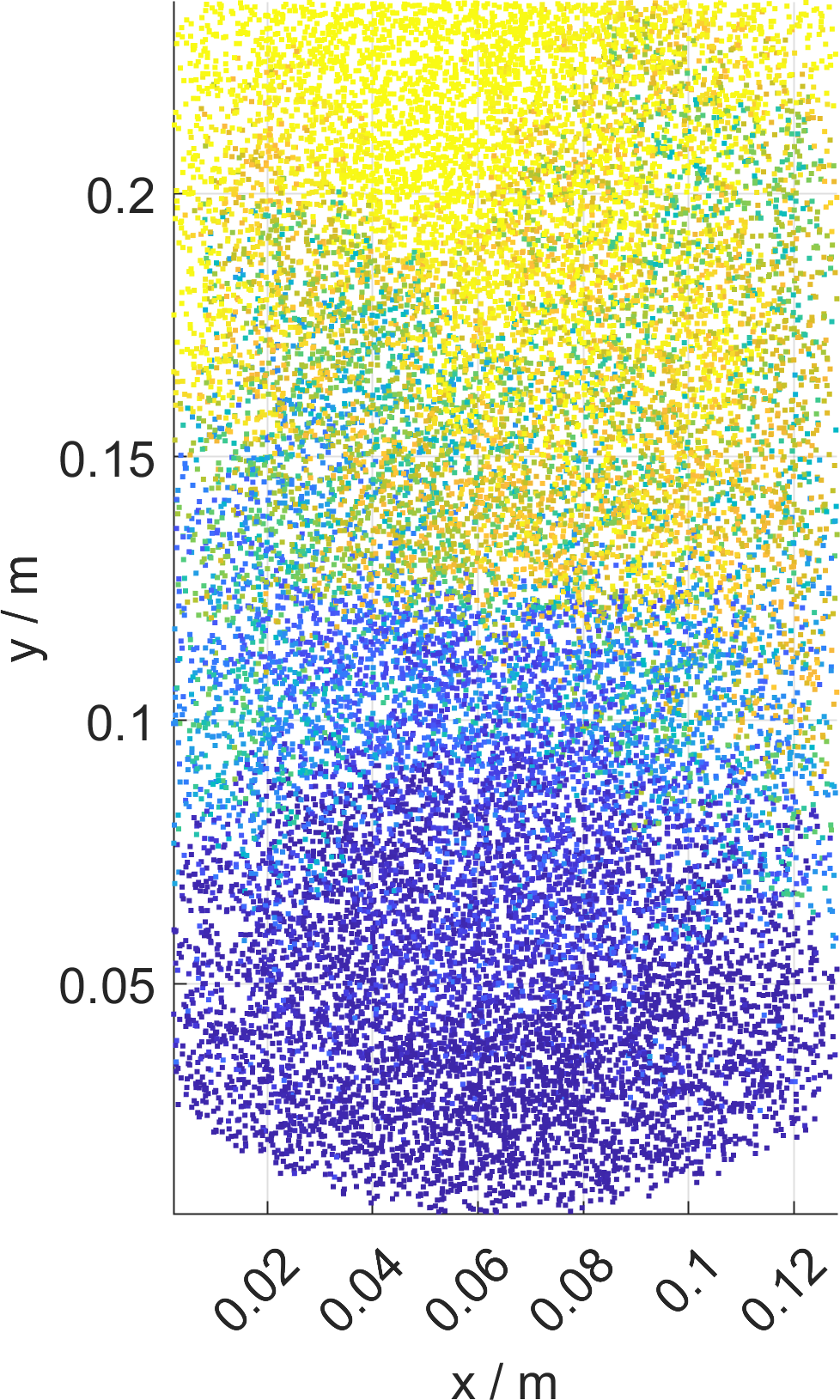}\\
\end{tabular}
\caption{Evolution of two differently colored fluids (top panel) for effective diffusion constants $D=2\cdot 10^{-9} \text{m}^2/\text{s}$ (second row) and  $D=1\cdot 10^{-5} \text{m}^2/\text{s}$ (third row), with the latter setting also applied to sparse trajectory data (bottom row).
Again the results after one, three and ten rotations of the stirrer are shown (cross-sectional view for $z=0$).}\label{fig:str_zk}
\end{figure}

Similarly to the previous examples, we study mixing of two differently colored fluid, where the yellow color is placed in the upper part of the reactor, above the median y-coordinate of all initial particles, and the blue color below, see Figure \ref{fig:str_zk} (top). Again the results after one, three and ten rotations of the stirrer are shown, for both the small effective diffusion constant  $D=2\cdot 10^{-9} \text{m}^2/\text{s}$, (second row) and the larger one (third row). For better visibility we plot the results on a two-dimensional plane, fixing $z=0$, i.e., cutting centrally through the stirrer shaft.
 The choice $D=1\cdot 10^{-5} \text{m}^2/\text{s}$ appears to be too small for the given data resolution and diffusion is suppressed since the trajectories are too far from each other to pass on the color concentration. For the larger diffusion constant one observes again regions at the top and bottom of the reactor that remain coherent and do not mix well with the surrounding fluid. Similar effects are also reported using a different analysis technique in \cite{Steuwe2022} and with mixing time experiments in the same reactor geometry \cite{Fitschen2021}. Experimental trajectory data from 4D particle tracking is sparse. In \cite{Steuwe2026} approximately $38,000$ tracers are observed at every time step. To mimic this experimental situation we sparsify the simulated STR data and consider only 45,002 trajectories. We repeat the mixing study for $D=1\cdot 10^{-5} \text{m}^2/\text{s}$ on this coarse data set. The results that are obtained in less than 70 seconds on a laptop, are shown in Figure \ref{fig:str_zk} (bottom row). They compare well to the complete-data case (third row), which is computationally much more involved, and demonstrate that the proposed framework is particularly suited for the practically relevant setting of undersampling.

Using the methods described in section \ref{sec:spacetimediff} we want to study the coherent behavior in more detail. We form the time-averaged matrix $\bm{Q}_{\epsilon}$ as in \eqref{eq:Qepsilon} by considering every third of the first 73 time slices $t_0, t_3, \ldots, t_{72}$ of the simulation, corresponding to approximately three rotations of the stirrer. In this specific step, we employ a smaller search radius $r=2\cdot \epsilon$ with $\epsilon=\sqrt{5\cdot10^{-6}}$. If we choose the same search radius $r$ and $\epsilon$ as before, $Q_{\epsilon}$ will be very large, but the resulting coherent sets look quite similar. We note that comparable coherent sets can also be obtained for the sparsified case of only 45,002 trajectories (not shown), where one could retain the original search radius and time stepping.

\begin{figure}[!ht]
\centering
\begin{tabular}{c}
\includegraphics[width=0.55\textwidth]{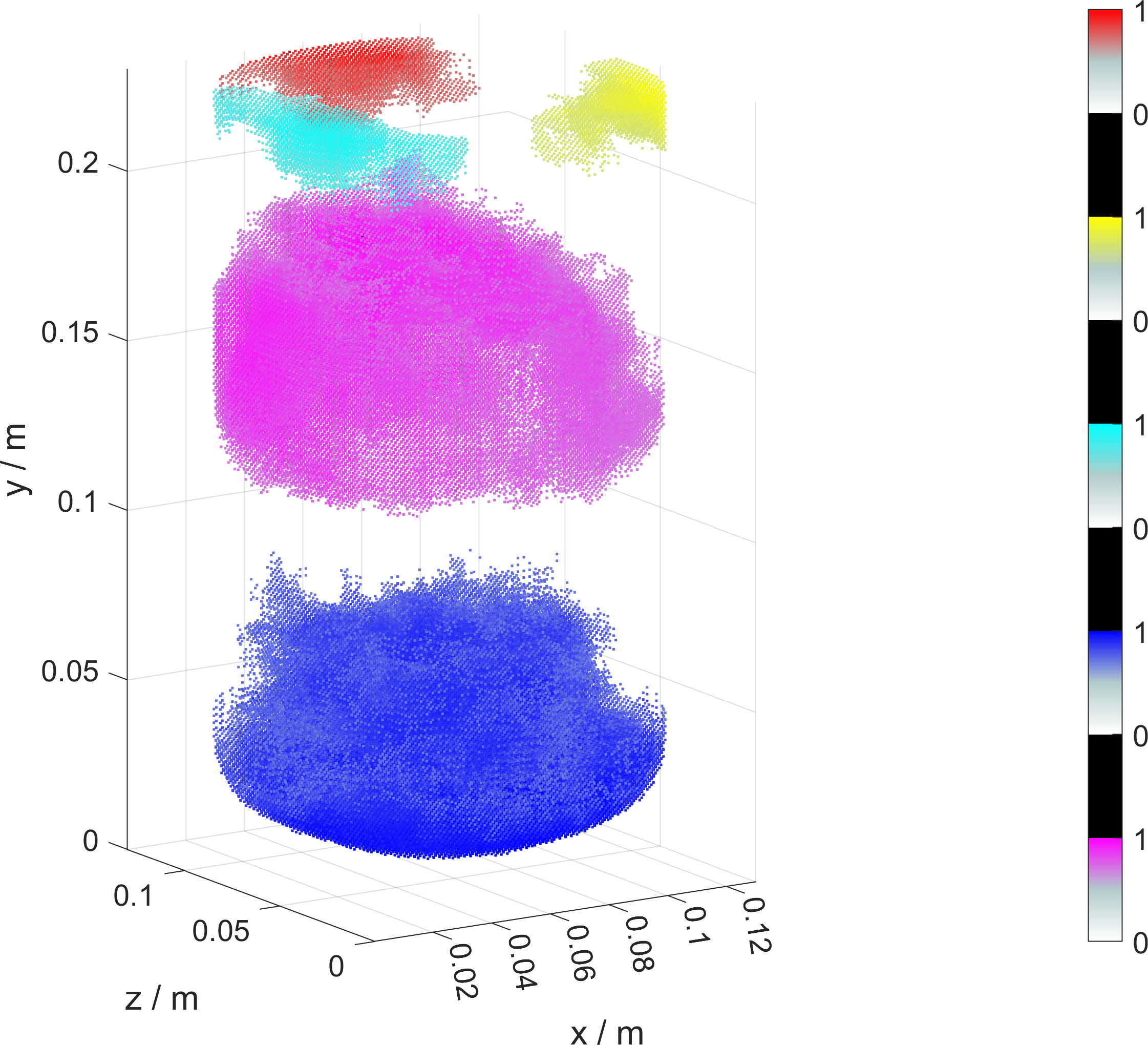}
\end{tabular}
\caption{Five sparse vectors highlighting the location of coherent sets shown in one plot. For better visibility only particles with membership values higher than 0.7 are plotted. }\label{fig:str_coh}
\end{figure}

There is a spectral gap after the fifth eigenvalue of $\bm{Q}_{\epsilon}$ and we extract five coherent sets based on the respective first five eigenvectors and a subsequent sparse eigenbasis approximation SEBA \cite{Froyland2019}. For each coherent set we obtain a sparse vector, where each entry gives the likelihood that the respective particle belongs to it for the time span under consideration, resulting in a soft-clustering of the data. We plot the particles and the respective vectors, neglecting particles that have a less than 70\% likelihood to belong to one of the sets, see Figure \ref{fig:str_coh}.  

\begin{figure}[!h]
\centering
\begin{tabular}{ccc}
\includegraphics[width=0.3\textwidth]{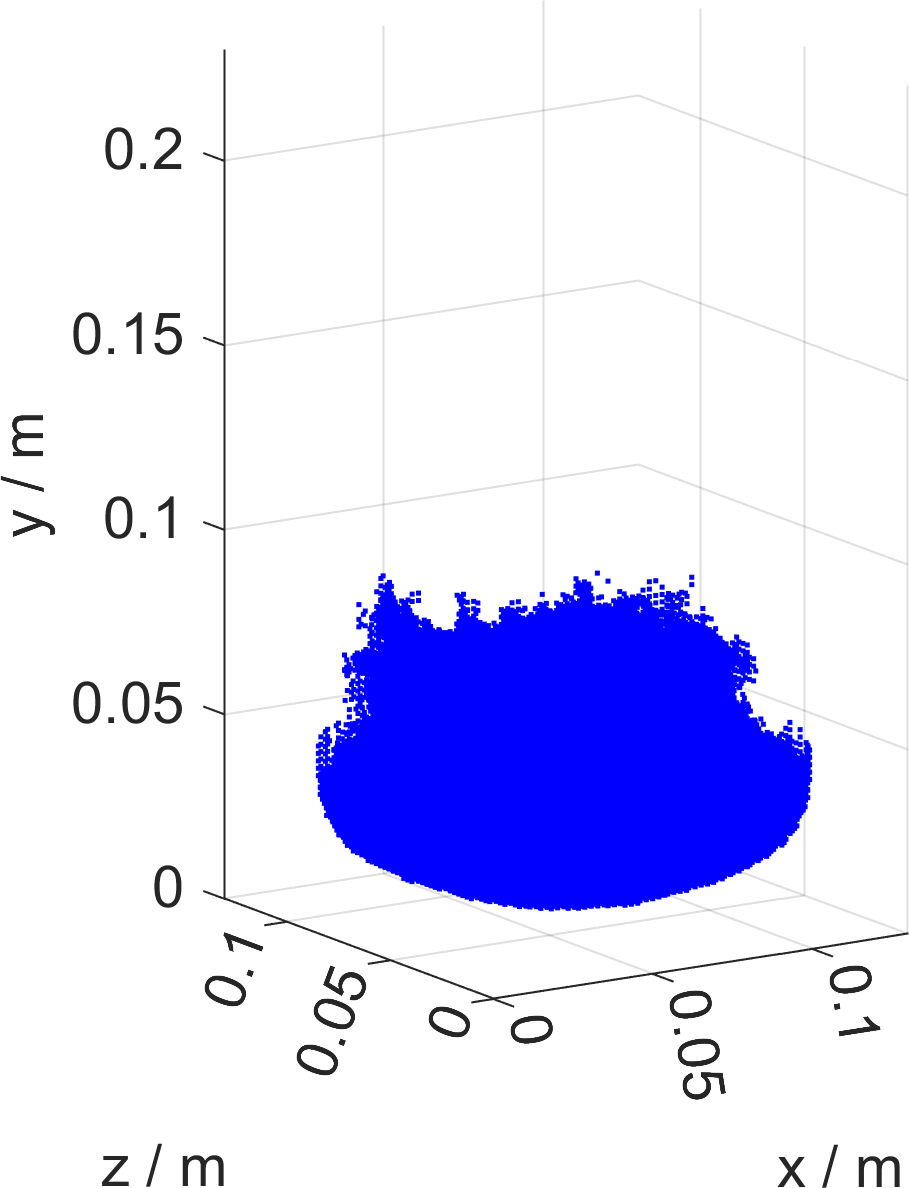}& 
\includegraphics[width=0.3\textwidth]{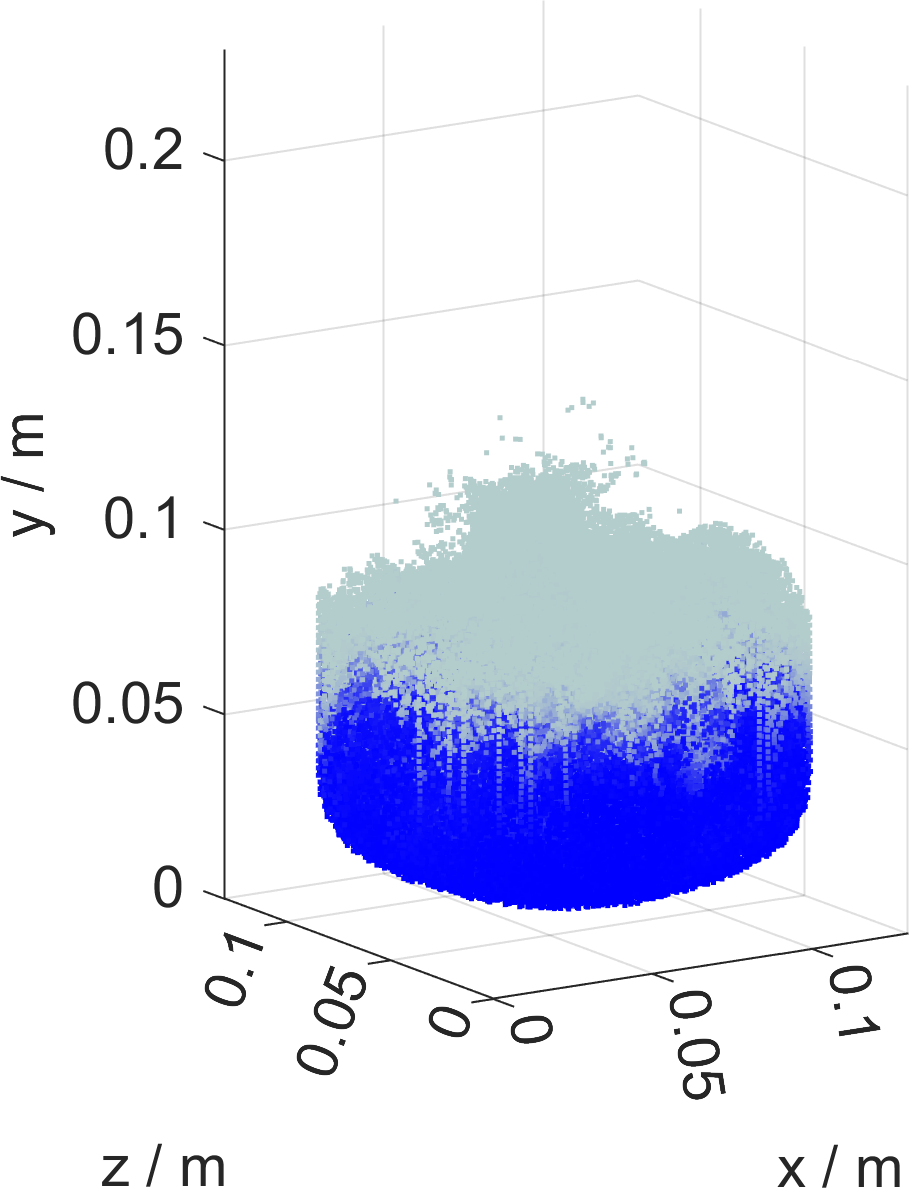}&
\includegraphics[width=0.3\textwidth]{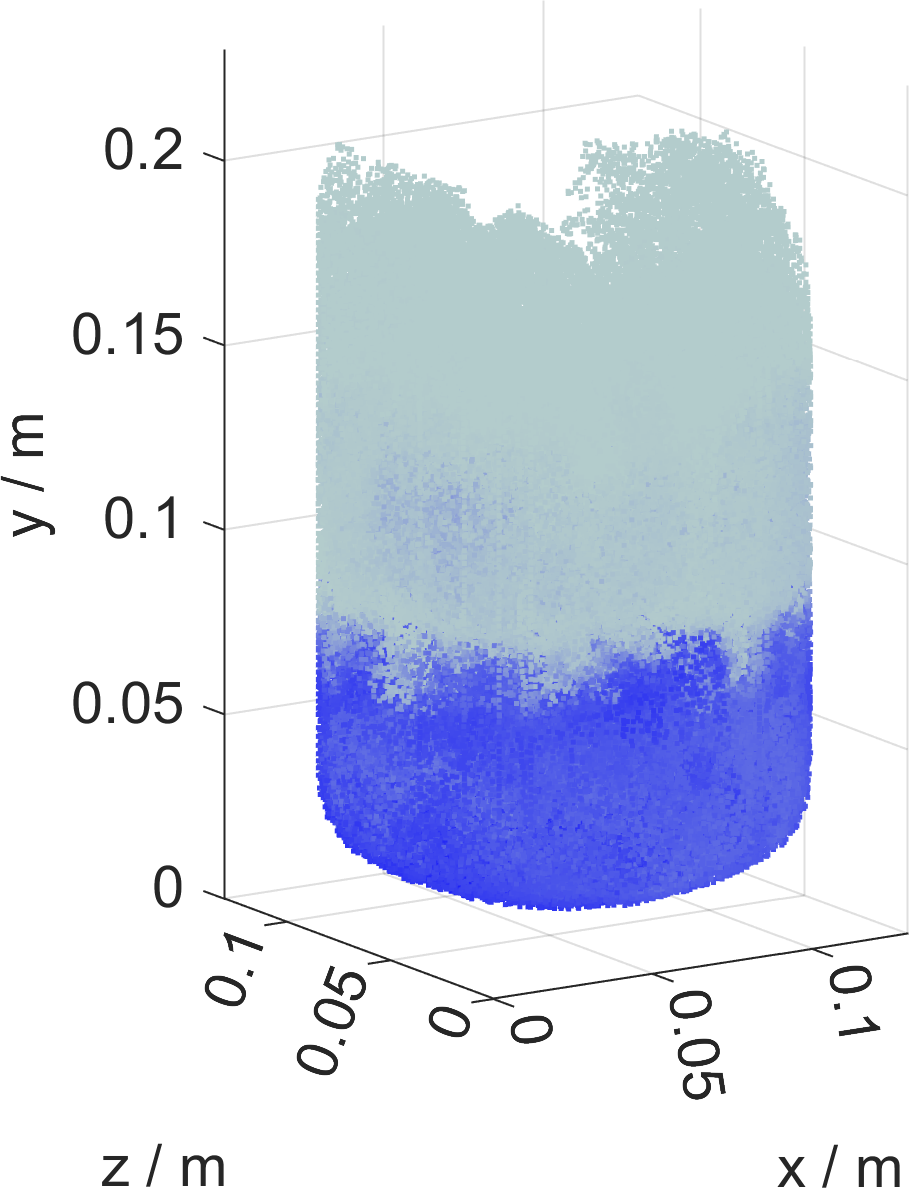}\\
&\multicolumn{2}{c}{\includegraphics[width=0.6\textwidth]{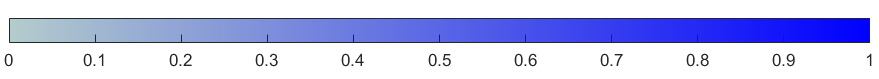}}
\end{tabular}
\caption{Evolution of the bottom coherent set (hard cluster, cut value 0.7) at initial time as well as at time slices closest to three and ten stirrer rotations for the choice $D=1\cdot 10^{-5} \text{m}^2/\text{s}$. Plotted are only particle positions $\bm{x}_i(t_k)$ for which $w_i^k>0.0001$. }\label{fig:str_bcoh}
\end{figure}

\begin{figure}[!ht]
\centering
\begin{tabular}{c}
\includegraphics[width=0.8\textwidth]{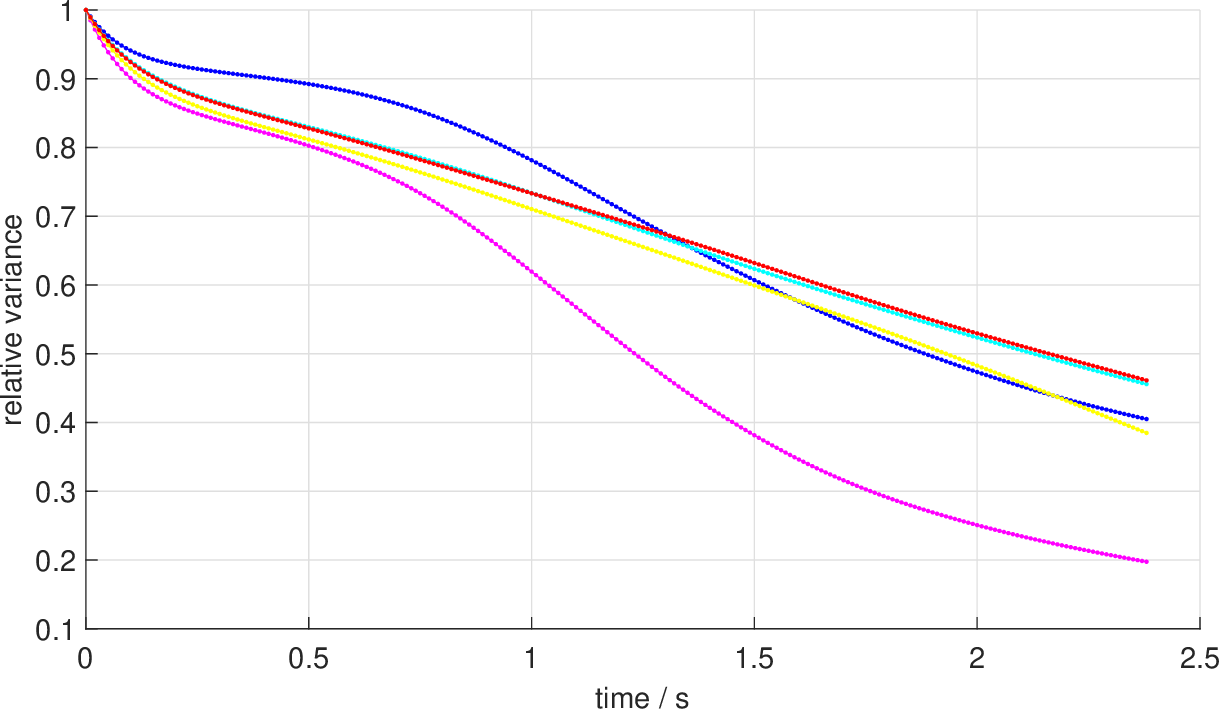}
\end{tabular}
\caption{Relative variances over time for coevolved vectors initialized in each of the five coherent sets (hard clusters, cut value 0.7) as shown in Figure \ref{fig:str_coh}.}\label{fig:str_coh_var}
\end{figure}

We study mixing when the dye is initialized in one of the coherent sets as shown in Figure \ref{fig:str_coh} (i.e.\ defined via the $0.7$ membership threshold). The vector $\bm{w}^0$ has entries $1$ in the respective coherent set and $0$ outside.
In Figure \ref{fig:str_bcoh} we show the results for the bottom coherent set. As this set has been computed for a time span of approximately three stirrer rotations, we observe that for this time horizon indeed there is very limited mixing with the surrounding fluid. After around ten stirrer rotations this picture has changed a bit, but still only low concentrations are observed outside the bottom part of the reactor.
 Finally, we quantify mixing over time for the five different coherent sets (Figure \ref{fig:str_coh_var}). As the sets have different volumes, we consider the relative variance (i.e.\ the time-dependent sample variances are divided by the respective initial variances). The upper and lower sets appear to be significantly more coherent than the central set (magenta), which is in accordance with our previous observations in Figure \ref{fig:str_b1}.

%%%%%%%%%%%%%%%%%%%%%%%%%%%%%%%%%%%%%%%%%%%%%%%%%%%%%%%%%%%%%%%%%%%%%%%%%%%%%%%%%%%%%

\section{Conclusion and outlook}\label{sec:conclusion}
We have introduced a computational framework inspired by deterministic particle methods and a space-time diffusion map approach to model the advective-diffusive evolution of scalar quantities in a purely data-based manner from Lagrangian particle trajectories. We have demonstrated the new approach in a number of two- and three-dimensional example flows of increasing complexity.
Comparisons with solutions of respective advection-diffusion equations showed very good agreement, even for low data resolution.

The main purpose of this work is to set the methodological foundations that allow us to study mixing resulting from advection and effective diffusion of a scalar quantity directly using given, possibly experimentally measured trajectories. Once the instantaneous transition matrices are computed, the proposed framework allows us to study many different initial conditions and to employ arbitrary mixing measures. Moreover, finite-time coherent sets can be identified by means of the transition matrices, providing the basis for investigating  the impact of such influential Lagrangian flow structures on scalar mixing. Unlike using pure dispersion measures, the concept of coherent sets can identify poor mixing between different fluid volumes even though each of these volumes moves and deforms as a whole.

While we have assumed a divergence-free flow for simplicity, the method itself is very robust as it is only applied to given tracer trajectories and does not use any properties of the underlying flow. In case of flows that are not divergence-free  tracers will no longer be uniformly distributed (not even approximately). The diffusive transport might be hindered by spatial gaps in data, a problem that can also occur for experimentally measured trajectories from time-resolved particle tracking (4D-PTV). One possible way out of this is to fill the gaps with artificial tracers in every time-step to enable a realistic diffusive spread. The values of the evolved scalar quantity for such additional tracers can be estimated using one of the approaches outlined in section \ref{sec:opensystem}.

In the future, we will apply the framework to experimental data from time-resolved particle tracking to study mixing in dependence of the initial conditions (i.e.\ the vector $\bm{w}^0$). The corresponding particle tracking experiments in a lab-scale stirred tank reactor are already finalized \cite{Steuwe2026}.
One important issue in this context is the estimation of an effective diffusion constant, which depends on the resolved scales. For this we have to evaluate the trajectory data with regard to the absolute dispersion or perform accompanying mixing time experiments to match our computational mixing studies. Moreover, we plan to consider more complex reactor geometries than that of the classical stirred tank reactor.

Another aspect will be to include chemical reactions. To this end, we will combine advection-diffusion with reactive dynamics by interaction of several coevolved vectors that represent the different reagent concentrations. In particular, we will study the impact of coherent flow structures on reaction yields.  

\subsection*{Acknowledgements}
The authors thank Christian Weiland, Yvonne Schade and Eike Steuwe for providing the trajectory data from a Lattice-Boltzmann simulation of the stirred tank reactor (see also \cite{weiland2023}), Thanh Tung Thai for producing Figure 9, and all of them for fruitful discussions. KPG further acknowledges insightful discussions with Gary Froyland.
This research is supported by the Deutsche Forschungsgemeinschaft (DFG, German Research Foundation) – CRC 1615 – 503850735.

\subsection*{Data availability}
The MATLAB code to reproduce the results is available at 
\url{https://gitlab.gwdg.de/anna.kluenker01/datamixing-paper}. \textbf{The repository will be
archived on Zenodo upon publication of this manuscript.} 
The code uses functions from the adcell package \cite{thiffeaultcode} and the SEBA algorithm \cite{Froyland2019}.
The stirred tank reactor data is available at \cite{weiland2026data}.

%%%%%%%%%%%%%%%%%%%%%%%%%%%%%%%%%%%%%


\begin{thebibliography}{10}

\bibitem{Allshouse2015}
Michael~R. Allshouse and Thomas Peacock.
\newblock Lagrangian based methods for coherent structure detection.
\newblock {\em Chaos}, 25(9):097617, 2015.

\bibitem{Haller2015}
George Haller.
\newblock Lagrangian coherent structures.
\newblock {\em Annual Review of Fluid Mechanics}, 47:137--162, 2015.

\bibitem{Hadjighasem2017}
Alireza Hadjighasem, Mohammad Farazmand, Daniel Blazevski, Gary Froyland, and
  George Haller.
\newblock A critical comparison of {Lagrangian} methods for coherent structure
  detection.
\newblock {\em Chaos}, 27(5):053104, 2017.

\bibitem{Badza2023}
Aleksandar Badza, Trent~W. Mattner, and Sanjeeva Balasuriya.
\newblock How sensitive are {Lagrangian} coherent structures to uncertainties
  in data?
\newblock {\em Physica D: Nonlinear Phenomena}, 444:133580, 2023.

\bibitem{Balasuriya2018}
Sanjeeva Balasuriya, Nicholas~T. Ouellette, and Irina~I. Rypina.
\newblock {Generalized Lagrangian coherent structures}.
\newblock {\em Physica D}, 372:31--51, 2018.

\bibitem{Huhn2012a}
Florian Huhn, Alexandra {von Kameke}, Silvia Allen-Perkins, Pedro Montero,
  Anabela Venancio, and Vicente Pérez-Muñuzuri.
\newblock {Horizontal Lagrangian transport in a tidal-driven estuary —
  Transport barriers attached to prominent coastal boundaries}.
\newblock {\em Continental Shelf Research}, 39-40:1--13, 2012.

\bibitem{Huhn2012b}
Florian Huhn, Alexandra von Kameke, Vicente Pérez-Muñuzuri, Maria~Josefina
  Olascoaga, and Francisco~Javier Beron-Vera.
\newblock {The impact of advective transport by the South Indian Ocean
  Countercurrent on the Madagascar plankton bloom}.
\newblock {\em Geophysical Research Letters}, 39(6), 2012.

\bibitem{Kelley2013}
Douglas~H. Kelley, Michael~R. Allshouse, and Nicholas~T. Ouellette.
\newblock Lagrangian coherent structures separate dynamically distinct regions
  in fluid flows.
\newblock {\em Phys. Rev. E}, 88:013017, Jul 2013.

\bibitem{Kameke2019}
Alexandra von Kameke, Sven Kastens, Sophie Rüttinger, Sonja Herres-Pawlis, and
  Michael Schlüter.
\newblock How coherent structures dominate the residence time in a bubble wake:
  An experimental example.
\newblock {\em Chemical Engineering Science}, 207:317--326, 2019.

\bibitem{Aksamit2024}
Nikolas~O. Aksamit, Alex~P. Encinas-Bartos, George Haller, and David~E. Rival.
\newblock Relative fluid stretching and rotation for sparse trajectory
  observations.
\newblock {\em Journal of Fluid Mechanics}, 996:A40, 2024.

\bibitem{FroylandPadberg2015}
Gary Froyland and Kathrin Padberg-Gehle.
\newblock A rough-and-ready cluster-based approach for extracting finite-time
  coherent sets from sparse and incomplete trajectory data.
\newblock {\em Chaos}, 25(8):087406, 2015.

\bibitem{weiland2023}
Christian Weiland, Eike Steuwe, Jürgen Fitschen, Marko Hoffmann, Michael
  Schlüter, Kathrin Padberg-Gehle, and Alexandra {von Kameke}.
\newblock {Computational study of three-dimensional Lagrangian transport and
  mixing in a stirred tank reactor}.
\newblock {\em Chemical Engineering Journal Advances}, 14:100448, 2023.

\bibitem{Schoeller2025}
Henry Schoeller, Robin Chemnitz, P\'eter Koltai, Maximilian Engel, and Stephan
  Pfahl.
\newblock Assessing {Lagrangian} coherence in atmospheric blocking.
\newblock {\em Nonlinear Processes in Geophysics}, 32(1):51--73, 2025.

\bibitem{Curbelo2023}
Jezabel Curbelo and Irina~I. Rypina.
\newblock A three dimensional {Lagrangian} analysis of the smoke plume from the
  2019/2020 {Australian} wildfire event.
\newblock {\em Journal of Geophysical Research: Atmospheres},
  128(21):e2023JD039773, 2023.

\bibitem{hadjighasem2016spectral}
Alireza Hadjighasem, Daniel Karrasch, Hiroshi Teramoto, and George Haller.
\newblock Spectral-clustering approach to {Lagrangian} vortex detection.
\newblock {\em Physical Review E}, 93(6):063107, 2016.

\bibitem{padberg2017network}
Kathrin Padberg-Gehle and Christiane Schneide.
\newblock Network-based study of {Lagrangian} transport and mixing.
\newblock {\em Nonlinear Processes in Geophysics}, 24(4):661--671, 2017.

\bibitem{schlueter2017}
Kristy~L. Schlueter-Kuck and John~O. Dabiri.
\newblock Coherent structure colouring: identification of coherent structures
  from sparse data using graph theory.
\newblock {\em J. Fluid Mech.}, 811:468--486, 2017.

\bibitem{froyland2018robust}
Gary Froyland and Oliver Junge.
\newblock Robust {FEM}-based extraction of finite-time coherent sets using
  scattered, sparse, and incomplete trajectories.
\newblock {\em SIAM Journal on Applied Dynamical Systems}, 17(2):1891--1924,
  2018.

\bibitem{Filippi2021}
Margaux Filippi, Irina~I. Rypina, Alireza Hadjighasem, and Thomas Peacock.
\newblock An optimized-parameter spectral clustering approach to coherent
  structure detection in geophysical flows.
\newblock {\em Fluids}, 6(1), 2021.

\bibitem{Schneide2022}
Christiane Schneide, Philipp~P. Vieweg, Jörg Schumacher, and Kathrin
  Padberg-Gehle.
\newblock {Evolutionary clustering of Lagrangian trajectories in turbulent
  Rayleigh–Bénard convection flows}.
\newblock {\em Chaos}, 32(1):013123, 2022.

\bibitem{Vieweg2024}
Philipp~P. Vieweg, Anna Klünker, Jörg Schumacher, and Kathrin Padberg-Gehle.
\newblock {Lagrangian studies of coherent sets and heat transport in constant
  heat flux-driven turbulent Rayleigh–Bénard convection}.
\newblock {\em European Journal of Mechanics - B/Fluids}, 103:69--85, 2024.

\bibitem{banisch2017understanding}
Ralf Banisch and P{\'e}ter Koltai.
\newblock Understanding the geometry of transport: diffusion maps for
  {Lagrangian} trajectory data unravel coherent sets.
\newblock {\em Chaos: An Interdisciplinary Journal of Nonlinear Science},
  27(3):035804, 2017.

\bibitem{coifman2006diffusion}
Ronald~R. Coifman and St{\'e}phane Lafon.
\newblock Diffusion maps.
\newblock {\em Applied and Computational Harmonic Analysis}, 21(1):5--30, 2006.

\bibitem{nadler2006diffusion}
Boaz Nadler, St{\'e}phane Lafon, Ronald~R. Coifman, and Ioannis~G. Kevrekidis.
\newblock Diffusion maps, spectral clustering and reaction coordinates of
  dynamical systems.
\newblock {\em Applied and Computational Harmonic Analysis}, 21(1):113--127,
  2006.

\bibitem{lafon2006diffusion}
Stephane Lafon and Ann~B. Lee.
\newblock Diffusion maps and coarse-graining: A unified framework for
  dimensionality reduction, graph partitioning, and data set parameterization.
\newblock {\em IEEE Transactions on Pattern Analysis and Machine Intelligence},
  28(9):1393--1403, 2006.

\bibitem{Froyland2015}
Gary Froyland.
\newblock Dynamic isoperimetry and the geometry of {Lagrangian} coherent
  structures.
\newblock {\em Nonlinearity}, 28(10):3587, 2015.

\bibitem{Vieweg2021}
Philipp~P. Vieweg, Christiane Schneide, Kathrin Padberg-Gehle, and J\"org
  Schumacher.
\newblock Lagrangian heat transport in turbulent three-dimensional convection.
\newblock {\em Phys. Rev. Fluids}, 6:L041501, Apr 2021.

\bibitem{Schanz2016}
Daniel Schanz, Sebastian Gesemann, and Andreas Schröder.
\newblock {Shake-The-Box: Lagrangian particle tracking at high particle image
  densities}.
\newblock {\em Exp Fluids}, 57:70, 2016.

\bibitem{Schroeder2023}
Andreas Schröder and Daniel Schanz.
\newblock {3D Lagrangian} particle tracking in fluid mechanics.
\newblock {\em Annual Review of Fluid Mechanics}, 55:511--540, 2023.

\bibitem{Tan2020}
Shiyong Tan, Ashwanth Salibindla, Ashik U.~M. Masuk, and Rui Ni.
\newblock {Introducing OpenLPT}: new method of removing ghost particles and
  high-concentration particle shadow tracking.
\newblock {\em Exp Fluids}, 61:47, 2020.

\bibitem{klunker2022open}
Anna Kl{\"u}nker, Kathrin Padberg-Gehle, and Jean-Luc Thiffeault.
\newblock Open-flow mixing and transfer operators.
\newblock {\em Philosophical Transactions of the Royal Society A},
  380(2225):20210028, 2022.

\bibitem{degond1989weighted}
Pierre Degond and Sylvie Mas-Gallic.
\newblock {The weighted particle method for convection-diffusion equations. I.
  The case of an isotropic viscosity}.
\newblock {\em Mathematics of Computation}, 53(188):485--507, 1989.

\bibitem{eldredge2002particles}
Jeff~D. Eldredge, Anthony Leonard, and Tim Colonius.
\newblock A general deterministic treatment of derivatives in particle methods.
\newblock {\em Journal of Computational Physics}, 180(2):686--709, 2002.

\bibitem{Chertock2017}
Alina Chertock.
\newblock Chapter 7 - a practical guide to deterministic particle methods.
\newblock In Rémi Abgrall and Chi-Wang Shu, editors, {\em Handbook of
  Numerical Methods for Hyperbolic Problems}, volume~18 of {\em Handbook of
  Numerical Analysis}, pages 177--202. Elsevier, 2017.

\bibitem{Neufeld2009}
Zoltán Neufeld and Emilio Hernández-García.
\newblock {\em Chemical and Biological Processes in Fluid Flows}.
\newblock Imperial College Press, 2009.

\bibitem{Bakunin2008}
Oleg~G. Bakunin.
\newblock {\em Turbulence and Diffusion: Scaling Versus Equations}.
\newblock Springer, Berlin, Heidelberg, 1 edition, 2008.

\bibitem{coifman2014changing}
Ronald~R. Coifman and Matthew~J. Hirn.
\newblock Diffusion maps for changing data.
\newblock {\em Applied and Computational Harmonic Analysis}, 36(1):79--107,
  2014.

\bibitem{belkin2003laplacian}
Mikhail Belkin and Partha Niyogi.
\newblock Laplacian eigenmaps for dimensionality reduction and data
  representation.
\newblock {\em Neural Computation}, 15(6):1373--1396, 2003.

\bibitem{vonLuxburg2007}
Ulrike von Luxburg.
\newblock A tutorial on spectral clustering.
\newblock {\em Statistics and Computing}, 17:396--416, 2007.

\bibitem{Froyland2019}
Gary Froyland, Christopher~P. Rock, and Konstantinos Sakellariou.
\newblock Sparse eigenbasis approximation: Multiple feature extraction across
  spatiotemporal scales with application to coherent set identification.
\newblock {\em Commun. Nonlinear Sci. Numer. Simul.}, 77:81--107, 2019.

\bibitem{mathew2005multiscale}
George Mathew, Igor Mezi{\'c}, and Linda Petzold.
\newblock A multiscale measure for mixing.
\newblock {\em Physica D: Nonlinear Phenomena}, 211(1-2):23--46, 2005.

\bibitem{Thiffeault2012}
Jean-Luc Thiffeault.
\newblock Using multiscale norms to quantify mixing and transport.
\newblock {\em Nonlinearity}, 25(2):R1, 2012.

\bibitem{Banisch2019}
Ralf Banisch, P\'eter Koltai, and Kathrin Padberg-Gehle.
\newblock Network measures of mixing.
\newblock {\em Chaos}, 29(6):063125, 2019.

\bibitem{Taylor1923}
Geoffrey~I Taylor.
\newblock {LXXV. On the decay of vortices in a viscous fluid}.
\newblock {\em The London, Edinburgh, and Dublin Philosophical Magazine and
  Journal of Science}, 46(274):671--674, 1923.

\bibitem{thiffeaultcode}
Jean-Luc Thiffeault.
\newblock adcell: {MATLAB} code for solving the advection-diffusion equation
  for a two-dimensional incompressible autonomous cellular flow.
\newblock https://github.com/jeanluct/adcell.

\bibitem{shadden2005definition}
Shawn~C. Shadden, Francois Lekien, and Jerrold~E. Marsden.
\newblock Definition and properties of {Lagrangian} coherent structures from
  finite-time {Lyapunov} exponents in two-dimensional aperiodic flows.
\newblock {\em Physica D: Nonlinear Phenomena}, 212(3-4):271--304, 2005.

\bibitem{Hofmann2022}
Sebastian Hofmann, Christian Weiland, Jürgen Fitschen, Alexandra {von Kameke},
  Marko Hoffmann, and Michael Schlüter.
\newblock {Lagrangian sensors in a stirred tank reactor: Comparing trajectories
  from 4D-Particle Tracking Velocimetry and Lattice-Boltzmann simulations}.
\newblock {\em Chemical Engineering Journal}, 449:137549, 2022.

\bibitem{Kuschel2021}
Maike Kuschel, Jürgen Fitschen, Marko Hoffmann, Alexandra von Kameke, Michael
  Schlüter, and Thomas Thmas~Wucherpfennig.
\newblock Validation of novel {Lattice Boltzmann} large eddy simulations {(LB
  LES)} for equipment characterization in biopharma.
\newblock {\em Processes}, 9(6):950, 2021.

\bibitem{Steuwe2022}
Eike Steuwe.
\newblock Investigation of transport barriers within a continuous stirred tank
  reactor using three-dimensional {Lagrangian} coherent structures.
\newblock Master's thesis, TU Hamburg, 2022.

\bibitem{Fitschen2021}
Jürgen Fitschen, Sebastian Hofmann, Johannes Wutz, Alexandra von Kameke, Marko
  Hoffmann, Thomas Wucherpfennig, and Michael Schlüter.
\newblock Novel evaluation method to determine the local mixing time
  distribution in stirred tank reactors.
\newblock {\em Chemical Engineering Science: X}, 10:100098, 2021.

\bibitem{Steuwe2026}
Eike Steuwe, Thanh~Tung Thai, Christian Weiland, Anna Kl\"unker, Jan~H. Nissen,
  Kathrin Padberg-Gehle, and Alexandra von Kameke.
\newblock Mixing analysis in a baffled stirred tank reactor based on
  {4D}-particle tracking experiments, 2026.
\newblock Preprint.

\bibitem{weiland2026data}
Christian Weiland and Yvonne Schade.
\newblock Simulated {L}agrangian trajectory data in a lab-scaled stirred tank
  reactor, 2026.
\newblock TUHH Open Research (TORE), https://doi.org/10.15480/882.16844.

\end{thebibliography}
\end{document}